\documentclass[journal,onecolumn]{IEEEtran}
\usepackage{cite}
\ifCLASSINFOpdf
  \usepackage[pdftex]{graphicx}
  \graphicspath{{Figure/}}
  \DeclareGraphicsExtensions{.pdf,.jpg,.png}
\else
  \usepackage[dvips]{graphicx}
  \graphicspath{{Figure}}
  \DeclareGraphicsExtensions{.eps}
\fi
\usepackage{amsmath}
\usepackage{array}
\usepackage{dblfloatfix}
\usepackage{url}
\newcommand{\pushright}[1]{\ifmeasuring@#1\else\omit\hfill$\displaystyle#1$\fi\ignorespaces}
\makeatother
\newcommand {\ie} {{\em i.e., }}
\newcommand {\eg} {{\em e.g., }}
\newcommand {\beq} {\begin{equation}}
\newcommand {\eeq} {\end{equation}}
\newcommand {\bequn} {\begin{equation*}}
\newcommand {\eequn} {\end{equation*}}
\newcommand {\bear} {\begin{eqnarray}}
\newcommand {\eear} {\end{eqnarray}}
\newcommand {\bearun} {\begin{eqnarray*}}
\newcommand {\eearun} {\end{eqnarray*}}
\newcommand {\fig}[1]{Fig.~\ref{#1}}

\newcommand {\Eqref}[1]{Eq.~(\ref{#1})}
\newcommand{\Eqrefs}[2]{Eqs.~(\ref{#1}) and~(\ref{#2})}

\usepackage[utf8]{inputenc}
\usepackage{amsfonts}
\usepackage{amssymb}
\usepackage{amsthm}
\usepackage{bm}
\usepackage{comment}
\usepackage{paralist}
\usepackage{enumitem}
\usepackage{color}

\usepackage{breakurl}
\usepackage[normalem]{ulem}
\usepackage{longtable}
\usepackage{subcaption}
\usepackage{algorithmic}
\usepackage{multirow}
\usepackage{pifont}

\usepackage{lipsum}
\usepackage{mathtools}
\usepackage{cuted}

\newtheorem{theorem}{Theorem}
\newtheorem{lemma}[theorem]{Lemma}

\newtheorem{proposition}[theorem]{Proposition}

\newtheorem{corollary}[theorem]{Corollary}

\makeatletter
\newcommand{\vast}{\bBigg@{3}}
\newcommand{\Vast}{\bBigg@{4}}
\makeatother

\newcommand{\useColor}{clean}  % Removes revision formatting
\usepackage{xstring}
\usepackage{soul,xcolor} 
\setstcolor{red}

\newcommand{\highlight}[2]{%
    \IfEqCase{#1}{%
        {color}{\textcolor{blue}{#2}}%
        {clean}{#2}%
    }[\PackageError{highlight}{Undefined option to highlight: #1}{}]%
}%

\newcommand{\strikenOut}[2]{%
    \IfEqCase{#1}{%
        {color}{\st{#2}}%
        {clean}{}%
    }[\PackageError{strikenOut}{Undefined option to strikenOut: #1}{}]%
}%

\newcommand{\onlineVersion}[2]{%
    \IfEqCase{#1}{%
        {notonline}{\textcolor{black}{#2}}%
        {clean}{}%
    }[\PackageError{onlineVersion}{Undefined option to onlineVersion: #1}{}]%
}%

\begin{document}
%\title{Minimum Bandwidth Solutions for Link-level Latency Constraints}
\title{On the Benefits of Traffic ``Reprofiling'' \\
The Single Hop Case}

\author{Jiayi~Song,~\IEEEmembership{}
        Jiaming~Qiu,~\IEEEmembership{Member,~IEEE,}
        Roch~Gu\'{e}rin,~\IEEEmembership{Fellow,~ACM,~IEEE,}
        and~Henry~Sariowan,~\IEEEmembership{Member,~IEEE}% <-this % stops a space
\thanks{J. Qiu and R. Gu\'{e}rin are with the Computer Science and Engineering department at Washington University in St.~Louis, Saint Louis, MO 63130, USA, e-mail: {\tt \{qiujiaming,guerin\}@wustl.edu}.}% <-this % stops a space
\thanks{J. Song was with the Computer Science and Engineering department at Washington University in St.~Louis and is now with ByteDance, Mountain View, CA 94041, USA, e-mail: {\tt jiayisong@wustl.edu}.}% <-this % stops a space
\thanks{H. Sariowan is with Google, Mountain View, CA 94043, USA, e-mail {\tt hsariowan@google.com}.}% <-this % stops a space
\thanks{An early version of the paper was presented at the 33rd International Teletraffic Congress (ITC'33) in September 2021.}% <-this % stops a space
\thanks{The work was supported by NSF grant CNS~2006530 and a gift from Google.}
}

\maketitle

\begin{abstract}
The need to guarantee hard delay bounds to traffic flows with deterministic traffic profiles, \eg token buckets, arises in a number of network settings.  Of interest are solutions that offer such guarantees while minimizing network bandwidth.  The paper explores a basic building block towards realizing such solutions, namely, a single hop configuration. The main results are in the form of optimal solutions for meeting local deadlines under schedulers of varying complexity and therefore cost.  The results demonstrate how judiciously modifying flows' traffic profiles, \ie \emph{reprofiling} them, can help simple schedulers reduce the bandwidth they require, often performing nearly as well as more complex ones.

\bigskip

\hspace{-1.3cm}\fbox{%
	\parbox{\textwidth}%
	 {\textcopyright 2024 IEEE.  Personal use of this material is permitted.  Permission from IEEE must be obtained for all other uses, in any current or future media, including reprinting/republishing this material for advertising or promotional purposes, creating new collective works, for resale or redistribution to servers or lists, or reuse of any copyrighted component of this work in other works.
	}%
}
\end{abstract}


% \fi
\IEEEpeerreviewmaketitle

\section{Introduction}\label{sec:intro}

The provision of deterministic delay guarantees to traffic flows is emerging as an important requirement in increasingly diverse settings.  They include automotive, avionics, and manufacturing applications, smart grids, and datacenters~\cite{automotive21,afdx,factory20,smartgrid23,aws22,google-netw21,msft21}.  This is reflected in standards such as Time Sensitive Networking (TSN) and Deterministic Networking (DetNet)~\cite{tsn18,parsons22,seol21,detnet} and in the Service Level Objectives/Agreements (SLOs/SLAs)\cite{sre16} of many service provider networks that are increasingly including latency targets, motivated in part by the rapid growth of edge computing offerings~\cite{sla22}.

In such settings, the traffic eligible for latency guarantees is commonly controlled using a traffic regulator~\cite{leboudec18} in the form of a token bucket $(r,b)$ that limits both the flow's long-term rate, $r$, and burstiness, $b$. A flow's token bucket parameters are typically determined using traces, and selected to ensure zero access delay~\cite{zhu17}.  The network's goal is then to ensure that  the latency guarantees of all such rate-controlled flows are met, preferably with as little bandwidth as possible.

This is the environment this paper assumes, with a focus on a basic building block, namely, delivering latency guarantees on a \emph{single link} (hop) with \emph{the least amount of bandwidth}. The answer obviously depends on the type of scheduler controlling access to the link, and the paper considers schedulers of different levels of complexity.  Of greater interest is whether, what the paper terms \emph{reprofiling}, can be beneficial. Reprofiling amounts to modifying a flow's original \highlight{\useColor}{(chosen by the user)} token bucket parameters \highlight{\useColor}{to make the flow ``easier'' to accommodate.  This concept was explored in WorkloadCompactor~\cite{zhu17} with one important difference, namely, the constraint that reprofiling should not introduce any delay.  In contrast, our reprofiling solutions impose an added delay in exchange for smoother flows. This in turn calls for tighter network latency bounds to ensure that the original delay targets are still met. The outcome of this trade-off depends on the level of reprofiling applied as well as the type of scheduler in use. Investigating when and how it is positive in a single hop setting is the focus of this paper. }

Specifically, the paper considers reprofiling of the form $(r,b)\xrightarrow{\text{reprofiling}}(r,b')$, where $b'\leq b$.  In other words, we reduce the flow's burstiness to make it easier to handle.  We note that more complex reprofiling solutions are possible.  Our motivations for focusing on burst reduction are two-fold.  First, we want to minimize any added complexity, and this reprofiling can be realized simply by modifying the burst parameter of the \emph{existing} token bucket.  Second, As shown in~Appendix~F, in simple configurations involving only two flows and a static priority scheduler, adjusting the burst size is sufficient to minimize the required bandwidth.  These motivations notwithstanding, more complex reprofilers, \eg adding a second token bucket that controls the peak rate, can be of benefit in more general settings.  We explore this extension in~\cite{multiple_nodes21} in the multiple hops setting.  

We note that the notion of reprofiling is closely tied to the definition of \emph{greedy shapers} of~\cite[Section 1.5]{nc}, with one important difference.  Specifically, depending on the scheduler, a reprofiler can be either non-work-conserving, \ie as a (greedy) shaper, or work-conserving.  The latter is only applicable when relying on dynamic priority schedulers such as earliest deadline first (edf) that can combine the local link deadline and the reprofiling delay when determining the order in which to send packets. 

The paper makes the following contributions when it comes to meeting latency targets in the single-hop case with traffic profiles in the form of token buckets:
\begin{itemize}[nosep]
\item Characterize the optimal (minimum bandwidth) solution, and show that a dynamic priority (edf) scheduler can realize it. The solution readily establishes that reprofiling yields no benefit with such a scheduler.
\item Identify optimal reprofiling solutions for static priority and fifo schedulers, and demonstrate how they allow those schedulers to closely approximate the performance of the more complex edf scheduler across a range of scenarios.
\end{itemize}

For ease of exposition, the results are derived and presented assuming a fluid model, which, therefore, implies a preemptive behavior.  As the discussion of~\cite[Section~1.1.1]{nc} highlights, extending the results to a packet setting is readily achievable from standard network calculus results.  For illustration purposes, Appendix~\ref{app:one_2flow} derives a solution for a static priority scheduler under a packet-based model, but the results do not contribute further insight.

The paper is structured as follows.  Section~\ref{sec:model} introduces our traffic model and optimization framework.  The next three sections present optimal solutions for schedulers of different complexity. Section~\ref{sec:one_dynamic} considers a general, dynamic priority scheduler, while Sections~\ref{sec:one_static} and~\ref{sec:one_fifo} assume simpler static priority and fifo schedulers. For the latter two, the benefits of reprofiling flows are also explored.  Section~\ref{sec:evaluation} quantifies performance for each scheduler, starting with two-flow configurations that help build intuition for the results, before considering more general multi-flow scenarios. Section~\ref{sec:related} reviews related works, while  Section~\ref{sec:conclusion} summarizes the paper's findings and their relevance to the multi-hop extension of~\cite{multiple_nodes21}.  Proofs and ancillary results are relegated to appendices.
\section{Model Formulation}\label{sec:model}

\begin{figure}[h]
%\vspace{-0.5cm}
\center\includegraphics[height = 5cm]{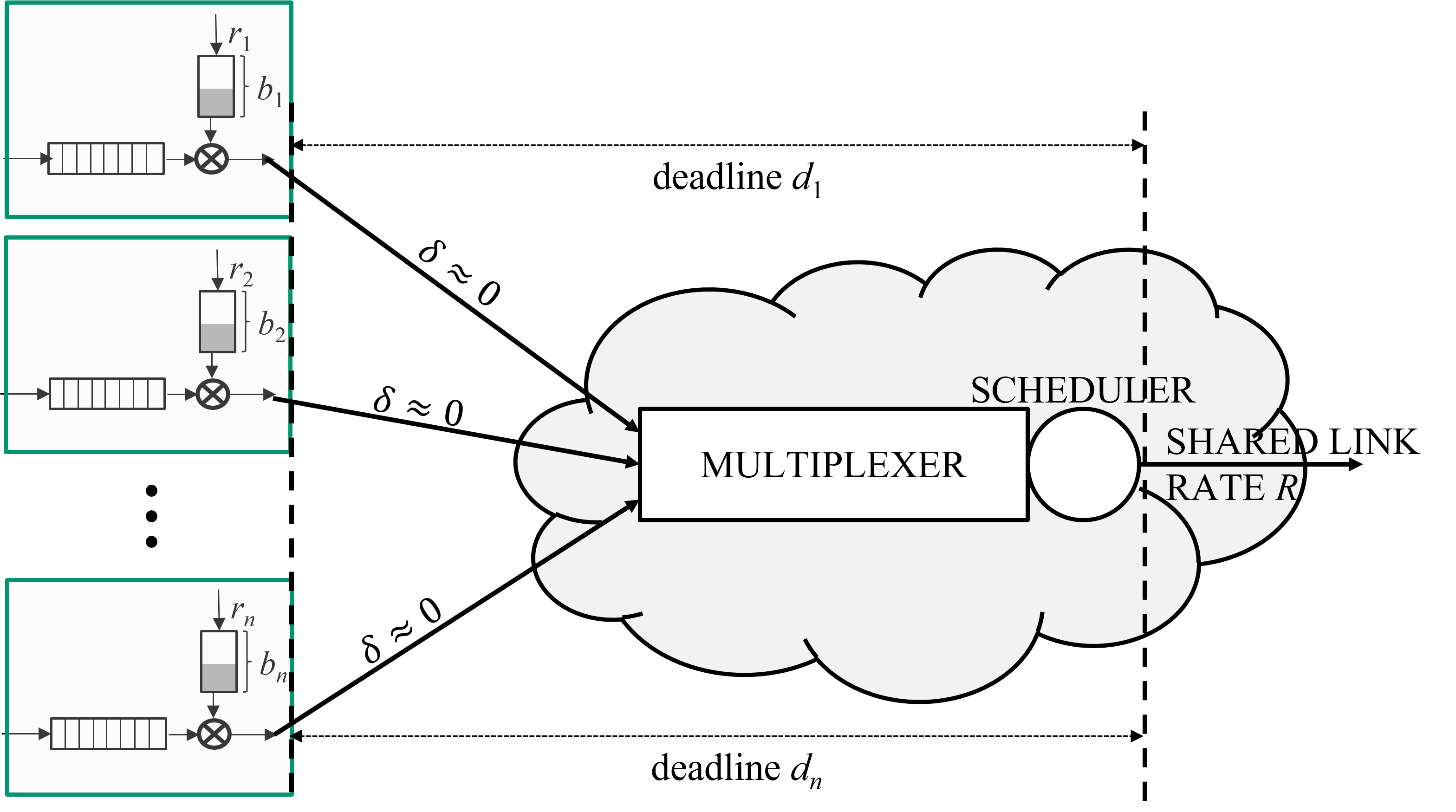}
\caption{A typical one-hop configuration with $n$ flows.}
\label{fig:n_flow} 
\end{figure}

Consider the configuration of \fig{fig:n_flow} with $n$~flows sharing a common link\footnote{For simplicity, we assume that enough buffering is available and that the link capacity is such that the system is stable and lossless.} of rate $R$. The traffic generated by flow~$i$ is rate-controlled using a two-parameter token bucket $(r_i, b_i)$~\cite{leboudec18}, its \emph{traffic profile}, where $r_i$ is the token rate and $b_i$ the bucket size. Flow~$i$ also has a local packet-level deadline $d_i$, where w.l.o.g. we assume $d_1 > d_2 > \ldots > d_n$ with $d_1<\infty$. Our goal is to meet the deadlines of all $n$ flows with the lowest possible link bandwidth~$R$.  In doing so, we further assume \emph{greedy sources}~\cite[Proposition 1.2.5]{nc} that fully realize the arrival curve associated with their token bucket.

In this setting,
let $\boldsymbol{r} =\left(r_1, r_2, \ldots, r_n \right)$, $\boldsymbol{b} =\left(b_1, b_2, \ldots, b_n \right)$, and $\boldsymbol{d} =\left(d_1, d_2, \ldots, d_n \right)$ be the vectors of rates, burst sizes, and deadlines of the flows sharing the link, respectively, and let $D^*_i(\boldsymbol{r}, \boldsymbol{b}, R)$ denote flow $i$'s \emph{worst-case} delay (queueing+transmission). Our bandwidth minimization problem can then be formulated as an optimization of the form:
\begin{equation*}
%\label{opt}
\begin{aligned}
&\text{\bf{OPT\_\ding{111}}} && \min R &\\
& \text{s.t}
& & D^*_i(\boldsymbol{r}, \boldsymbol{b}, R) \leq d_i, & \forall i, 1 \leq i \leq n  \\
\end{aligned}
\end{equation*}
where $R$ is the optimization variable and \ding{111} denotes the scheduler type, which for notational simplicity has been omitted in the expression of $D^*_i(\boldsymbol{r}, \boldsymbol{b}, R)$. As mentioned in Section~\ref{sec:intro}, one of our goals is to evaluate the trade-off between (bandwidth) efficiency and complexity across different schedulers.  

Another goal is to investigate the potential benefits of \emph{reprofiling} flows prior to forwarding their traffic to the scheduler.  Reprofiling amounts to applying a different, typically ``smaller'', traffic profile to each flow before forwarding them to the scheduler.  This can\footnote{When the reprofiler operates in a non-work-conserving manner.} introduce an up-front reprofiling delay, but may lower the bandwidth required to meet overall latency goals if it makes flows ``easier'' to handle.

More formally, given a scheduler ``\textbf{\ding{111}}'' and $n$ flows sharing a link, where flow~$i, 1\leq i\leq n,$ has traffic profile $(r_i,b_i)$ and deadline $d_i$, the goal of reprofiling is to identify smaller burst sizes $b'_i\leq b_i, 1\leq i\leq n,$ that minimize the link bandwidth $R$ needed to meet the flows' deadlines, inclusive of any resulting reprofiling delay (the smaller burst $b'_i$ introduces a reprofiling delay of $\frac{b_i-b'_i}{r_i}$).  We note that we restrict reprofiling options to only reducing the burst size, rather than also considering adding a ``peak rate'' shaper. This is in part to simplify the resulting optimization \textbf{OPT\_R\ding{111}}, and also because, as shown in Appendix~\ref{app:one_2flow}, this is sufficient in simple configurations with only two flows.  This translates into a modified optimization problem \textbf{OPT\_R\ding{111}} of the form  
\begin{equation*}
%\label{opt}
\begin{aligned}
&\text{\bf{OPT\_R\ding{111}}} && \min_{
%\strikenOut{\useColor}{(r',b')}
\bm{b'} } R &\\
& \text{s.t}
& & D^*_i(\bm{r}, \bm{b},\bm{b'}, R) \leq d_i, & \forall i, 1 \leq i \leq n  \\
\end{aligned}
\end{equation*}
where $R$ and $\bm{b'}$ are the optimization variables.  The latter denotes the vectors of updated (reprofiled) burst sizes of the $n$ flows, and $D^*_i(\bm{r}, \bm{b},\bm{b'}, R), 1\leq i\leq n,$ are the worst case delays, accounting for reprofiling delays, of the $n$ flows under scheduler \textbf{\ding{111}} and a link bandwidth of $R$. The optimization explores the extent to which making flows smoother (smaller bursts) can facilitate meeting their delay targets with less bandwidth in spite of the access delay that reprofiling adds.

The next three sections explore solutions to \textbf{OPT\_\ding{111}} (and \textbf{OPT\_R\ding{111}}) for different schedulers, namely, dynamic priority, static priority, and fifo (\textbf{OPT\_DP, OPT\_SP}, and \textbf{OPT\_F}).

\section{Dynamic Priorities}\label{sec:one_dynamic}
We start with the most powerful but most complex scheduler, dynamic priorities, with priorities derived from service curves assigned to flows as a function of their profile (deadline and traffic envelope). We first solve \textbf{OPT\_DP} by characterizing the service curves that achieve the lowest bandwidth while meeting all deadlines in the absence of any reprofiling.  

To derive the result, we first specify a service-curve assignment $\Gamma_{sc}$ that satisfies all deadlines, identify the minimum link bandwidth $R^*$ required to realize $\Gamma_{sc}$, and show that any scheduler requires at least $R^*$. We then show that an  earliest deadline first scheduler realizes $\Gamma_{sc}$ and, therefore, meets all the flow deadlines under $R^*$. e note that this then implies that reprofiling is of no benefit when an edf scheduler is available.

\begin{proposition}\label{prop:shifted_ac}
Consider a link shared by $n$ token bucket controlled flows, where flow $i, 1 \leq i \leq n$, has a traffic profile $(r_i, b_i)$ and a deadline $d_i$, with $d_1 > d_2 > ... > d_n$ and $d_1<\infty$. Consider a service-curve assignment $\Gamma_{sc}$ that allocates flow $i$ a service curve of 
\beq\label{eq:sc}
SC_i(t) = \left\{
\begin{aligned}
& 0 &\text{ when } t < d_i, \\
& b_i + r_i(t - d_i) &\text{otherwise}.
\end{aligned}
\right.
\eeq
Then 
\begin{enumerate}
\item For any flow $i, 1 \leq i \leq n$, $SC_i(t)$ ensures a worst-case end-to-end delay no larger than $d_i$.
\item Realizing $\Gamma_{sc}$ requires a link bandwidth of at least
\beq\label{eq:global_min_R}
R^* = \max_{1 \leq h \leq n}\left\{\sum_{i = 1}^n r_i, \frac{\sum_{i = h}^n b_i + r_i (d_h - d_i)}{d_h} \right\}.
\eeq
\item Any scheduling mechanism capable of meeting all the flows' deadlines requires a bandwidth of at least $R^*$.
\end{enumerate}
\end{proposition}

The proof of Proposition~\ref{prop:shifted_ac} is in Appendix~\ref{app:one_hop_dynamic}. The optimality of $\Gamma_{sc}$ is intuitive. Recall that a service curve is a lower bound on the service received by a flow.  \Eqref{eq:sc} assigns service to a flow at a rate exactly equal to its input rate, but delayed by its deadline, \ie provided at the latest possible time. 
%% Note that if the server transfers data only at its deadline, it will miss no flow's deadline, and by time $t$ it will transfer cumulatively $SC_i(t)$ amount of data for flow $1 \leq i \leq n$. 
Conversely, any mechanism $\widehat{\Gamma}$ that meets all flows' deadlines must by time $t$ have provided flow~$i$ a cumulative service at least equal to the amount of data that flow~$i$ may have generated by time $t-d_i$, which is exactly $SC_i(t)$.  Hence the mechanism must offer flow~$i$ a service curve $\widehat{SC}_i(t) \geq SC_i(t), \forall t$. 
%% Thus, realizing $\hat{\Gamma}$ requires a bandwidth greater than $R^*$.

Next, we identify at least one mechanism capable of realizing the services curves of \Eqref{eq:sc} under $R^*$, and consequently providing a solution to \textbf{OPT\_DP} for schedulers that support dynamic priorities.
\begin{proposition}\label{prop:edf}
Consider a link shared by $n$ token bucket controlled flows, where flow $i, 1 \leq i \leq n$, has traffic profile $(r_i, b_i)$ and deadline $d_i$, with $d_1 > d_2 > ... > d_n$ and $d_1<\infty$. The earliest deadline first (edf) scheduler realizes $\Gamma_{sc}$ under a link bandwidth of $R^*$.
\end{proposition}

The proof of Proposition~\ref{prop:edf} is in Appendix~\ref{app:edf}.
We note that the optimality of edf is intuitive, as minimizing the required bandwidth is the dual problem to maximizing the schedulable region for which edf's optimality is known~\cite{Georgiadis97}. 

As previously mentioned and as the next proposition formally states, reprofiling does not reduce the minimum required bandwidth $R^*$ of \Eqref{eq:global_min_R}.  Consequently, it affords no benefits with edf schedulers capable of meeting the deadlines under $R^*$.  This is expected given the optimality of edf schedulers.

\begin{proposition}\label{prop:sc_noshaper}
Consider a link shared by $n$ token bucket controlled flows, where flow $i, 1 \leq i \leq n$, has traffic profile $(r_i, b_i)$ and deadline $d_i$, with $d_1 > d_2 > ... > d_n$ and $d_1<\infty$. Reprofiling flows will not decrease the minimum bandwidth required to meet the flows' deadlines.
\end{proposition}
The proof is in Appendix~\ref{app:sc_noshaper}.

Note that $\Gamma_{sc}$ specifies a non-linear (piece-wise-linear) service curve for each flow. Given the popularity and simplicity of linear service curves, \ie rate-based schedulers, it is tempting to investigate whether such schedulers, \eg GPS~\cite{parekh93}, could be used instead.  Unfortunately, it is easy to find scenarios where linear service curves perform worse. 

Consider a link shared by two flows with traffic profiles $(r_1, b_1)= (1,45)$ and $(r_2, b_2) = (1,5)$, and deadlines $d_1=10$ and $d_2=1$. A rate-based scheduler must allocate a bandwidth of $\max\left\{\frac{b}{d}, r\right\}$ to a flow with traffic profile $(r, b)$ to meet its deadline $d$ of.  Applying this to flow~$2$ that has the tighter deadline calls for a bandwidth of $5$ to meet its deadline. After $1.25$ units of time (the time to clear the initial burst of~$5$ and the additional data that accumulated during its transmission), flow~$2$'s bandwidth usage drops down to $r_2=1$. The remaining $4$~units then become available to flow~$1$. This means that the initial dedicated bandwidth needed by flow~$1$ to meet its deadline of~$10$ given its burst size of $b_1=45$ is simply its token rate $r_1=1$\footnote{Clearing the burst of flow~$1$ by its deadline $d_1=10$ calls for a bandwidth $x$ such that $45-\frac{5}{4}x-(x+4)\left(10-\frac{5}{4}\right)\leq 0$, which yields $x\geq 1$.}, for a total network bandwidth of $6$ units. In contrast, \Eqref{eq:global_min_R} tells us that $\Gamma_{sc}$, only requires a bandwidth of $R^* = 5.9$.  

The next two sections consider simpler static priority and fifo schedulers, and quantify the bandwidth they require to meet flows' deadlines.  Both schedulers are considered either alone or with ``reprofilers'' that first modify the flows' traffic profiles before they are allowed to access the scheduler. 
\section{Static Priorities}\label{sec:one_static}

Though edf schedulers are efficient and increasingly realizable~\cite{Sivaraman16, saeed19, Sharma20}, they are expensive and may not be practical in all environments. It is, therefore, of interest to explore simpler alternatives while quantifying the trade-off they entail between efficacy and complexity. For that purpose, we consider next a static priority scheduler where each flow is assigned a fixed priority as a function of its deadline. 

As before, we consider $n$ flows with traffic profiles $(r_i, b_i)$ and deadlines $d_i, 1\leq i\leq n$, sharing a common link. The question we first address is how to assign (static) priorities to each flow given their deadlines and \textbf{OPT\_SP}'s goal of minimizing link bandwidth? The next proposition offers a partial and somewhat intuitive answer to this question by establishing that the minimum link bandwidth can be achieved by giving flows with shorter deadlines a higher priority. Formally, 
\begin{proposition}\label{prop:order}
Consider a link shared by $n$ token bucket controlled flows, where flow $i, 1 \leq i \leq n$, has traffic profile $(r_i, b_i)$ and deadline $d_i$, with $d_1 > d_2 > ... > d_n$ and $d_1<\infty$. Under a static-priority scheduler, there exists an assignment of flows to priorities that minimizes link bandwidth while meeting all flows deadlines such that flow~$i$ is assigned a priority strictly greater than that of flow~$j$ only if $d_i<d_j$.
\end{proposition}
The proof is in Appendix~\ref{app:static_order}. We note that while Proposition~\ref{prop:order} states that link bandwidth can be minimized by assigning flows to priorities in the order of their deadline, it neither rules out other mappings nor does it imply that flows with different deadlines should always be mapped to distinct priorities.  For example, large enough deadlines can all be met by a link bandwidth equal to the sum of the flows' average rates, \ie $R^*=\sum_{i=1}^n r_i$. 
In this case, priorities and their ordering are irrelevant.  More generally, grouping flows with different deadlines in the same priority class can often result in a lower bandwidth than mapping them to distinct priority classes\footnote{We illustrate this in 
Appendix~\ref{app:one_static_merge} for the case of two flows sharing a static priority scheduler.}. Nevertheless, motivated by Proposition~\ref{prop:order}, we propose a simple assignment rule that strictly maps lower deadline flows to higher priorities, and evaluate its performance.

\subsection{Static Priorities without Reprofiling}
\label{sec:one_prio_noshaper}

From~\cite[Proposition~$1.3.4$]{nc} we know that when $n$~flows with traffic profiles $(r_i,b_i), 1\leq i\leq n$, share a link of bandwidth $R\geq \sum_{i=1}^n r_i$ with flow~$i$ assigned to priority~$i$ (priority~$n$ is the highest), then, under a static-priority scheduler, the worst case delay of flow~$h$ is upper-bounded by $\frac{\sum_{i = h}^n b_i}{R -  \sum_{i = h+1}^n r_i}$ (recall that under our notation, priority~$n$ is the highest). As a result, the minimum link bandwidth $\widetilde{R}^*$ to ensure that flow~$h$'s deadline $d_h$ is met for all $h$, \ie solving {\bf OPT\_SP}, is given by:
\beq\label{eq:R_p}
\widetilde{R}^* = \max_{1 \leq h \leq n}\left\{\sum_{i=1}^n r_i, \frac{\sum_{i = h}^n b_i}{d_h} + \sum_{i = h+1}^n r_i \right\}
\eeq

Towards evaluating the performance of a static priority scheduler 
versus that of an edf scheduler, we compare $\widetilde{R}^*$ with $R^*$ through their relative difference, \ie $\frac{\widetilde{R}^* - R^*}{R^*}$. For ease of comparison, we rewrite $R^*$ as 
\beq
\label{eq:new_r*}
R^* = \max_{1 \leq h \leq n}\left\{\sum_{i = 1}^n r_i, \frac{\sum_{i = h}^n b_i}{d_h} + \sum_{i = h+1}^n r_i \left( 1 - \frac{d_i }{d_h} \right)\right\}
\eeq
Comparing \Eqrefs{eq:R_p}{eq:new_r*} shows that $R^* = \widetilde{R}^*$ iff $\widetilde{R}^* = \sum_{i = 1}^n r_i$, \ie $\frac{\sum_{i = h}^n b_i}{d_h} \leq \sum_{i = 1}^h r_i, \forall \ 1 \leq h \leq n$. In other words, static priority and edf schedulers perform equally well (yield the same minimum bandwidth), when flow bursts are small and deadlines relatively large so that they can be met with a link bandwidth equal to the sum of the token rates. However, when $\widetilde{R}^* \neq \sum_{i = 1}^n r_i$, a static priority scheduler can require a much larger bandwidth. 

Consider a scenario where $R^*$ is achieved at $h^*$, \ie $R^* = \frac{\sum_{i = h^*}^n b_i}{d_{h^*}} + \sum_{i = h^*+1}^n r_i \left( 1 - \frac{d_i }{d_{h^*}} \right)$. Though $\widetilde{R}^*$ may not be realized at the same $h^*$ value, this still provides a lower bound for $\widetilde{R}^*$, namely, $\widetilde{R}^*\geq \frac{\sum_{i = h^*}^n b_i}{d_{h^*}} + \sum_{i = h^*+1}^n r_i$. Thus, the relative difference between $\widetilde{R}^*$ and $R^*$ is no less than 
%\begin{strip}
\begin{align}
\nonumber
&\frac{ \frac{\sum_{i = h^*}^n b_i}{d_{h^*}} + \sum_{i = h^*+1}^n  r_i}{\frac{\sum_{i = h^*}^n b_i}{d_{h^*}} + \sum_{i = h^*+1}^n r_i \left( 1 - \frac{d_i }{d_{h^*}} \right)}- 1 \\
&= \frac{\sum_{i = h^*+1}^n d_i r_i}{\sum_{i = h^*}^n b_i\ + \sum_{i = h^*+1}^n r_i \left( d_{h^*}  - d_i  \right)}
\label{Eq:diff_static}
\end{align}
%\end{strip}
As the right-hand-side of \Eqref{Eq:diff_static} increases with $d_i$ for all $i \geq h^*$, it is maximized for $d_i = d_{h^*}-\epsilon_i, \forall i > h^*$, for arbitrarily small $\epsilon_{h^*+1}<\ldots<\epsilon_n$, so that its supremum is equal to $\frac{\sum_{i = h^*+1}^n r_i d_{h^*}}{\sum_{i = h^*}^n b_i }$. Note that this is intuitive, as when flows have arbitrarily close deadlines, they should receive equal service shares, which is in direct conflict with a strict priority ordering. 

Under certain flow profiles, the above supremum can be large.   
In a two-flow scenario, basic algebraic manipulations give a supremum of $\frac{r_2}{r_1 + r_2}$, which is achieved at $d_2 = d_1 = \frac{b_2 + b_1}{r_1 + r_2}$. Since $ \frac{r_2}{r_1 + r_2} \to 1$ as $\frac{r_1}{r_2} \to 0$, the optimal static priority scheduler in the two-flow case could require twice as much bandwidth as the optimal edf scheduler.

\subsection{Static Priorities with Reprofiling}\label{sec:one_shaper}
Static priorities can require significantly more bandwidth than $R^*$ mostly because they are a rather blunt instrument when it comes to fine-tuning the allocation of transmission opportunities as a function of packet deadlines.  In particular, they often result in some flows experiencing a delay much lower than their target deadline. 

This is intrinsic to the static structure of the scheduler and to our choice of an assignment that maps distinct deadlines to different priorities, but can be mitigated by anticipating and leveraging the ``slack'' in the delay of some flows. One such option is to use this slack towards reprofiling those flows, \ie make them ``smoother''.  Of interest then, is how to reprofile flows to maximize any resulting link bandwidth reduction?

Consider the trivial example of a single link shared by two flows with traffic profiles $(r_1, b_1) = (1, 5)$ and $(r_2, b_2) = (4, 5)$ and deadlines $d_1= 1.4, d_2=1.25$. A strict static-priority scheduler requires a bandwidth $\widetilde{R}^* = 11.14$. Assume next that we reprofile flow~$2$ to $(r_2, b'_2) = (4, 0)$ before it enters the scheduler.  The added reprofiling delay of $(b_2-b'_2)/r_2=1.25$ reduces the scheduling delay budget down to $0$, but eliminates all burstiness. As a result, we only need a bandwidth of~$7.57$ (under a fluid model) to meet both flows' deadlines (a bandwidth of $4=r_2$ is still consumed by flow~$2$, but the remaining $3.57$ is sufficient to allow flow~$1$ to meet its deadline).  In other words, reprofiling flow~$2$ yields a bandwidth decrease of more than $30\%$. This simple example illustrates the benefits that judicious reprofiling can afford.  

The next few propositions characterize the optimal reprofiling solution and the resulting bandwidth gains for a static priority scheduler and a set of flows and deadlines.  We first derive expressions for flows' reprofiling and scheduling delays under static priorities, before obtaining the optimal reprofiling solution and the resulting minimum link bandwidth $\widetilde{R}^*_R$.

Specifically, given $n$ flows with initial traffic profiles $(r_i,b_i), 1\leq i\leq n,$ deadlines $d_1> d_2>\ldots> d_n,$ a reprofiling solution $(r_i,b'_i), 1\leq i\leq n,$ and a link of bandwidth $R$, 
Proposition~\ref{nflow:delay} characterizes the worst case delay (reprofiling plus scheduling) of each flow, when  a static priority scheduler assigns flow~$i$ priority~$i$ (shorter deadlines have higher priority). The result is used to formulate an optimization problem, {\bf OPT\_RSP}, that seeks to minimize the link bandwidth required to meet individual flows' deadlines.  The variables of the optimization are the reprofiling solution and the link bandwidth.  Proposition~\ref{nflow:R} characterizes the minimum bandwidth $\widetilde{R}^*_R$ that {\bf OPT\_RSP} can achieve, while Proposition~\ref{nflow:b} provides the optimal reprofiling solution.

Let $\boldsymbol{b}' = (b'_1, b'_2, b'_3, ..., b'_n)$ be the vector of reprofiled flow bursts, with $B'_i = \sum_{j = i}^n b'_j$ and $R_i = \sum_{j = i}^n r_j$, the sum of the reprofiled bursts and rates of flows with priority greater than or equal to $i, 1 \leq i \leq n$, where $B'_i = 0$ and $R_i = 0$ for $i > n$. Flow $i$'s worst-case end-to-end delay is characterized next.
%%%%%%%%%%%%%%%%%%%%%%%%%%%%%%%%%%%%%%%%%%%%%%%%%%%%%%%%%%%%%%%%%%%%%%
%%%%%%%%%%%%%%%%%%%%%%% worst-case delay %%%%%%%%%%%%%%%%%%%%%%%%%%%%%
%%%%%%%%%%%%%%%%%%%%%%%%%%%%%%%%%%%%%%%%%%%%%%%%%%%%%%%%%%%%%%%%%%%%%%
\begin{proposition}\label{nflow:delay}
Consider a link shared by $n$ token bucket controlled flows, where flow $i, 1 \leq i \leq n$, has traffic profile $(r_i, b_i)$.
%%and a deadline of $d_i$, with $d_1 > d_2 > ... > d_n$ and $d_1<\infty$. 
Assume a static priority scheduler that assigns flow $i$ a priority of $i$, where priority $n$ is the highest priority, and reprofiles flow $i$ to $(r_i, b'_i)$, where $0 \leq b'_i \leq b_i$. Given a link bandwidth of $R \geq \sum_{j = 1}^n r_j$, the worst-case delay for flow $i$ is 
\beq\label{eq:worst_case_delay}
D^*_i = \max\left\{\frac{b_i + B'_{i+1}}{R - R_{i+1}}, \ \frac{b_i - b'_i}{r_i} + \frac{B'_{i+1}}{R-R_{i+1}} \right\}.
\eeq 
\end{proposition}

The proof is in Appendix~\ref{app:static+reshape}. Note that \Eqref{eq:worst_case_delay} states that flow $i$'s worst-case delay is realized by the last bit of its burst. The two terms of \Eqref{eq:worst_case_delay} capture the cases when this bit arrives before or after the end of flow $i$'s last busy period at the link, respectively, as this determines the extent to which it is affected by the reprofiling delay. 
%%%%%%%%%%%%%%%%%%%%%%%%%%%%%%%%%%%%%%%%%%%%%%%%%%%%%%%%%%%%%%%%%%%%%%%%%%%%%%%%
%% RG:  Removing this, as it is basically trying to informally duplicate the proof. I'm not sure of the benefit, as the interested reader can simply check the proof.
%%%%%%%%%%%%%%%%%%%%%%%%%%%%%%%%%%%%%%%%%%%%%%%%%%%%%%%%%%%%%%%%%%%%%%%%%%%%%%%%
%% Assume the last bit arrives at the shaper at $0$, w.l.o.g, and arrives at the shared link at $t$. If $t$ is during flow $i$'s busy period, then to generate the worst-case delay for the last bit, higher-priority flows should send a burst of $B'_{i+1}$ at time $0$, and then send at a constant rate of $R_{i+1}$. Thus, at $t$ there is $b_i + B'_{i+1} + R_{i+1}t - Rt $ amount of data from flows whose priority is no smaller than $i$ waiting to be served, which takes a time of $\frac{b_i + B'_{i+1}}{R-R_{i+1}} - t$ to clear. Thus, the last bit has an end-to-end delay of $\frac{b_i + B'_{i+1}}{R-R_{i+1}}$. 
%% Conversely, if $t$ is after the busy period, then in the worst case higher priority flows will send a burst of $B'_{i+1}$ at $t$, and will then send at a rate of $R_{i+1}$. Thus, the bit needs to wait for a time of $\frac{B'_{i+1}}{R-R_{i+1}}$ at the shared link. As the (re)shaping delay $t$ is at most $\frac{b_i - b'_i}{r_i}$, the last bit has an end-to-end delay of $\frac{b_i - b'_i}{r_i} + \frac{B'_{i+1}}{R-R_{i+1}}$.

Observe also that $D^*_i$ is independent of $b'_1$ for $2 \leq i \leq n$, and decreases with $b'_1$ when $i = 1$. This is intuitive as flow~$1$ has the lowest priority so that reprofiling it can neither decrease the worst-case end-to-end delay of other flows, nor consequently reduce the minimum link bandwidth required to meet specific deadlines for each flow. Formally,

\begin{corollary}\label{cor:b1}
Consider a link shared by $n$ token bucket controlled flows, where flow $i, 1 \leq i \leq n$, has traffic profile $(r_i, b_i)$ and deadline $d_i$, with $d_1 > d_2 > ... > d_n$ and $d_1<\infty$.  Assume a static priority scheduler that assigns flow $i$ a priority of $i$, where priority $n$ is the highest priority, and reprofiles flow $i$ to $(r_i, b'_i)$, where $0 \leq b'_i \leq b_i$. Given a link bandwidth of $R \geq \sum_{j = 1}^n r_j$, reprofiling flow~$1$ cannot reduce the minimum required bandwidth.
\end{corollary}

Combining Proposition~\ref{nflow:delay} and Corollary~\ref{cor:b1} with \textbf{OPT\_SP} gives the following optimization \textbf{OPT\_RSP} for a link shared by $n$ flows and relying on a static priority scheduler preceded by reprofiling. Note that since the minimum link bandwidth needs to satisfy $R \geq \sum_{i = 1}^n r_i$, combining this condition with $R_i$'s definition gives $\sum_{i = 1}^n r_i = R_1 \leq R$.

\bequn\label{nflow:opt}
\begin{aligned}
&\text{\textbf{OPT\_RSP}} \quad \min_{\boldsymbol{b}'} R \quad \text{s.t} \\
%& & \frac{b_1 + B'_2}{R - R_2} \leq d_1, & \\
&\!\max\left\{\frac{b_i + B'_{i+1}}{R - R_{i+1}}, \frac{b_i - b'_i}{r_i} + \frac{B'_{i+1}}{R-R_{i+1}} \right\} \leq d_i, \ \forall \ 1 \leq i \leq n,\\
&R_1 \leq R, \quad  b'_1 = b_1, \quad 0 \leq b'_i \leq b_i,\ \forall \ 2 \leq i \leq n. 
\end{aligned}
\eequn

The solution of \textbf{OPT\_RSP} is characterized in Propositions~\ref{nflow:R} and~\ref{nflow:b} whose proofs are in Appendix~\ref{sec:opt-rsp}.  Proposition~\ref{nflow:R} gives the optimal bandwidth $\widetilde{R}^*_R$ based only on flow profiles, and while it is too complex to yield a closed-form expression, it offers a feasible numerical procedure to compute $\widetilde{R}^*_R$.  

\begin{proposition}\label{nflow:R}
For $1 \leq i \leq n$, denote $H_i = b_i - d_i r_i$, $\Pi_i(R) = \frac{r_i + R- R_{i+1}}{R-R_{i+1}}$ and $V_i(R) = d_i(R-R_{i+1}) - b_i$. Define $\mathbb{S}_1(R) = \left\{ V_1(R) \right\}$, and $\mathbb{S}_i(R) = \mathbb{S}_{i-1}(R) \bigcup \left\{ V_i(R) \right\} \bigcup \left\{\frac{ s - H_i}{\Pi_i(R)} \ | \ s \in \mathbb{S}_{i-1}(R) \right\}$ for $2 \leq i \leq n$. Then we have $\widetilde{R}^*_R = \max\left\{R_1, \inf \{ R \ | \ \forall s \in \mathbb{S}_n(R), s \geq 0 \} \right\}$.
\end{proposition}

Computing $\widetilde{R}^*_R$ requires solving polynomial inequalities of degree $(n-1)$, so that a closed-form expression is not feasible except for small $n$. However, as $\mathbb{S}_i(R)$ relies only on flow profiles and $\mathbb{S}_j(R)$, $\forall j < i$, we can recursively construct $\mathbb{S}_n(R)$ from $\mathbb{S}_1(R)$. Hence, since $R_1 \leq \widetilde{R}^*_R \leq \widetilde{R}^*$, we can use a binary search to compute $\widetilde{R}^*_R$ from the relation $\widetilde{R}^*_R = \max\left\{R_1, \inf \{ R \ | \ \forall s \in \mathbb{S}_n(R), s \geq 0 \} \right\}$ in Proposition~\ref{nflow:R}.

Next, Proposition~\ref{nflow:b} gives a constructive procedure to obtain the optimal reprofiling burst sizes $\boldsymbol{b}'^*$ given $\widetilde{R}^*_R$ and the original flow profiles.

\begin{proposition}\label{nflow:b}
The optimal reprofiling solution $\boldsymbol{b}'^*$ satisfies
\beq
\label{eq:static_opt_b}
b'^*_i = \left\{
\begin{aligned}
& \max\{0, b_n - r_n d_n \}, &i = n; \qquad\quad\,\,\,\\
& \max \left\{0, b_i - r_i d_i + \frac{r_i B'^*_{i+1}}{\widetilde{R}^*_R-R_{i+1}} \right\}, & 2 \leq i \leq n-1.
\end{aligned}
\right.
\eeq
\end{proposition}
\noindent
where we recall that $b'^*_1=b_1$ and $B'^*_i = \sum_{j = i}^n b'^*_j$. 
 
Note that the optimal reprofiling burst size $b'^*_i$ of flow $i, 1 < i < n$ relies only on the optimal link bandwidth $\widetilde{R}^*_R$ and the reprofiling burst sizes of higher priority flows. Hence, we can recursively characterize $b'^*_i$ from $b'^*_n$ given $\widetilde{R}^*_R$.
\section{Basic FIFO with Reprofiling}\label{sec:one_fifo}

In this section, we consider a simple first-in-first-out (fifo) scheduler that serves data in the order in which it arrives.  For conciseness and given the benefits of reprofiling demonstrated in Section~\ref{sec:one_shaper}, we directly assume that flows are reprofiled prior to being scheduled.  Considering again a link shared by $n$~flows with traffic profiles 
$(r_i,b_i), 1\leq i\leq n,$ and deadlines $d_1> d_2>\ldots> d_n,$ our goal is to find a reprofiling solution $(r_i,b'_i),$ $1\leq i\leq n,$ to minimize the link bandwidth required to meet  the flows' deadlines. 

Towards answering this question, we first proceed to characterize the worst case delay across $n$~flows sharing a link of bandwidth $R$ equipped with a fifo scheduler when the flows have initial traffic profiles $(r_i,b_i), 1\leq i\leq n,$ and are reprofiled to $(r_i,b'_i), 1\leq i\leq n,$ prior to being scheduled.  Using this result, we then identify the reprofiled burst sizes $b'_i, 1\leq i\leq n,$ that minimize the link bandwidth required to ensure that all deadlines $d_1>d_2>\ldots>d_n,$ and $d_1<\infty$ are met.  As with other configurations, we only state the results with proofs relegated to Appendix~\ref{app:one_fifo}.

\begin{proposition}\label{prop:delay_fifo}
Consider a system with $n$ token bucket controlled flows with traffic profiles $(r_i, b_i), 1 \leq i\leq n$, sharing a fifo link with bandwidth $R \geq R_1 = \sum_{j=1}^n r_j$. Assume that the system reprofiles flow $i$ to $(r_i, b'_i)$. The worst-case delay for flow $i$ is then
\beq\label{eq:fifo_delay}
\widehat{D}^*_i = \max\left\{\frac{b_i - b'_i}{r_i}+ \frac{\sum_{j \neq i} b'_j}{R}, \frac{\sum_{j = 1}^n b'_j}{R} + \frac{(b_i - b'_i)R_1}{r_i R} \right\}.
\eeq
\end{proposition}
The proof of Proposition~\ref{prop:delay_fifo} is in Appendix~\ref{app:fifo+reshape}.

With the result of Proposition~\ref{prop:delay_fifo} in hand, we can formulate a corresponding optimization problem, {\bf OPT\_RF}, for computing the optimal reprofiling solution that minimizes the link bandwidth required to meet the deadlines $d_1>d_2>\ldots>d_n,$ and $d_1<\infty$ of the $n$ flows.  Specifically, combining Proposition~\ref{prop:delay_fifo} with \textbf{OPT\_F} gives the following optimization \textbf{OPT\_RF} for a link shared by~$n$ flows when relying on a fifo scheduler preceded by reprofiling. As before, $\sum_{i = 1}^n r_i = R_1 \leq R$.
\beq\label{fifo:opt}
\begin{aligned}
&\text{\textbf{OPT\_RF}} \quad \min_{\boldsymbol{b}'} R \quad \text{s.t} \quad \forall \ 1 \leq i \leq n \\
& \max\left\{\frac{b_i - b'_i}{r_i}+ \frac{\sum_{j \neq i} b'_j}{R}, \frac{\sum_{j = 1}^n b'_j}{R} + \frac{(b_i - b'_i)R_1}{r_i R} \right\} \leq d_i, \\
& R_1 \leq R, \quad 0 \leq b'_i \leq b_i,  \forall \ 1 \leq i \leq n. \\
\end{aligned}
\eeq
The solution of \textbf{OPT\_RF} is characterized in Propositions~\ref{fifo:R} and~\ref{fifo:B} with proofs in
Appendix~\ref{sec:opt-rf}.  As with a static priority scheduler, Proposition~\ref{fifo:R} gives a numerical procedure to compute the optimal bandwidth $\widehat{R}^*_R$ given the flows' profiles, while Proposition~\ref{fifo:B} gives the optimal reprofiling solution $\boldsymbol{\widehat{b}}'^*$ given $\widehat{R}^*_R$ and the original flows' profiles. 
\begin{proposition}\label{fifo:R}
For $1 \leq i \leq n$, define $H_i = b_i - d_i r_i$, $\widehat{B}_i = \sum_{j = 1}^i b_j$, and $\mathbb{Z}_i = \{ 1 \leq j \leq i \mid j \in \mathbb{Z}\}$. Denote 
$$
X_F(R) = \!\!\!\max_{\substack{P_1, P_2 \subseteq \mathbb{Z}_n,\\P_2 \neq \mathbb{Z}_n, P_1 \bigcap P_2 = \emptyset}}\!\!\!\frac{\sum_{i \in P_1} \frac{R H_i}{R+r_i} + \sum_{i \in P_2} \left(b_i - \frac{r_i d_i R}{R_1} \right)}{1 - \sum_{i \in P_1}\frac{r_i}{R+r_i} - \sum_{i \in P_2} \frac{r_i}{R_1}}
$$ 
and 
$$
\begin{aligned}
Y_F(R) &= \min_{1 \leq i \leq n-1}
    \vast\{\widehat{B}_n, Rd_n, \\
    \min_{\substack{P_1, P_2 \subseteq \mathbb{Z}_i,\\ P_1 \bigcap P_2 = \emptyset,\\ P_1 \bigcup P_2 \neq \emptyset}}
    & \left.\left\{\frac{\widehat{B}_i - \sum_{j \in P_1} \frac{R H_j}{R+r_j} -  \sum_{j \in P_2} \left( b_j - \frac{r_j d_j R}{R_1}\right)}{\sum_{j \in P_1}\frac{r_j}{R+r_j} +  \sum_{j \in P_2} \frac{r_j}{R_1}} \right\} \right\}.
\end{aligned}
$$ 
Then the optimal solution for \textbf{OPT\_RF} is
$$\widehat{R}^*_R = \max \left\{R_1,  \frac{\widehat{B}_n R_1}{\sum_{i = 1}^n r_i d_i}, \min\{R \mid X_F(R) \leq Y_F(R) \} \right\}.$$  
\end{proposition}
As $\max \left\{R_1,  \frac{\widehat{B}_n R_1}{\sum_{i = 1}^n r_i d_i}\right\} \leq \widehat{R}^*_R \leq \widehat{R}^* = \max\left\{ R_1, \frac{\widehat{B}_n}{d_n} \right\}$, where $\widehat{R}^*$ is the minimum required bandwidth achieved by a base (no reprofiling) fifo system, we can use Proposition~\ref{fifo:R} and a binary search to compute $\widehat{R}^*_R$. Once $\widehat{R}^*_R$ is known, Proposition~\ref {fifo:B} gives the optimal reprofiling solution.
\begin{proposition}\label{fifo:B}
For $1 \leq i \leq n$, define $T_i(\widehat{B}'_n, R) = \max \left\{0, \frac{R}{R+r_i}\left(H_i + \frac{r_i}{R}\widehat{B}'_n\right), b_i + \frac{r_i(\widehat{B}'_n - R d_i)}{R_1} \right\}$. The optimal reprofiling solution $\boldsymbol{\widehat{b}}'^*$ of \textbf{OPT\_RF}'s is given by $\widehat{b}'^*_1 = \widehat{B}'^*_1$, and $\widehat{b}'^*_i = \widehat{B}'^*_i - \widehat{B}'^*_{i-1}$, $2 \leq i \leq n$, where $\boldsymbol{\widehat{B}}_i'^*$ satisfy 
\beq
\left\{
\begin{aligned}
& \widehat{B}'^*_n = X_F(\widehat{R}^*_R), \\
& \widehat{B}'^*_i = \max\left\{\sum_{j = 1}^i T_j(\widehat{B}'^*_n, \widehat{R}^*_R), \widehat{B}'^*_{i+1} - b_{i+1} \right\},\\
& \text{ when } 1 \leq i \leq n-1.
\end{aligned}
\right.
\eeq
\end{proposition}
Note that $\widehat{B}'^*_n$ relies only on $\widehat{R}^*_R$ and flows' profiles. Whereas when $1 \leq i \leq n-1$, $\widehat{B}'^*_i$ relies only on $\widehat{R}^*_R$, $\widehat{B}'^*_n$, $\widehat{B}'^*_{i+1}$ and flows' profiles. Hence, we can recursively characterize $\widehat{B}'^*_i$ from $\widehat{B}'^*_n$ given $\widehat{R}^*_R$.

\section{Evaluation}\label{sec:evaluation}

In this section, we explore the relative benefits of the solutions developed in the previous three sections. Of interest is assessing the ``cost of simplicity,'' namely, the amount of additional bandwidth required by simpler schedulers such as static priority or fifo compared to an edf scheduler.  Also of interest is the magnitude of the improvements that reprofiling affords with static priority and fifo schedulers. To that end, the evaluation proceeds with a number of \emph{pairwise comparisons} to quantify the relative (bandwidth) cost of each alternative.  

The evaluation first focuses (Section~\ref{sec:perf_2flow}) on scenarios with just two flows.  Closed-form expressions are then available for the minimum bandwidth of each configuration, which make formal comparisons possible.  Section~\ref{sec:perf_nflows} extends this to more ``general'' scenarios involving multiple flows with different combinations of deadlines and traffic profiles.

In the initial two-flow comparisons of Section~\ref{sec:perf_2flow}, we first select a pair of representative traffic profiles (token buckets), and then vary the flows' respective deadlines over a wide range of values.  For each such combination, we explicitly compute the relative differences in bandwidth required by the different schedulers (with and without reprofiling, as applicable) using expressions derived from the propositions obtained in the previous sections.  The results are presented in the form of ``heat-maps'' across the range of deadline combinations.

For the more general scenarios involving multiple flows (Section~\ref{sec:perf_nflows}), we first generate a set of flow profiles, \ie token buckets and deadlines, by randomly selecting them from within specified ranges.  For each such combination, the amount of bandwidth required to meet the flows' deadlines are then computed using again results from the propositions derived in the previous sections.  Finally, for each pair of schedulers, we report statistics (means, standard deviations and the $95\%$ confidence intervals of the means) of the relative bandwidth differences across those random selections.

\subsection{Basic Two-Flow Configurations}
\label{sec:perf_2flow}

Recalling our earlier notation for the minimum bandwidth in each configuration, \ie $R^{*}$ (edf); $\widetilde{R}^*$ (static priority); $\widetilde{R}^{*}_R$ (static priority w/ reprofiling); $\widehat{R}^*$ (fifo); and $\widehat{R}^*_R$ (fifo w/ reprofiling), and specializing \Eqref{eq:global_min_R} to a configuration with two flows, $(r_1,b_1)$ and $(r_2,b_2)$, the absolute minimum bandwidth to meet the flows' deadlines $d_1$ and $d_2$ is given by
\beq
\label{eq:2flow_dyn}
R^{*} = \max\left\{r_1 + r_2, \ \frac{b_2}{d_2},\  \frac{b_1 + b_2 - r_2d_2}{d_1} + r_2 \right\},
\eeq
which is then also the bandwidth required by the edf scheduler.

Similarly, if we consider a static priority scheduler, from \Eqref{eq:R_p}, its bandwidth requirement $\widetilde{R}^*$ (in the absence of any reprofiling) for the same two-flow configuration is of the form
\beq
\label{eq:2flow_stat}
\widetilde{R}^* = \max \left\{r_1 + r_2,\  \frac{b_2}{d_2}, \ \frac{b_1 + b_2}{d_1}+ r_2 \right\};
\eeq
If (optimal) reprofiling is introduced, specializing Proposition~\ref{nflow:R} to two flows, the minimum bandwidth $\widetilde{R}^{*}_R$ reduces to
\beq
\label{eq:2flow_stat+s}
\begin{aligned}
& \max\left\{r_1 + r_2, \frac{b_2}{d_2}, \frac{b_1 + b_2 - r_2d_2}{d_1} + r_2 \right\}, \text{ when }  \frac{b_2}{r_2} \geq \frac{b_1}{r_1} \\
& \max\left\{r_1 + r_2, \frac{b_2}{d_2},\frac{b_1 + \max \left\{ b_2 - r_2d_2, 0 \right\}}{d_1} + r_2 \right\},\\ & \text{ otherwise}; 
\end{aligned}
\eeq
Finally, specializing the results of Propositions~\ref{fifo:R} and~\ref{fifo:B} to two flows, we find that the minimum required bandwidth $\widehat{R}^*$ under fifo without reprofiling is
\beq
\label{eq:2flow_fifo}
\widehat{R}^* = \max \left\{r_1 + r_2,\  \frac{b_1 + b_2}{d_2} \right\};
\eeq
and that when (optimal) reprofiling is used, $\widehat{R}^*_R$ is given by \Eqref{eq:2flow_fifo+s}.
\begin{figure*}[h]
\beq
\label{eq:2flow_fifo+s}
\widehat{R}^*_R = \max \left\{ r_1 + r_2, \frac{b_2}{d_2}, \frac{(b_1 + b_2)(r_1 + r_2)}{d_1 r_1 + d_2 r_2}, \frac{b_1 + b_2 - d_1 r_1 + \sqrt{(b_1 + b_2 - d_1 r_1 )^2 + 4 r_1 d_2 b_2}}{2d_2} \right\}.
\eeq
\end{figure*}
With these expressions in hand, we can now assess the relative benefits of each option in this two-flow scenario.  

Specifically, we consider next combinations consisting of two flows with representative token bucket parameters $(r_1, b_1) = (4,10)$ and $(r_2, b_2) = (10,18)$, and systematically vary their respective deadlines $d_1\geq d_2$ over a range of values.  The bandwidth required to meet the deadlines is then compared for different pairs of schedulers using the expressions reported in \Eqref{eq:2flow_dyn}, \Eqref{eq:2flow_stat+s}, and \Eqref{eq:2flow_fifo+s}.

%%%%%%%%%%%%%%%%%%%%%%%%%%%%%%%%%%%%%%%%%%%%%%%%%%%%%%%%%%%%%%%%%%%%%%%
%%%%%%%%%%%%%%%%%%%%%%%%%%%%%%%%%%%%%%%%%%%%%%%%%%%%%%%%%%%%%%%%%%%%%%%

\subsubsection{The Impact of Scheduler Complexity}
\label{sec:2flow_eval_sched_complex}

We first evaluate the impact of relying on schedulers of decreasing complexity, when those schedulers are coupled with an optimal reprofiling solution. In other words, we compare the bandwidth requirements of an edf scheduler to those of static priority and fifo schedulers combined with an optimal reprofiler.  The comparison is in the form of relative differences (improvements realizable from more complex schedulers), \ie $\frac{\widetilde{R}^*_R - R^*}{\widetilde{R}^*_R}$,  $\frac{\widehat{R}^*_R - R^*}{\widehat{R}^*_R}$, and $\frac{\widehat{R}^*_R - \widetilde{R}^*_R}{\widehat{R}^*}$.\\

%%%%%%%%%%%%%%%%%% sc vs priority + sh %%%%%%%%%%%%%%%%%%%%%%%%%%%%%%
\noindent
\emph{edf vs.~static priority w/ optimal reprofiling}. 

We start with comparing an edf scheduler with a static priority scheduler plus optimal reprofiling. \Eqrefs{eq:2flow_dyn}{eq:2flow_stat+s} then state that $R^{*} < \widetilde{R}^{*}_R$ iff $\frac{b_2}{r_2} < d_2 \leq d_1 < \frac{b_1}{r_1}$. 

\begin{figure}
\centering
\begin{subfigure}{0.32\textwidth}
  \centering
  \includegraphics[width=0.95\linewidth]{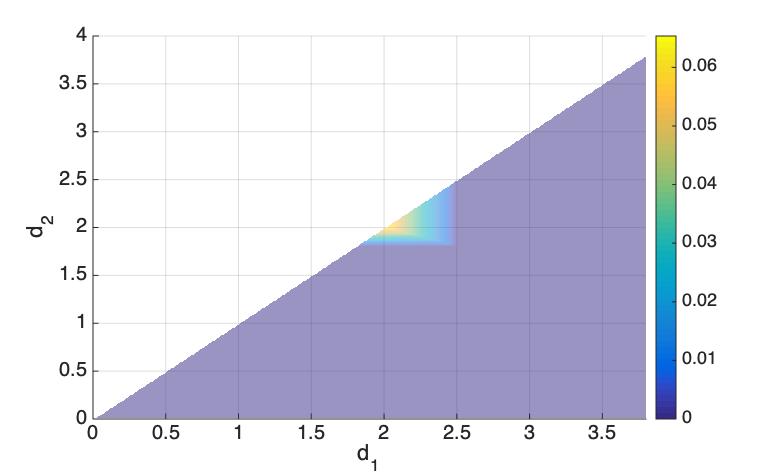}
  \caption{Dyn. prio. vs.~stat. prio. + reprofiling}
  \label{fig:sub11}
\end{subfigure}
\begin{subfigure}{0.32\textwidth}
  \centering
  \includegraphics[width=0.95\linewidth]{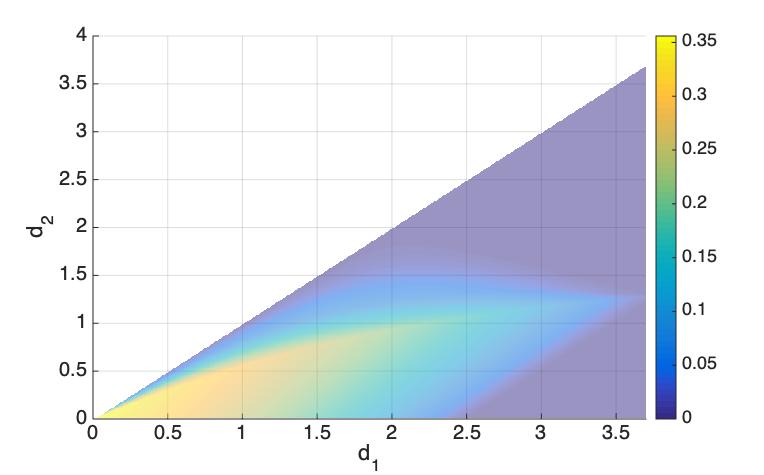}
  \caption{Dyn. prio. vs.~fifo + reprofiling}
  \label{fig:sub13}
\end{subfigure}
\begin{subfigure}{0.32\textwidth}
  \centering
  \vspace{10pt}
  \includegraphics[width=0.95\linewidth]{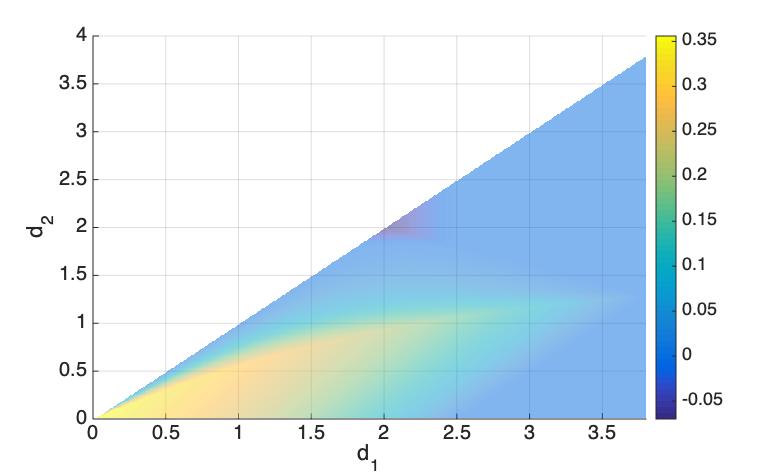}
  \caption{fifo + reprofiling vs.~stat. prio. + reprofiling}
  \label{fig:sub12}
\end{subfigure}
\caption{Relative bandwidth increases for $(r_1, b_1) = (4,10)$ and $(r_2, b_2) = (10,18)$, as a function of $d_1$ and $d_2<d_1$.
The figure is in the form of a heat-map. Darker colors (purple) correspond to smaller increases than lighter ones (yellow).}
\label{fig:relative_diff}
\end{figure}

The results are reported in \fig{fig:sub11}, and, as mentioned, are in the form of a heat-map of the relative bandwidth differences as the flows' respective deadlines vary. As shown in the figure, a static priority scheduler, when combined with reprofiling, performs as well as an edf scheduler, except for a relatively small (triangular) region where $d_1$ and $d_2$ are close to each other and both of intermediate values\footnote{(i) When $d_2$ and $d_1$ are close and small, the bandwidth required to meet the deadlines is very large under either edf or static priority schedulers.  This ensures that both produce similar transmissions' orders. Consider, for example, a low-priority (larger deadline) burst that arrives $(d_1-d_2)$ before a high-priority (smaller deadline) one. It has higher priority under edf, and the speed of the link ensures it is transmitted before the arrival of the high-priority burst, which ensures no difference between edf and a static priority scheduler. 
\\
(ii) When $d_2$ and $d_1$ are close but large, both schedulers meet their deadlines with the same bandwidth, \ie the sum of the flows' average rates.}. Towards better characterizing this range, \ie $d_2 > \frac{b_2}{r_2}$ and $d_1 < \frac{b_1}{r_1}$, we see that the supremum of $\frac{\widetilde{R}^{*}_R - R^*}{\widetilde{R}^{*}_R}$ is achieved at $d_1 = d_2 = \frac{b_1+ b_2}{r_1+r_2}$, with $\widetilde{R}^{*}_R = \frac{b_1}{d_1} + r_2$, and $R^* = r_1 + r_2$. The relative difference in bandwidth between the two schemes is then of the form
$$
\frac{\widetilde{R}^{*}_R - R^*}{\widetilde{R}^{*}_R} = 1- \frac{1}{\frac{b_1}{b_1 + b_2} + \frac{r_2}{r_1 + r_2}},
$$
which can be shown to be upper-bounded by $0.5$.  In other words, in the two-flow case, the (optimal) edf scheduler can result in a bandwidth saving of at most $50\%$ when compared to a static priority scheduler with (optimal) reprofiling.  This happens when the deadlines of the two flows are very close to each other, a scenario unlikely in practice.
\\

%%%%%%%%%%%%%%%%%% sc vs fifo +sh %%%%%%%%%%%%%%%%%%%%%%%%%%%%%
\noindent
\emph{edf vs. fifo w/ optimal reprofiling}

Next, we compare an edf scheduler and a fifo scheduler plus optimal reprofiling.  \Eqrefs{eq:2flow_dyn}{eq:2flow_fifo+s} state that $\widehat{R}^*_R > R^*$ iff $d_1 - \frac{b_1}{r_1} < d_2 < \frac{b_1 + b_2 - d_1 r_1}{r_2}$. We illustrate the corresponding relative differences in \fig{fig:sub13} using the same two-flow combination as before. From the figure, we see that fifo + reprofiling performs poorly relative to an edf scheduler when neither $d_1$ nor $d_2$ are large. As with static priorities, such configurations may not be common in practice.

We note that the supremum of $\frac{\widehat{R}^*_R - R^*}{\widehat{R}^*_R }$ is achieved when $0 < d_2 < \frac{b_1 + b_2 + r_2 d_1 - \sqrt{(b_1 + b_2 + r_2 d_1)^2 - 4 r_2 b_2 d_1}}{2r_2}$, with \Eqref{eq:2flow_dyn} defaulting to $R^* = \frac{b_2}{d_2}$ and \Eqref{eq:2flow_fifo+s} to $\widehat{R}^{*}_R  = \frac{b_1 + b_2 - d_1 r_1 + \sqrt{(b_1 + b_2 - d_1 r_1 )^2 + 4 r_1 d_2 b_2}}{2d_2}$. Hence, the relative difference becomes 
\bequn
\begin{aligned}
\frac{\widehat{R}^{*}_R - R^{*}}{\widehat{R}^{*}_R } &= 1 \\
- &\frac{2b_2}{b_1 + b_2 - d_1 r_1 + \sqrt{(b_1 + b_2 - d_1 r_1 )^2 + 4 r_1 d_2 b_2}},
\end{aligned}
\eequn
which increases with $d_2$. Thus, its supremum is achieved as $d_2 \to d_1$. Similarly, one easily shows that $1 - \frac{2b_2}{b_1 + b_2 - d_1 r_1 + \sqrt{(b_1 + b_2 - d_1 r_1 )^2 + 4 r_1 d_2 b_2}}$ decreases with $d_1$. Hence, the supremum of the relative difference is achieved as $d_1 \to 0$, and is of the form $\frac{b_1}{b_1 + b_2}$, which goes to~$1$ as $\frac{b_1}{b_2} \to \infty$.  In other words, an edf scheduler can yield a~$100\%$ improvement over a fifo scheduler with optimal reprofiling.\\

%%%%%%%%%%%%%%%%%% priority + sh vs fifo +sh %%%%%%%%%%%%%%%%%%%%%%%%%%%%%
\noindent
\emph{Fifo vs.~static priority both w/ optimal reprofiling}

Finally, we compare fifo and static priority schedulers when both rely on optimal reprofiling.   \Eqrefs{eq:2flow_stat+s}{eq:2flow_fifo+s}
%(FIFO with reshaping  vs.~static priority with reshaping) 
give that $\widehat{R}^{*}_R > \widetilde{R}^{*}_R$ iff $\max\left\{\frac{b_2}{r_2}, \frac{(b_1 + b_2)(r_1 + r_2)}{r_2(b_1 /d_1 + r_2)} \right\} < d_1 < \frac{b_1}{r_1}$.
\fig{fig:sub12} illustrates the difference, again relying on a heat-map for the same two-flow combination as the two previous scenarios. 

The figure shows that the benefits of priority are maximum when $d_2$ is small and $d_1$ is not too large. This is intuitive in that a small $d_2$ calls for affording maximum protection to flow~$2$, which a priority structure offers more readily than a fifo.  Conversely, when $d_1$ is large, flow~$1$ can be reprofiled to eliminate all burstiness, which limits its impact on flow~$2$ even when both flows compete in a fifo scheduler.

The figure also reveals that a small region exists (when $d_1$ and $d_2$ are close to each other and both are of intermediate value) where fifo outperforms static priority. As alluded to in the discussion following Proposition~\ref{prop:order} and as expanded in Appendix~\ref{app:one_static_merge}, this is because a \emph{strict} priority ordering of flows as a function of their deadlines needs not always be optimal. For instance, it is easy to see that two otherwise identical flows that only differ infinitesimally in their deadlines should be treated ``identically.'' This is more readily accomplished by having them share a common fifo queue than assigned to two distinct priorities.

To better understand differences in performance between the two schemes, we characterize the supremum and the infimum of $\frac{\widehat{R}^{*}_R - \widetilde{R}^{*}_R}{\widehat{R}^{*}_R }$. Basic algebraic manipulations show that the supremum is achieved as $d_1 = d_2 \to 0$, where \Eqref{eq:2flow_fifo+s} defaults to $\widehat{R}^{*}_R  = \frac{b_1 + b_2 - d_1 r_1 + \sqrt{(b_1 + b_2 - d_1 r_1 )^2 + 4 r_1 d_2 b_2}}{2d_2} $ and \Eqref{eq:2flow_stat+s} to $\widetilde{R}^{*}_R = \frac{b_2}{d_2}$, so that their relative difference is ultimately of the form
\bequn
\frac{\widehat{R}^{*}_R - \widetilde{R}^{*}_R}{\widehat{R}^{*}_R } = \frac{b_1}{b_1 + b_2},
\eequn
which goes to~$1$ as $\frac{b_1}{b_2} \to \infty$, \ie a maximum penalty of $100\%$ for fifo with reprofiling over static priorities with reprofiling.

Conversely, the infimum is achieved at $d_1 = d_2 = \frac{b_1+ b_2}{r_1+r_2}$, with \Eqrefs{eq:2flow_stat+s}{eq:2flow_fifo+s} defaulting to $\widetilde{R}^{*}_R = \frac{b_1}{d_1} + r_2$ and $\widehat{R}^*_R = r_1 + r_2$, and a relative difference of the form
\bequn
\frac{\widehat{R}^{*}_R - \widetilde{R}^{*}_R}{\widehat{R}^{*}_R } = \frac{r_1}{r_1 + r_2} - \frac{b_1}{b_1 + b_2},
\eequn
which increases with $\frac{r_1}{r_2}$ and decreases with $\frac{b_1}{b_2}$. When $\frac{r_1}{r_2} \to 0$ and $\frac{b_1}{b_2} \to \infty$, it achieves an infimum of~$-1$, \ie a maximum penalty of $100\%$ but now for static priorities with reprofiling over fifo with reprofiling.  In other words, when used with reprofiling, both fifo and static priority can require twice as much bandwidth as the other.  

Ensuring that static priority always outperforms fifo calls for determining when flows should be grouped in the same priority class rather than assigned to separate classes. Such grouping can be identified in simple scenarios with two or three flows, \eg see Appendix~\ref{app:one_static_merge}, but a general solution appears challenging.   However, as we shall see in Section~\ref{sec:perf_nflows}, the simple strict priority assignment on which we rely performs well in practice across a broad range of flow configurations.
%%%%%%%%%%%%%%%%%%%%%%%%%%%%%%%%%%%%%%%%%%%%%%%%%%%%%%%%%%%%%%%%%%%%%%%
%%%%%%%%%%%%%%%%%%%%%%%%%%%%%%%%%%%%%%%%%%%%%%%%%%%%%%%%%%%%%%%%%%%%%%%
\subsubsection{The Benefits of Reprofiling}
\label{sec:2flow_eval_benef_shaping}

In this section, we evaluate the benefits afforded by (optimally) reprofiling flows with static priority and fifo schedulers.  This is done by computing for both schedulers the minimum bandwidth required to meet flows' deadlines without and with reprofiling, and evaluating the resulting relative differences, \ie $\frac{\widetilde{R}^* - \widetilde{R}^*_R}{\widetilde{R}^*}$ and $\frac{\widehat{R}^* - \widehat{R}^*_R}{\widehat{R}^*}$.

\begin{figure}
\centering
\begin{subfigure}{0.48\textwidth}
  \centering
  \includegraphics[width=0.95\linewidth]{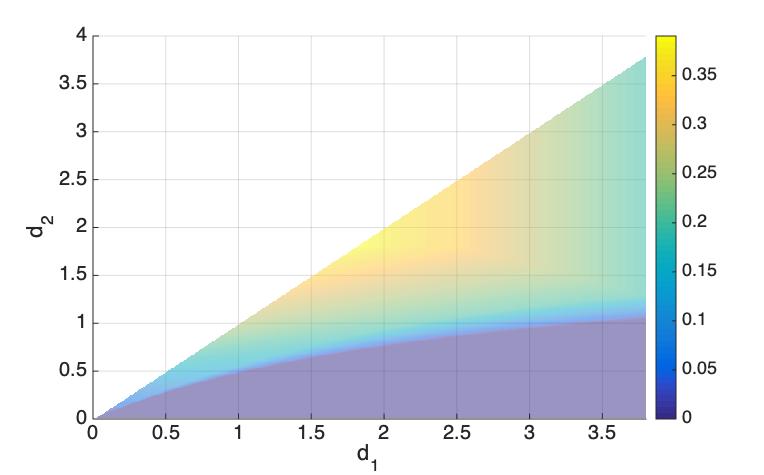}
  \caption{Stat. prio. without vs.~with reprofiling}
  \label{fig:sub21}
\end{subfigure}
\begin{subfigure}{0.48\textwidth}
  \centering
  \includegraphics[width=0.95\linewidth]{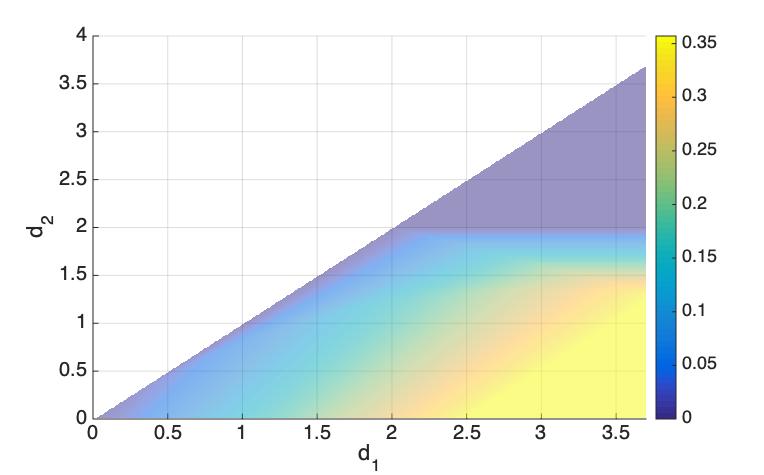}
  \caption{fifo without vs.~with reprofiling}
  \label{fig:sub22}
\end{subfigure}
\caption{Relative bandwidth increases for $(r_1, b_1) = (4,10)$ and $(r_2, b_2) = (10,18)$ as a function of $d_1$ and $d_2<d_1$.
The figure is in the form of a heat-map. Darker colors (purple) correspond to smaller increases than lighter ones (yellow).}
\label{fig:relative_diff2}
\end{figure}

For a static priority scheduler, \Eqrefs{eq:2flow_stat}{eq:2flow_stat+s} indicate that $\widetilde{R}^*_R < \widetilde{R}^*$ iff $\widetilde{R}^*  = \frac{b_1 + b_2}{d_1} + r_2 > \max \left\{r_1 + r_2, \frac{b_2}{d_2} \right\}$, \ie for a static priority scheduler, reprofiling\footnote{Recall from Corollary~\ref{cor:b1} that the flow with the largest deadline, flow~$1$ in the two-flow case, is never reprofiled.} decreases the required bandwidth only when $d_1$, the larger deadline, is not too large and $d_2$, the smaller deadline, is not too small. This is intuitive. When $d_1$ is large, the low-priority flow~$1$ can meet its deadline even without any mitigation of the impact of flow~$2$. Conversely, a small $d_2$ offers little to no opportunity for reprofiling flow~$2$ as the added delay it introduces would need to be compensated by an even higher link bandwidth.  This is illustrated in \fig{fig:sub21} for the same two-flow combination as in \fig{fig:relative_diff}, \ie $(r_1, b_1) = (4,10)$ and $(r_2, b_2) = (10,18)$. The intermediate region where ``$d_1$ is not too large and $d_2$ is not too small'' corresponds to the yellow triangular region where the benefits of reprofiling can reach~$40\%$.
\label{sec:reshaping_benefits}

Similarly, \Eqrefs{eq:2flow_fifo}{eq:2flow_fifo+s} indicate that $\widehat{R}^*_R < \widehat{R}^*$ iff $d_2 < \frac{b_1 + b_2}{r_1 + r_2}$, \ie for a fifo scheduler, reprofiling decreases the required bandwidth only when $d_2$, the smaller deadline, is small. This is again intuitive as a large $d_2$ means that the deadline can be met even without reprofiling flow~$1$\footnote{Note that under static priority the smaller deadline flow is reprofiled, while in the fifo case it is the larger deadline flow that is reprofiled to minimize its impact on the one with the tighter deadline.}. \fig{fig:sub22} presents the relative gain in bandwidth for again the same 2-flow combination. As in the static priority case, the figure shows that for a fifo scheduler the benefits of reprofiling can reach about~$40\%$ in the example under consideration.  

The next section explores scenarios involving combinations of multiple flow profiles. Based on those results it appears that, unsurprisingly, a fifo scheduler stands to generally benefit more from reprofiling than a static priority one.

\subsection{Relative Performance -- Multiple Flows}
\label{sec:perf_nflows}

In this section, we extend the investigation to configurations with more than two flows, using both synthetic flow profiles and profiles derived from datacenter traffic traces.  The evaluation relies on generating a set of flow profiles, \ie token buckets plus deadlines, and for each combination compute the bandwidth required to meet the deadlines using the results derived in the paper.  The main difference with the 2-flow configurations of Section~\ref{sec:perf_2flow} is that, as described next, we now consider a wider range of token bucket parameters with different possible combinations of deadlines.  Additionally, unlike the 2-flow configurations for which the amount of bandwidth required could be obtained from explicit expressions, \ie \Eqref{eq:2flow_dyn}, \Eqref{eq:2flow_stat+s}, and \Eqref{eq:2flow_fifo+s}, computing the required bandwidth now typically involves  numerical procedures, as documented in the propositions derived in the paper.

\subsubsection{Synthetic Flow Profiles}
\label{sec:synthetic_nflows}

We assign flows to \emph{ten} different deadline classes with a dynamic range of~$10$, \ie with minimum and maximum deadlines of~$0.1$ and~$1$, respectively, and consider different spreads in that range for the~$10$ deadlines classes.  Specifically, we select three different possible types of spreads for deadline classes, namely,\\
\emph{Even} deadline spread: 
    \begin{enumerate}[nosep]
    \item $\boldsymbol{d}_{11} = (1, 0.9, 0.8, 0.7, 0.6, 0.5, 0.4, 0.3, 0.2, 0.1)$;
    \end{enumerate}
\emph{Bi-modal} deadline spread: 
    \begin{enumerate}[nosep]
    \setcounter{enumi}{1}
    \item $\boldsymbol{d}_{21} = (1, 0.95, 0.9, 0.85, 0.8 \| 0.3, 0.25, 0.2, 0.15, 0.1)$, 
    \item $\boldsymbol{d}_{22} = (1, 0.96, 0.93, 0.9, 0.86, 0.83, 0.8 \| 0.2, 0.15, 0.1)$, 
    \item $\boldsymbol{d}_{23} = (1, 0.95, 0.9 \| 0.3, 0.26, 0.23, 0.2, 0.16, 0.13, 0.1)$; 
    \end{enumerate}
\emph{Tri-modal} deadline spread: 
    \begin{enumerate}[nosep]
    \setcounter{enumi}{4}
    \item $\boldsymbol{d}_{31} = (1, 0.95, 0.9 \| 0.6, 0.55, 0.5, 0.45 \| 0.2, 0.15, 0.1)$, 
    \item $\boldsymbol{d}_{32} = (1 \| 0.68, 0.65, 0.62, 0.6, 0.57, 0.55, 0.53, 0.5 \| 0.1)$, 
    \item $\boldsymbol{d}_{33} = (1 \| 0.6 \| 0.28, 0.25, 0.23, 0.2, 0.17, 0.15, 0.12, 0.1)$,  
    \item $\boldsymbol{d}_{34} = (1, 0.97, 0.95, 0.93, 0.9, 0.88, 0.85, 0.82 \| 0.6 \| 0.1)$.
    \end{enumerate}

Those three types of spreads translate into different groupings of deadlines, which affect the relative numbers of deadlines in close proximity to each others.

Each of the above eight groupings is used across $1,000$ experiments, where an experiment consists of randomly selecting a ``flow's'' traffic profile for each of the ten deadline classes. Note that what we denote by a flow, in practice maps to the aggregate of all individual flows assigned to the corresponding deadline class (individual flow profiles add up).  Flow profiles are generated by independently drawing ten (aggregate) flow burst sizes $b_1$ to $b_{10}$ from $U(1,10)$, and ten (aggregate) rates $r_1$ to $r_{10}$ from $U(0, r_{\max})$.  The upper bound $r_{\max}$ corresponds to a rate value beyond which a fifo scheduler always performs as well as the optimal solution even without reprofiling\footnote{This happens when the sum of the rates is large enough to alone clear the aggregate burst before the smallest deadline, \ie $10r_{\max}=\frac{\sum_{i = 1}^{10} b_i}{0.1}$.}.

The primary purpose of those synthetic experiments is to allow a systematic exploration of the performance of the different schemes across a broad range of configurations.  The results can then be used to assess the expected performance of each scheme for individual configurations of practical interest.

The results of the experiments are summarized in Table~\ref{table:cmp_sc}, which gives the mean, standard deviation, and the mean's $95\%$ confidence interval for the \emph{relative savings} in link bandwidth, first for edf over static priority with reprofiling, followed by edf over fifo with reprofiling, and then static priority over fifo both with reprofiling.  As mentioned, bandwidth values are computed numerically for each configuration using results from the previously derived propositions.
\begin{table}[ht]
\begin{center}
\caption{Bandwidth savings from scheduler choice.\\ Synthetic flow profiles.}
%%\\$R^*$: dynamic priority; $\widetilde{R}^*$: static priority; $\widetilde{R}^*_R$: static prioriyu w/ reprofiling;  $\widehat{R}^*$: fifo; $\widehat{R}^*_R$ fifo w/ reprofiling}
%difference between $R^*$ and $\widetilde{R}^*_R$, $R^*$ and $\widehat{R}^*_R$, and $\widehat{R}^*_R$ and $\widetilde{R}^*_R$}
\label{table:cmp_sc}
  \begin{tabular}{c|c|c|c|c}
    \hline
{\hspace{-1pt}\textbf{Comparisons}\hspace{-1pt}} & {\hspace{-1pt}\textbf{Scenario}\hspace{-1pt}} & {\hspace{-1pt}\textbf{Mean}\hspace{-1pt}} &{\hspace{-1pt}\textbf{Std. Dev.}\hspace{-1pt}} & {\textbf{$95\%$ Conf.}}  \\ \hline
\multirow{10}{*}{$\stackrel{\substack{\mbox{edf vs.}\\ \mbox{static+reprofiling}\\ \\ R^* \mbox{ vs. } \widetilde{R}^*_R}}{\left(\frac{\widetilde{R}^*_R - R^*}{\widetilde{R}^*_R} \right)}$} & $\boldsymbol{d}_{11}$ & $1.2\%$ & $2.3\%$ & $[1.02\%, 1.31\%]$ \\ \cline{2-5} 
& $\boldsymbol{d}_{21}$ & $1.5\%$ & $2.7\%$ & $[1.35\%, 1.69\%]$ \\ \cline{2-5} 
& $\boldsymbol{d}_{22}$ & $1.1\%$ & $2.7\%$ & $[1.01\%, 1.28\%]$ \\ \cline{2-5} 
& $\boldsymbol{d}_{23}$ & $2.9\%$ & $4.2\%$ & $[2.59\%, 3.12\%]$ \\ \cline{2-5} 
& $\boldsymbol{d}_{31}$ & $1.4\%$ & $2.5\%$ & $[1.2\%, 1.51\%]$ \\ \cline{2-5} 
& $\boldsymbol{d}_{32}$ & $1.0\%$ & $2.1\%$ & $[0.84\%, 1.1\%]$ \\ \cline{2-5} 
& $\boldsymbol{d}_{33}$ & $6.2\%$ & $6.5\%$ & $[5.76\%, 6.58\%]$ \\ \cline{2-5} 
& $\boldsymbol{d}_{34}$ & $0.7\%$ & $1.7\%$ & $[0.6\%, 0.81\%]$ \\ \hline
\multirow{10}{*}{$\stackrel{\substack{\mbox{edf vs.}\\ \mbox{fifo+reprofiling}\\ \\ R^* \mbox{ vs. } \widehat{R}^*_R}}{\left(\frac{\widehat{R}^*_R- R^*}{\widehat{R}^*_R} \right)}$} & $\boldsymbol{d}_{11}$ & $1.7\%$ & $6.5\%$ & $[1.13\%, 2.11\%]$ \\ \cline{2-5} 
& $\boldsymbol{d}_{21}$ & $3.2\%$ & $8.7\%$ & $[2.68\%, 3.76\%]$ \\ \cline{2-5} 
& $\boldsymbol{d}_{22}$ & $1.7\%$ & $6.2\%$ & $[1.26\%, 2.03\%]$ \\ \cline{2-5} 
& $\boldsymbol{d}_{23}$ & $8.0\%$ & $12.8\%$ & $[7.24\%, 8.82\%]$ \\ \cline{2-5} 
& $\boldsymbol{d}_{31}$ & $2.5\%$ & $7.8\%$ & $[2.06\%, 3.03\%]$ \\ \cline{2-5} 
& $\boldsymbol{d}_{32}$ & $0.8\%$ & $4.6\%$ & $[0.54\%, 1.11\%]$]\\ \cline{2-5} 
& $\boldsymbol{d}_{33}$ & $12.0\%$ & $14.1\%$ & $[11.15\%, 12.9\%]$ \\ \cline{2-5} 
& $\boldsymbol{d}_{34}$ & $0.4\%$ & $3.2\%$ & $[0.2\%, 0.6\%]$ \\ \hline
\multirow{10}{*}{$\stackrel{\substack{\mbox{static vs. fifo}\\ \mbox{both}\\ \mbox{w/ reprofiling}\\ \\ \widetilde{R}^*_R \mbox{ vs. } \widehat{R}^*_R}}{\left(\frac{\widehat{R}^*_R- \widetilde{R}^*_R}{\widehat{R}^*_R} \right)}$} & $\boldsymbol{d}_{11}$ & $0.6\%$ & $6.5\%$ & $[0.16\%, 0.95\%]$ \\ \cline{2-5} 
& $\boldsymbol{d}_{21}$ & $1.8\%$ & $8.3\%$ & $[1.26\%, 2.28\%]$ \\ \cline{2-5} 
& $\boldsymbol{d}_{22}$ & $0.5\%$ & $6.1\%$ & $[0.12\%, 0.88\%]$ \\ \cline{2-5} 
& $\boldsymbol{d}_{23}$ & $5.5\%$ & $11.3\%$ & $[4.84\%, 6.24\%]$ \\ \cline{2-5} 
& $\boldsymbol{d}_{31}$ & $1.2\%$ & $7.5\%$ & $[0.76\%, 1.69\%]$ \\ \cline{2-5} 
& $\boldsymbol{d}_{32}$ & -0.2\% & $4.5\%$ & [-0.43\%, 0.13\%] \\ \cline{2-5} 
& $\boldsymbol{d}_{33}$ & $6.6\%$ & $11.2\%$ & $[5.92\%, 7.3\%]$ \\ \cline{2-5} 
& $\boldsymbol{d}_{34}$ & -0.3\% & $3.3\%$ & [-0.53\%, -0.12\%] \\ \hline
\hline
  \end{tabular}
\end{center}
\end{table}

The first conclusion one can draw from Table~\ref{table:cmp_sc} is that while an edf scheduler affords some benefits, they are on average smaller than the maximum values of Section~\ref{sec:perf_2flow}. Average improvements over static priority with reprofiling hover around $1\%$ and did not exceed about $6\%$.  Improvements are a little higher when considering fifo with reprofiling, where they reach $12\%$, but those values are still significantly less than the worst case scenarios of Section~\ref{sec:perf_2flow}.

Table~\ref{table:cmp_sc} also reveals that, somewhat surprisingly, static priority and fifo perform similarly when both are afforded the benefit of reprofiling (the largest difference observed in the experiments is $5.5\%$).  Static priority has an edge on average even if, as discussed in Section~\ref{sec:perf_2flow}, a few scenarios exist where a fifo scheduler outperforms static priority when both are combined with reprofiling, \eg $d_{32}$ and $d_{34}$. Recall that this is because we strictly map smaller deadlines to higher priority. The differences are, however, small, \ie $0.2\%$ and $0.3\%$, respectively, for the two scenarios where fifo outperforms static priority on average. 
%%%%%%%%%%%%%%%%%%%%%%%%%%%
\begin{table}[ht]
\begin{center}
\caption{Benefits of reprofiling for static priority \& fifo.\\ Synthetic flow profiles.}
%%difference between $\widetilde{R}^*$ and $\widetilde{R}^*_R$,  and $\widehat{R}$ and $\widehat{R}^*_R$}
\label{table:cmp_sh}
  \begin{tabular}{ c | c | c | c | c }
    \hline
{\hspace{-1pt}\textbf{Comparisons}\hspace{-1pt}} & {\hspace{-1pt}\textbf{Scenario}\hspace{-1pt}} & {\hspace{-1pt}\textbf{Mean}\hspace{-1pt}} &{\hspace{-1pt}\textbf{Std. Dev.}\hspace{-1pt}} & {\textbf{$95\%$ Conf.}}  \\ \hline
\multirow{10}{*}{$\stackrel{\substack{\mbox{static}\\ \mbox{w/ \& w/o} \\ \mbox{reprofiling}\\ \\ \widetilde{R}^*_R \mbox{ vs. } \widetilde{R}^*}}{\left(\frac{\widetilde{R}^*- \widetilde{R}^*_R}{\widetilde{R}^*} \right)}$} & $\boldsymbol{d}_1$ & $8.43\%$ & $4.50\%$ & $[8.15\%, 8.71\%$] \\ \cline{2-5} 
& $\boldsymbol{d}_{21}$ & $8.11\%$ & $4.19\%$ & $[7.85\%, 8.37\%]$ \\ \cline{2-5} 
& $\boldsymbol{d}_{22}$ & $8.42\%$ & $4.52\%$ & $[8.14\%, 8.71\%]$\\ \cline{2-5} 
& $\boldsymbol{d}_{23}$ & $9.38\%$ & $4.80\%$ & $[9.08\%, 9.67\%]$ \\ \cline{2-5} 
& $\boldsymbol{d}_{31}$ & $8.24\%$ & $4.33\%$ & $[7.97\%, 8.51\%]$\\ \cline{2-5} 
& $\boldsymbol{d}_{32}$ & $9.49\%$ & $5.07\%$ & $[9.18\%, 9.81\%]$\\ \cline{2-5} 
& $\boldsymbol{d}_{33}$ & $15.97\%$ & $4.78\%$ & $[15.67\%, 16.27\%]$\\ \cline{2-5} 
& $\boldsymbol{d}_{34}$ & $8.83\%$ & $4.94\%$ & $[8.53\%, 9.14\%]$ \\ \hline
\multirow{10}{*}{$\stackrel{\substack{\mbox{fifo}\\ \mbox{w/ \& w/o} \\ \mbox{reprofiling}\\ \\ \widehat{R}^*_R \mbox{ vs. } \widehat{R}^*}}{\left(\frac{\widehat{R}^* - \widehat{R}^*_R}{\widehat{R}^*} \right)}$} & $\boldsymbol{d}_1$ & $49.52\%$ & $8.17\%$ & $[49.01\%, 50.03\%]$ \\ \cline{2-5} 
& $\boldsymbol{d}_{21}$ & $48.71\%$ & $7.62\%$ & $[48.24\%, 49.18\%]$\\ \cline{2-5} 
& $\boldsymbol{d}_{22}$ & $49.53\%$ & $8.27\%$ & $[49.02\%, 50.05\%]$\\ \cline{2-5} 
& $\boldsymbol{d}_{23}$ & $45.78\%$ & $6.52\%$ & $[45.37\%, 46.18\%]$\\ \cline{2-5} 
& $\boldsymbol{d}_{31}$ & $49.08\%$ & $7.88\%$ & $[48.59\%, 49.57\%]$\\ \cline{2-5} 
& $\boldsymbol{d}_{32}$ & $49.95\%$ & $8.59\%$ & $[49.42\%, 50.49\%]$\\ \cline{2-5} 
& $\boldsymbol{d}_{33}$ & $42.47\%$ & $6.19\%$ & $[42.08\%, 42.85\%]$ \\ \cline{2-5} 
& $\boldsymbol{d}_{34}$ & $50.13\%$ & $8.84\%$ & $[49.59\%, 50.68\%]$ \\ \hline
\hline
  \end{tabular}
\end{center}
\end{table}

Towards gaining a better understanding of reprofiling and the extent to which it is behind the somewhat unexpected good performance of fifo, Table~\ref{table:cmp_sh} reports its impact for both static priority and fifo.  As Table~\ref{table:cmp_sc}, it gives the mean, standard deviation, and the mean's $95\%$ confidence interval, but now of the relative gains in bandwidth that reprofiling affords over no reprofiling for both fifo and static priority schedulers.

The data from Table~\ref{table:cmp_sh} highlights that while both static priority and fifo benefit from reprofiling, the magnitude of the improvements is significantly higher for fifo. Specifically, improvements from reprofiling are systematically above $40\%$ and often close to $50\%$ for fifo, while they exceed $10\%$ only once for static priority (at $15\%$ for scenario $d_{33}$) and are typically around $8\%$.  As alluded to earlier, this is not surprising given that static priority offers some ability to discriminate flows based on their deadlines, while fifo does not. 

\subsubsection{Application Derived Flow Profiles}
\label{sec:realistic_nflows}

The benefits of synthetic flow profiles in allowing a systematic investigation notwithstanding, it is also of interest to target configurations more directly representative of traffic mixes as they arise in practice. To that end, we rely on a methodology similar to that used in~\cite[Section VIII-B2]{multiple_nodes21}, and construct a set of flow profiles derived from traffic data reported in~\cite{roy15}.

\begin{figure}
\centering
  \includegraphics[width=0.7\linewidth]{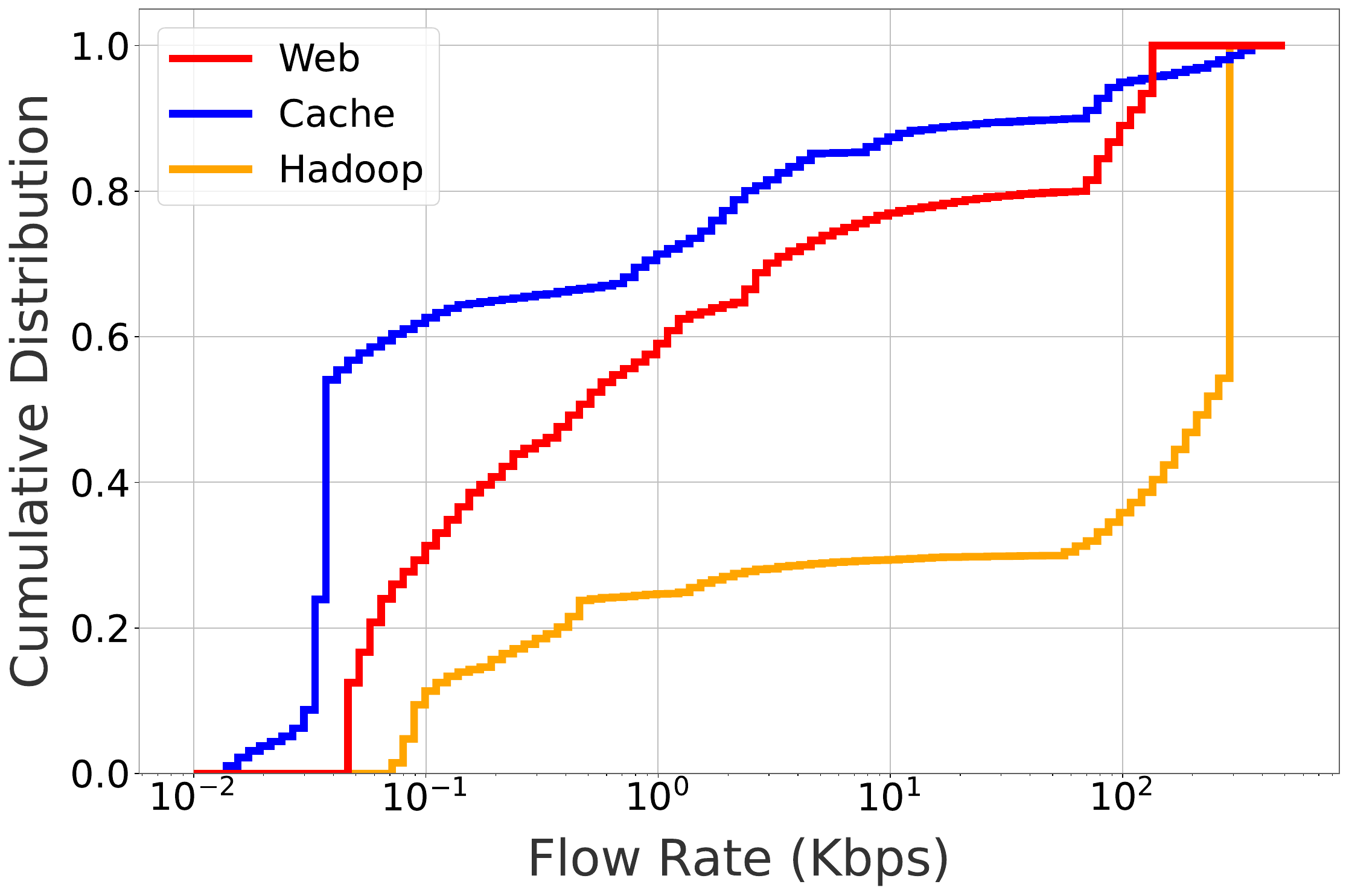}
\caption{CDF of flow rates for Web, Cache, \& Hadoop applications from~\cite{roy15} assuming correlated flow durations and sizes.}
\label{fig:rate_cdf}
\end{figure}

Specifically, \cite{roy15} investigates the traffic flowing through the network of one of Facebook's large datacenter, and reports, among other things, the distribution of flow sizes and durations (Figs.~6 \&~7 of~\cite{roy15}) for three representative applications:  Web (W), Cache read and replacement (C), and Hadoop (H). We rely on these data to generate sample traffic profiles $(r,b)$ for flows from those three applications as follows:
\begin{enumerate}[nosep]
\item For a given application, we generate flow size+duration tuples by sampling the corresponding distributions assuming they are perfectly positively correlated.  In other words, we assume that larger flows last longer. 
\item A flow's token rate $r$ is then obtained by dividing the flow size by its duration.  \fig{fig:rate_cdf} shows the resulting cumulative distributions of flow rates
for all three applications.
\item Generating token bucket sizes $b$ involves an additional step and associated assumption:
\begin{enumerate}[nosep]
\item The smallest flow sizes from Fig.~6 of~\cite{roy15} are assumed representative of a single transmission burst.  This yields burst sizes $S_W=0.15$Kbytes, $S_C=0.4$Kbytes, and $S_H=0.3$Kbytes for our three sample applications.
\item As bucket sizes are typically chosen to accommodate consecutive bursts, we leverage the claim in~\cite{roy15} that all three applications are ``internally bursty'' with Cache significantly burstier
than Hadoop, and Web in between, to randomly select bucket sizes in $[0, 20S_C], [0,10S_W]$ and $[0, 2S_H]$, respectively.  We note that these values yield relatively small buckets, and, therefore, maximum burst sizes. 
\end{enumerate}
\end{enumerate}

The resulting profiles have relatively low rates and burstiness, at least when it comes to individual flows, with Hadoop's profile typical of bandwidth hungry applications, and Web and Cache representative of more interactive applications.  This maps to the types of services that~\cite{roy15} mentions as relying on those three applications, \ie Web search, user data query, and offline analysis (\eg data mining).  As a result, we assign deadlines to each application that broadly reflect those services, with three deadline classes set to $10$ms, $50$ms, and $200$ms for Web, Cache, and Hadoop respectively. 

In evaluating performance, we consider four ``traffic mixes'' that differ in their relative proportion of flows from each application.  Specifically, the corresponding four scenarios sample our W, C, H applications in the proportions: 1:1:1, 3:9:1, 9:3:1, and 9:9:1, respectively.  For each scenario, we randomly sample $100$ flow profiles in those proportions.  The $100$ flows are then grouped according to their deadline class, which yields a set of three aggregates to schedule on the shared link according to their deadlines.  This procedure is repeated $1000$ times, and the relative link bandwidth requirements across schedulers and reprofiling options are given in Table~\ref{table:cmp_real}.

\begin{table}[ht]
\begin{center}
\caption{Bandwidth savings \& benefits of reprofiling. \\ Application-derived flow profiles.}
\label{table:cmp_real}
  \begin{tabular}{ c | c | c | c | c }
    \hline
{\hspace{-1pt}\textbf{Comparisons}\hspace{-1pt}} & {\hspace{-1pt}\textbf{Scenario}\hspace{-1pt}} & {\hspace{-1pt}\textbf{Mean}\hspace{-1pt}} &{\hspace{-1pt}\textbf{Std. Dev.}\hspace{-1pt}} & {\textbf{$95\%$ Conf.}}  \\ \hline
\multirow{4}{*}{$\stackrel{R^* \mbox{ vs. } \widetilde{R}^*_R}{\left(\frac{\widetilde{R}^*_R - R^*}{\widetilde{R}^*_R} \right)}$}
& 1:1:1 & $0.0\%$ & $0.0\%$ & $[0\%,0\%]$ \\ \cline{2-5}
& 3:9:1 & $0.0\%$ & $0.0\%$ & $[0\%,0\%]$ \\ \cline{2-5}
& 9:3:1 & $0.0\%$ & $0.0\%$ & $[0\%,0\%]$ \\ \cline{2-5}
& 9:9:1 & $0.0\%$ & $0.0\%$ & $[0\%,0\%]$ \\ \hline

\multirow{4}{*}{$\stackrel{R^* \mbox{ vs. } \widehat{R}^*_R}{\left(\frac{\widehat{R}^*_R- R^*}{\widehat{R}^*_R} \right)}$}
& 1:1:1 & $41.82\%$ & $17.04\%$ & $[40.76\%, 42.88\%]$ \\ \cline{2-5}
& 3:9:1 & $72.36\%$ & $3.43\%$ & $[72.15\%, 72.58\%]$\\ \cline{2-5}
& 9:3:1 & $56.84\%$ & $8.43\%$ & $[56.32\%, 57.37\%]$\\ \cline{2-5}
& 9:9:1 & $69.70\%$ & $4.10\%$ & $[69.45\%, 69.96\%]$\\ \hline

$\stackrel{\,}{\widetilde{R}^*_R \mbox{ vs. } \widehat{R}^*_R}$ & \multicolumn{4}{c}{See above}\\ \hline

\multirow{4}{*}{$\stackrel{\widetilde{R}^*_R \mbox{ vs. } \widetilde{R}^*}{\left(\frac{\widetilde{R}^*- \widetilde{R}^*_R}{\widetilde{R}^*} \right)}$}
& 1:1:1 & $0.01\%$ & $0.22\%$ & $[0.00\%, 0.03\%]$ \\ \cline{2-5}
& 3:9:1 & $1.74\%$ & $0.73\%$ & $[1.70\%, 1.79\%]$\\ \cline{2-5}
& 9:3:1 & $1.02\%$ & $2.34\%$ & $[0.87\%, 1.16\%]$\\ \cline{2-5}
& 9:9:1 & $4.06\%$ & $1.19\%$ & $[3.98\%, 4.13\%]$\\ \hline

\multirow{4}{*}{$\stackrel{\widehat{R}^*_R \mbox{ vs. } \widehat{R}^*}{\left(\frac{\widehat{R}^* - \widehat{R}^*_R}{\widehat{R}^*} \right)}$}
& 1:1:1 & $22.74\%$ & $9.17\%$ & $[22.17\%, 23.31\%]$ \\ \cline{2-5}
& 3:9:1 & $21.61\%$ & $8.48\%$ & $[21.09\%, 22.14\%]$\\ \cline{2-5}
& 9:3:1 & $14.93\%$ & $8.93\%$ & $[14.37\%, 15.48\%]$\\ \cline{2-5}
& 9:9:1 & $19.55\%$ & $8.62\%$ & $[19.02\%, 20.08\%]$\\ \hline
\hline
  \end{tabular}
\end{center}
\end{table}

The first conclusion from Table~\ref{table:cmp_real} is that static priority with reprofiling performs just as well as edf for all four scenarios (top four rows).  This is not surprising.  The large gaps between the deadlines of the three classes of applications ensure that once high-priority bursts are cleared, the residual bandwidth is sufficient to transmit lower priority bursts before their deadline.  As with synthetic profiles, reprofiling is instrumental in realizing this outcome, even if the wide gaps between deadlines together with the limited burstiness of the applications produce a smaller gain ($\widetilde{R}^*_R$ vs.~$\widetilde{R}^*$).

The benefits of reprofiling are again more apparent with fifo, even if it significantly under-performs both edf and static priority.  The lack of discrimination across flows that fifo suffers from is exacerbated by the large gaps between deadlines, and compensating for it calls for an average of about $50\%$ more bandwidth across all four scenarios. However, without reprofiling this bandwidth increase ($\widehat{R}^*_R$ vs.~$\widehat{R}^*$) is around $20\%$ larger.  This again demonstrates the extent to which reprofiling can help simpler schedulers.

\subsection{Summary Discussion}

Several common themes emerge between the evaluations of Sections~\ref{sec:synthetic_nflows} and~\ref{sec:realistic_nflows}.  The first is that reprofiling can help a static priority scheduler perform nearly as well as an edf scheduler. Second, while its inability to discriminate between flows puts fifo at a clear disadvantage, reprofiling is again capable of partially mitigating its handicap.  Finally, while with static priority the benefits of reprofiling are realized by reprofiling high-priority flows to limit their impact on low-priority ones, the opposite holds for fifo.

The differences between the results of Sections~\ref{sec:synthetic_nflows} and~\ref{sec:realistic_nflows} also revealed a number of intuitive findings brought about by the differences in deadline spreads in the two scenarios.  In particular, the large gaps between deadlines present in Section~\ref{sec:realistic_nflows} make it easier for a static priority scheduler to perform nearly as well as an edf scheduler.  This holds with and without reprofiling, even if reprofiling remains useful.  Conversely, more closely packed deadlines offer additional opportunities for reprofiling to be useful, as closer deadlines amplify the need for fine tuning of a flow's profile relative to its deadline and impact on other flows.

%Fluctuation Smoothing Policies
\section{Related Works}\label{sec:related}
The question of meeting deadlines for a set of rate-limited flows is one that has received much attention in the scheduling literature.  It is not our intent to provide an exhaustive review of those works.  Instead, we limit ourselves to highlighting works whose results are closest to ours or that offered early insight into the problem, including the benefits of adjusting flows' profiles (reprofiling) that is one of the foci of this paper.

\paragraph{Packet-level shaping and scheduling} 
Scheduling flows with deterministic traffic profiles was investigated in~\cite{Georgiadis97} that considered both buffer and delay requirements.  In particular, the paper established\footnote{A similar result was reported in~\cite{liebeherr96}.} the optimality of the edf policy in terms of maximizing the schedulable region.  This is the ``dual'' of the bandwidth minimization problem investigated in this paper, and the result parallels that of Proposition~\ref{prop:edf}.  Static priority and fifo schedulers were, however, not investigated, and neither was the impact of reprofiling flows.

The aspect of minimizing the resources required to meet the latency targets of token bucket-controlled flows was explored in~\cite{schmitt02}.  The paper relied on service curves with high and low rates and sought to identify the earliest possible time for switching to the lower rate. The focus was, however, on minimizing resources required by each flow individually rather than in aggregate, as in this paper.  In addition, the potential impact of reprofiling flows was not addressed.

\paragraph{Shaping bulk data transfers}
Minimizing bandwidth (cost) through reprofiling (reshaping) flows has been investigated for bulk data transfers where transfer completion times rather than packet-level deadlines are the targets\cite{thyaga12,amoeba15,pretium16,trafficshaper18,butler19,cascara21}. The problem stems from non-linear bandwidth costs, \eg based on the $95^{th}$ percentile, so that judiciously adjusting (shaping) the transmission rates of bulk transfers can yield significant savings. Rate shaping is, however, at a time-granularity of minutes rather than at the packet-level.  The optimization frameworks of those papers are, therefore, not applicable to our problem.  Their solutions, can, however, complement ours by leveraging the fluctuations in link utilization inherent in delivering hard, packet-level delay bounds, as we do.

\paragraph{Deterministic networking}
The deterministic traffic profiles and delay bounds of the TSN and DetNet standards have also given rise to related investigations as documented in recent surveys~\cite{tsn_survey22,walrand23}.  In particular, the optimization framework that underlies many of those studies have connections to the problem we address. However, like most prior similar works, traffic profiles are assumed fixed and the impact of reprofiling is not considered.

\paragraph{Datacenter solutions}
The emergence of traffic profiles and latency targets in datacenter networks motivated~\cite{qjump15}.  It targets a multi-hop network, but calls on topological properties of typical datacenter networks to collapse its model to a single hop, thereby aligning with the scope of this paper. Similarities extend to considering a static priority scheduler, but traffic profiles differ. Rather than a token bucket, \cite{qjump15} relies on the notion of network ``epochs'' to bound packet bursts.  Delay bounds are then expressed as a function of the network fan-in and a ``throughput factor'' that reflects the number of transmission opportunities sources can have per network epoch.  Also absent from the paper are exploring bandwidth minimization and the potential benefits of reprofiling flows.

\paragraph{Reprofiling investigations}  Meeting packet-level latency constraints with a static priority scheduler while minimizing costs through reprofiling of flows is the focus of WorkloadCompactor~\cite{zhu17}.  The reprofiling decisions of~\cite{zhu17} are, however, focused on selecting token bucket parameters from among a family of feasible regulators\footnote{Regulators above the \emph{r-b} curve using the terminology of~\cite{zhu17}.} that do not introduce additional delay.  In contrast, our reprofiling allows for an added delay that the scheduler must then compensate for.  Exploring when and how this trade-off is of benefit is our main contribution and what makes this work {\it complementary} to the approach from WorkloadCompactor. 

Specifically, WorkloadCompactor considers traffic/workload traces for which it seeks to first identify \emph{feasible} token bucket parameter pairs $\langle r,b\rangle$ that result in \emph{zero} access delay for those traces. WorkloadCompactor's main contribution is in realizing that multiple such $\langle r,b\rangle$ pairs are possible (the $r$-$b$ curve of~\cite{zhu17}), and that \emph{``jointly optimizing the choice of $\langle r,b\rangle$ rate limit parameters for each workload to better compact workloads onto servers''} can reduce the required server capacity. This is where the contribution of WorkloadCompactor ends, and where that of this paper actually starts.

More precisely, once the optimization of WorkloadCompactor completes, the set of $\langle r,b\rangle$ pairs it produces can be used, together with the associated target latency bounds, as inputs to Proposition~\ref{nflow:b}. Proposition~\ref{nflow:b} explores, for a static priority scheduler, how to best reprofile token bucket-controlled flows to meet their deadlines with the least bandwidth.  This reprofiling is beyond that suggested by WorkloadCompactor\footnote{Again~\cite{zhu17} focuses on regulators that ensure zero access delay for a given trace, while we investigate how a non-zero access delay can be of benefit.}, and explores how trading-off access delay to further smooth flows can yield additional benefits.  In other words, the approach proposed in the paper complements that of WorkloadCompactor, in that it can be applied to any set of token bucket profiles produced by WorkloadCompactor, and modify them to yield further reductions in system resources (bandwidth or server capacity) while still meeting latency targets.

We also note that, because WorkloadCompactor considers the problem of selecting token bucket parameters for traffic traces (to ensure zero access delay), it addresses an aspect that this paper does not consider since we assume that token bucket profiles are given.  This is yet another aspect in which the two papers are complementary.

\paragraph{Early works}
Finally, we note that exploring the trade-off between making traffic smoother and end-to-end performance is not unique to packet networks. It is present in the early ``fluctuation smoothing'' scheduling policies of~\cite{kumar94} that sought to reduce processing time in manufacturing plants, and more recently in the reshaping of parallel I/O requests to improve the scalability of database systems~\cite{li22}.

\section{Conclusion and Future Work}\label{sec:conclusion}
The paper investigated the question of minimizing the bandwidth needed to guarantee worst case latencies to a set of token bucket-controlled flows sharing a single link.  The investigation was carried for schedulers of different complexity.  

The paper first characterized the minimum required bandwidth independent of schedulers, and showed that an edf scheduler could realize all flows' deadlines under such bandwidth.  Motivated by the need for lower complexity solutions, the paper then explored simpler static priority and fifo schedulers.  It derived the minimum required bandwidth for both, but more interestingly established how to optimally reprofile flows to reduce the bandwidth needed while still meeting all deadlines. The relative benefits of such an approach were illustrated numerically for a number of different flow combinations, which showed how reprofiling can enable simpler schedulers to perform nearly as well a more complex ones across a range of configurations.

The obvious direction in which to extend the paper is to a multi-hop setting. In~\cite{multiple_nodes21}, we build on the results of Proposition~\ref{prop:shifted_ac} and provide initial results for the multi-hop case under the assumption that (service curve) edf schedulers are available at each hop. Extending the investigation to static priority and fifo schedulers is under way.

Another aspect of interest with static priority schedulers is relaxing the assumption that flows with different deadlines map to distinct priority classes, and allow multiple deadlines to be assigned to the same class.  Not only does it enhance scalability, but it can also improve performance\footnote{See Appendix~\ref{app:one_static_merge} for the simple case of a two-flow configuration.}. Last but not least, extensions to statistical rather than deterministic delay guarantees are also of practical relevance.

% or


\begin{comment}
\section*{Acknowledgment}

The work of the first three authors was supported in part by a gift from Google and by NSF grant CNS 2006530.
\end{comment}

\ifCLASSOPTIONcaptionsoff
  \newpage
\fi



\bibliographystyle{IEEEtran}
\bibliography{IEEEabrv,ref}

\appendices
\clearpage
\onecolumn
\section{Summary of Notation Used in the Paper}
\begin{center}
  \begin{longtable}{ c | c }
    \hline
{\textbf{Notation}} & {\textbf{Definition}} \\ \hline
$n$ & number of flows inside the network  \\ \hline
$b_i$ & token bucket size of flow $i$ \\ \hline
$\boldsymbol{b}$ & vector of token bucket sizes across $n$ flows: $(b_1, b_2, \ldots, b_n)$  \\ \hline
$b'_i$ & reprofiled token bucket size of flow $i$  \\ \hline
$\boldsymbol{b}'$ & vector for all reprofiled token bucket sizes across $n$ flows: $(b'_1, b'_2, \ldots, b'_n)$ \\ \hline
$b'^*_i$ & optimal bucket size of flow $i$'s reprofiler under static priority \\ \hline
$\boldsymbol{b}'^*$ & vector for all optimal reprofiled token bucket sizes across $n$ flows: $(b'^*_1, b'^*_2, \ldots, b'^*_n)$ under static priority\\ \hline
$\widehat{b}'^*_i$ & optimal bucket size of flow $i$'s reprofiler under fifo \\ \hline
$\widehat{\boldsymbol{b}}'^*$ & vector for all optimal reprofiled token bucket sizes across $n$ flows: $(\hat{b}'^*_1, \hat{b}'^*_2, \ldots, \hat{b}'^*_n)$ under fifo \\ \hline
$B'_i$ & cumulative reprofiled token bucket size for flows with a priority no smaller than $i$, \ie $\sum_{j = i}^n b'_j$  \\ \hline
$\widehat{B}_i$ & cumulative token bucket burst size for flows~$1$ to~$i$, \ie $\sum_{j = 1}^i b_j$  \\ \hline
$\widehat{B}'_i$ & cumulative reprofiled token bucket size for flows from~$1$ to~$i$, \ie $\sum_{j = 1}^i b'_j$  \\ \hline
$d_i$ & end-to-end deadline for flow $i$ \\ \hline
$\boldsymbol{d}$ & vector of end-to-end deadlines for $n$ flows: $(d_1, d_2, \ldots, d_n)$ \\ \hline
$D^*_i$ & worst-case end-to-end delay for flow $i$ under priority+reprofiling \\ \hline
$\widehat{D}^*_i$ & worst-case end-to-end delay for flow $i$ under fifo+reprofiling \\ \hline
$r_i$ & token bucket rate of flow $i$  \\ \hline
$\boldsymbol{r}$ & vector for all token bucket rates across $n$ flows: $(r_1, r_2, \ldots, r_n)$ \\ \hline
$(r_i, b_i)$ & traffic profile of flow $i$ \\ \hline
%$(r'_i, b'_i)$ & flow $i$'s reshaping parameters \\ \hline
%$\boldsymbol{R}$ & vector for bandwidths of all the shared links \\ \hline
$R_i$ & cumulative token bucket rates for flows with a priority no smaller than $i$, \ie $\sum_{j = i}^n r_j$  \\ \hline
$R$ & shared link bandwidth \\ \hline
$R^*$ & optimal minimum required bandwidth \\ \hline
$\widetilde{R}^*$ & minimum required link bandwidth in the absence of reprofiling  under static priority \\ \hline
$\widetilde{R}^*_R$ & minimum required link bandwidth with reprofiling under static priority \\ \hline
$\widehat{R}^*$ & minimum required link bandwidth in the absence of reprofiling under fifo \\ \hline
$\widehat{R}^*_R$ & minimum required link bandwidth with reprofiling under fifo \\ \hline
$t$ & time  \\ \hline
$H_i$ & $b_i - d_i r_i$ \\ \hline
$\Pi_i(R)$ & $\frac{r_i + R - R_{i+1}}{R-R_{i+1}}$ \\ \hline
$V_i(R)$ & $d_i(R - R_{i+1} - b_i)$ \\ \hline
$\mathbb{S}_1(R)$ & ${V_1(R)}$ \\ \hline
$\mathbb{S}_i(r)$ & $\mathbb{S}_{i-1}(R) \bigcup {V_i(R)}\bigcup   \left\{ \frac{s-H_i}{\Pi_i(R) } | s \in \mathbb{S}_{i-1}(R) \right\}$ \\ \hline
$\mathbb{Z}_i$ & the set of integers from~$1$ to~$i$, \ie $\left\{ 1 \leq i \leq j | j \in \mathbb{Z} \right\}$ \\ \hline
$X_F(R)$ & $\max_{P_1, P_2 \subseteq \mathbb{Z}_n,P_2 \neq \mathbb{Z}_n, P_1 \bigcap P_2 = \emptyset} \frac{\sum_{i \in P_1} \frac{R H_i}{R+r_i} + \sum_{i \in P_2} \left(b_i - \frac{r_i d_i R}{R_1} \right)}{1 - \sum_{i \in P_1}\frac{r_i}{R+r_i} - \sum_{i \in P_2} \frac{r_i}{R_1}}$ \\ \hline
$Y_F(R)$ & $\min_{1 \leq i \leq n-1}\left\{\widehat{B}_n, Rd_n, 
\min_{P_1, P_2 \subseteq \mathbb{Z}_i, P_1 \bigcap P_2 = \emptyset, P_1 \bigcup P_2 \neq \emptyset} \left\{\frac{\widehat{B}_i - \sum_{j \in P_1} \frac{R H_j}{R+r_j} -  \sum_{j \in P_2} \left( b_j - \frac{r_j d_j R}{R_1}\right)}{\sum_{j \in P_1}\frac{r_j}{R+r_j} +  \sum_{j \in P_2} \frac{r_j}{R_1}} \right\} \right\}$ \\ \hline
$T_i(\widehat{B}'_n, R)$ & $\max \left\{0, \frac{R}{R+r_i}\left(H_i + \frac{r_i}{R}\widehat{B}'_n\right), b_i + \frac{r_i(\widehat{B}'_n - R d_i)}{R_1} \right\}$ \\ \hline
$\Gamma_{sc}$ & a service curve assignment that gives each flow $i$ a service curve of $SC_i(t)$ \\ \hline
\textbf{OPT\_\ding{111}} & general optimization where \textbf{\ding{111}=S,F} for static priority and fifo  \\ \hline
\textbf{OPT\_R\ding{111}} & optimization with reprofiling where \textbf{\ding{111}=S,F} for static priority and fifo \\ \hline
  \end{longtable}
\end{center}

\begin{comment}
Notations used only in the Appendix
\begin{center}
  \begin{tabular}{ c | c }
    \hline
{\textbf{Notation}} & {\textbf{Definition}} \\ \hline
$t$ & system time \\ \hline
$D_1(t)$ & virtual delay inside the (re)shaper \\ \hline
$D_2(t)$ & virtual delay at the shared link \\ \hline
$D(t)$ & virtual delay inside the system with (re)shaper \\ \hline
$D'(t)$ & virtual delay inside the system without (re)shaper \\ \hline
$\alpha(t)$ & general arrival curve \\ \hline
$\beta(t)$ & general service curve \\ \hline
$\delta(t)$ & general maximum service curve \\ \hline
$\gamma_(r,b)(t)$ & arrival curve of token-bucket flow $(r,b)$ \\ \hline
$\beta_{R,T}(t)$ & service curve of a rate-latency server \\ \hline
$R^*_i$ & optimal solution for {\bf OPT\_P}'s $i$th suboptimization \\ \hline
$L_i$ & Lagrangian function for {\bf OPT\_P}'s $i$th suboptimization  \\ \hline
$\lambda_i$ & Lagrangian multiplier\\
\hline
  \end{tabular}
\end{center}
\end{comment}
\newpage
\section{Proofs for Dynamic Priority Scheduler}
\label{app:one_dyn}
%%%%%%%%%%%%%%%%%%%%%%%%%%%%%%%%%%%%%%%%%%%%%%%%%%%%%%%%%%%%%%%%%%%%%%%
%%%%%%%%%%%%%%%%%%%%%%%%%%% prop: shifted ac%%%%%%%%%%%%%%%%%%%%%%%%%&&
%%%%%%%%%%%%%%%%%%%%%%%%%%%%%%%%%%%%%%%%%%%%%%%%%%%%%%%%%%%%%%%%%%%%%%%
\subsection{Proof for Proposition~\ref{prop:shifted_ac}}\label{app:one_hop_dynamic}
For the reader's convenience, we restate Proposition~\ref{prop:shifted_ac}.

\begin{flushleft}
P{\scriptsize ROPOSITION}~\ref{prop:shifted_ac}. \emph{Consider a link shared by $n$ token bucket controlled flows, where flow $i, 1 \leq i \leq n$, has a traffic profile $(r_i, b_i)$ and a deadline $d_i$, with $d_1 > d_2 > ... > d_n$ and $d_1<\infty$. Consider a service-curve assignment $\Gamma_{sc}$ that allocates flow $i$ a service curve of 
\beq\tag{\ref{eq:sc}}
SC_i(t) = \left\{
\begin{aligned}
& 0 &\text{ when } t < d_i, \\
& b_i + r_i(t - d_i) &\text{otherwise}.
\end{aligned}
\right.
\eeq
Then 
\begin{enumerate}
\item For any flow $i, 1 \leq i \leq n$, $SC_i(t)$ ensures a worst-case end-to-end delay no larger than $d_i$.
\item Realizing $\Gamma_{sc}$ requires a link bandwidth of at least
\beq\tag{\ref{eq:global_min_R}}
R^* = \max_{1 \leq h \leq n}\left\{\sum_{i = 1}^n r_i, \frac{\sum_{i = h}^n b_i + r_i (d_h - d_i)}{d_h} \right\}.
\eeq
\item Any scheduling mechanism capable of meeting all the flows' deadlines requires a bandwidth of at least $R^*$.
\end{enumerate}}
\end{flushleft}

\begin{proof}
We first show that under $\Gamma_{sc}$ each flow meets its deadline, and then show that a bandwidth of $R^*$ is enough to accommodate all the service curves defined in $\Gamma_{sc}$. Next, we show that no mechanism exists than can meet the deadlines with a bandwidth strictly smaller than $R^*$. 
\begin{itemize}
\item For any flow $1 \leq i \leq n$, its token bucket constrained arrival curve is of the form 
$$AC_i(t) = \left\{ 
\begin{aligned}
& 0 & \text{ when } t = 0\\
& b_i + r_i t & \text{ otherwise}.
\end{aligned}
\right.$$
Combining it with flow $i$'s service curve $SC_i(t)$, the worst-case end-to-end delay for flow $i$ is of the form\footnote{See THEOREM 1.4.2 in~\cite{nc}, page 23.},
$$D^*_i = \sup_{t \geq 0} \inf_{\tau \geq 0} \left\{AC_i(t) \leq SC_i(t + \tau) \right\}  = d_i.$$

\item 

%\newtext{
To accommodate all the service curves in $\Gamma_{sc}$, the system needs a bandwidth $R$ such that $Rt \geq \sum_{i = 1}^n SC_i(t)$ for all $t > 0$, \ie

\beq\label{proof:opt_sc}
R \geq \sup_{t > 0} \frac{\sum_{i = 1}^n SC_i(t)}{t}.
\eeq

Towards establishing that the minimum link bandwidth $R^{*} = \sup_{t > 0} \frac{\sum_{i = 1}^n SC_i(t)}{t}$ is captured by \Eqref{eq:global_min_R}, we first introduce another proposition:

\begin{proposition}\label{prop:sup_linar}
Assume that $PL(t)$ is a wide-sense increasing, piecewise linear, and right continuous function defined by a finite set of $k$ linear segments:
$$
pl_i(t) = c_i + s_it, \ t \in  [t_{i}, t_{i+1}), \ 1 \leq i \leq k
$$
where $s_i \geq 0$ and $c_i$ are the slopes and intercepts of segment $pl_i(t)$ respectively, and $t_{i} < t_{i+1},\, \forall i, 1\leq i\leq k,$
%% RG:  I don't think the next statement is needed here, and it was also somewhat confusing as the index $i$ can represent either a left or right boundary...
%%
%is the left boundary of each linear segment, 
with $t_1 = 0$ and $t_{k+1} = \infty$.  Furthermore, $PL(0)=0$ so that $c_1 = 0$.

Then, to compute $\sup_{t > 0} \frac{PL(t)}{t}$, it is sufficient to consider values of $\frac{PL(t)}{t}$ at the following times $t$:
\begin{enumerate}[nosep]
    \item $t \rightarrow t_{k+1} = \infty$
    \item interval boundaries $t_{i}, \ 2 \leq i \leq k,$ when one of the following conditions is met:
    \begin{enumerate}
        \item $PL(t)$ is continuous and the slope of $PL(t)$ decreases, \ie $s_{i-1} > s_{i}$.
        \item $PL(t)$ is discontinuous.
    \end{enumerate}
\end{enumerate}
\end{proposition}

\begin{proof}
We readily know that to find $\sup_{t > 0} \frac{PL(t)}{t}$, it is sufficient to consider interval boundaries $t_{i}$s, since $\frac{pl_i(t)}{t} = \frac{c_i}{t} + s_i$, $\frac{PL(t)}{t}$ is a decreasing (increasing) function of $t$ within any interval $[t_i, t_{i+1})$ when $c_i \geq 0$ $(c_i < 0)$.

Consider first an intermediate boundary $t_{i}, 2 \leq i \leq k$, we first argue that if $PL(t)$ is \emph{continuous}, then $\sup_{t > 0} \frac{PL(t)}{t}$ cannot be achieved at any boundary $t_{i}$ for which $s_{i-1} < s_i$, \ie a boundary where the slope of $PL(t)$ increases.  Towards establishing the result, consider the following two cases:
\begin{enumerate}[nosep]
    \item $c_{i} < 0$.  In this case, irrespective of $s_i$, $\frac{PL(t)}{t}$ is increasing in $[t_i, t_{i+1})$ and $\sup_{t \in[t_i,t_{i+1})} \frac{PL(t)}{t}=\frac{PL(t_{i+1}^{-})}{t_{i+1}^{-}}$, so that $\sup_{t > 0} \frac{PL(t)}{t}$ cannot be achieved at $t_{i}$.
    \item $c_{i} \geq 0$.  Since we have assumed that $PL(t)$ is continuous, we have that $pl_{i-1}(t_i) = c_{i-1} + s_{i-1}t_i = pl_{i}(t_i) = c_{i} + s_{i}t_i$.  This implies $c_{i-1} = c_{i} + (s_{i} - s_{i-1})t_i > c_{i} \geq 0$ since $s_{i}> s_{i-1}$.  Since both $c_{i}$ and $c_{i-1}$ are non-negative, $\frac{PL(t)}{t}$ is a decreasing function throughout $[t_{i-1},t_{i+1})$.  Hence, $\sup_{t > 0} \frac{PL(t)}{t}$ cannot be realized at $t_{i}$.
\end{enumerate}

Turning next to the case where $PL(t)$ is discontinuous at some $t_i$, then $\lim_{t \rightarrow t_i^-}\frac{PL(t)}{t} < \lim_{t \rightarrow t_i^+}\frac{PL(t)}{t}$ since $PL(t)$ is wide-sense increasing.  Since $PL(t)$ is right continuous, $\frac{PL(t_i)}{t_i} = \lim_{t \rightarrow t_i^+}\frac{PL(t)}{t}$, which implies it is sufficient to consider $t_i$ for computing the supremum.

Finally, we consider the two extreme boundaries $t_1 = 0$ and $t_{k+1} = \infty$. For $t_1=0$, since $c_1=0$, we have 
$\frac{PL(t)}{t} = s_1, \forall\, t\in[t_1=0,t_2).$  Since $t_2$ is one of the intermediate boundaries covered by the first part of our proof, we do not need to consider $t_1$ when computing the supremum of $\frac{PL(t)}{t}$
%$\lim_{t\rightarrow0}\frac{PL(t)}{t} = s_1 = \lim_{t \rightarrow t_2^-}\frac{PL(t)}{t}$.  
For $t_{k+1} = \infty$, $\lim_{t \rightarrow \infty}\frac{PL(t)}{t} = s_k$.  
%Hence, $\sup_{t > 0} \frac{PL(t)}{t}$ may be realized at $t_{k+1} = \infty$, or some intermediate boundaries $t_{i}, \ 2 \leq i \leq k$.

Combining the fact that $\sup_{t > 0} \frac{PL(t)}{t}$ can only be realized at an interval boundary with the above establishes that it is realized at either an interval boundary $t_{i}$ where $PL(t)$ is continuous and experiencing a slope decrease, \ie $s_{i-1} > s_i$, or at an interval boundary where $PL(t)$ is discontinuous, or at $t_{k+1}$, \ie $t\rightarrow\infty$.
\end{proof}

Returning to the proof of Proposition~\ref{prop:shifted_ac}, the aggregate service curve $\sum_{i = 1}^n SC_i(t)$ is a wide-sense increasing, piecewise linear function since each $SC_i(t)$ is, by definition, wide-sense increasing and piecewise linear.  From Proposition~\ref{prop:sup_linar} we know that the supremum of $\frac{\sum_{i = 1}^n SC_i(t)}{t}$ is either $\lim_{t \rightarrow \infty}\frac{\sum_{i = 1}^n SC_i(t)}{t}=\sum_{i = 1}^n r_i$ or is achieved at a boundary value $d_k \in\left\{d_1, d_2, \ldots , d_n \right\}$. 
Combining this with the expressions for the individual service curves $SC_i(t), i=1,\ldots,n,$ gives
$$
R \geq \max_{1 \leq h \leq n} \left\{\sum_{i = 1}^n r_i,\ \frac{\sum_{i = h}^n b_i + r_i (d_h - d_i)}{d_h} \right\} = R^*.
$$
Thus, $R^*$ is sufficient to accommodates all service curves in the service curves assignment $\Gamma_{sc}$.

%}
%%%%%%%%%%%%%%%%%%%%%%%%%%%%%%%%%%%%%%%%%%%%%%%%%%%%%%%%%%%%%%%%
%% RG: Old proof - commented out
%%%%%%%%%%%%%%%%%%%%%%%%%%%%%%%%%%%%%%%%%%%%%%%%%%%%%%%%%%%%%%%%
\begin{comment}
\sout{To accommodate all the service curves defined in $\Gamma_{sc}$, the system needs a bandwidth $R$ such that 1) $R \geq \sum_{i = 1}^n r_i$, which guarantees a finite worst-case end-to-end delay for any flow, and 2) for all $t > 0$, $Rt \geq \sum_{i = 1}^n SC_i(t)$, \ie}

% \beq\label{proof:opt_sc}
\beq
R \geq \max\left\{\sum_{i = 1}^n r_i, \ \sup_{t > 0} \frac{\sum_{i = 1}^n SC_i(t)}{t} \right\}.
\eeq

\sout{As $\frac{SC_i(t)}{t}$ equals~$0$ when $t < d_i$, and decreases with $t$ when $t \geq d_i$, we know that the supremum of $\frac{\sum_{i = 1}^n SC_i(t)}{t}$ is achieved only at values among $\left\{d_1, d_2, \ldots , d_n \right\}$. Combining this with the expression for $SC_i(t)$ gives}
$$
R \geq \max_{1 \leq h \leq n} \left\{\sum_{i = 1}^n r_i,\ \frac{\sum_{i = h}^n b_i + r_i (d_h - d_i)}{d_h} \right\} = R^*.
$$
\sout{Thus, $R^*$ is enough to accommodates all service curves defined in $\Gamma_{sc}$.}
%%%%%%%%%%%%%%%%%%%%%%%%%%%%% bound %%%%%%%%%%%%%%%%%%%%%%%%%%%%%%%%%%
\end{comment}
\item Next, we show that $R^*$ is a lower bound for the minimum required bandwidth. Note that $R^* \geq \sum_{i = 1}^n r_i$, so that if the deadlines can be met with $R^* = \sum_{i = 1}^n r_i$ no improvement is feasible. Below we consider the case when $R^* > \sum_{i = 1}^n r_i$, \ie there exists $1 \leq \hat{h} \leq n$, such that $R^* = \frac{\sum_{i = \hat{h}}^n b_i + r_i ( d_{\hat{h}} - d_i)}{d_{\hat{h}}}> \sum_{i = 1}^n r_i$

Suppose there exists a mechanism achieving a minimum required bandwidth $R' < R^*$. Next we construct an arrival pattern consistent with each flow's token bucket arrival constraints, such that $R'$ cannot satisfy all flows' deadlines.

Consider the arrival pattern such that for all $1 \leq i \leq n$, flow $i$ sends $b_i$ at $t=0$ (where the system restarts the clock), and then constantly sends at a rate of $r_i$. 
By time $d_{\hat{h}}$, to satisfy the deadlines of all flows~$i$ with $d_i \leq d_{\hat{h}}$, \ie $i\geq \hat{h}$, the link must have transmitted at least $b_i + r_i(d_{\hat{h}} - d_i)$ amount of data for each such flow. Consequently, by $d_{\hat{h}}$ the link should have cumulatively transmitted at least $\sum_{i = \hat{h}}^n b_i + r_i(d_{\hat{h}} - d_i)$. As $\sum_{i = \hat{h}}^n b_i + r_i(d_{\hat{h}} - d_i) = d_{\hat{h}} R^* > d_{\hat{h}} R'$, a bandwidth of $R'$ must violate some flows' deadlines. 
\end{itemize}
\end{proof}

%%%%%%%%%%%%%%%%%%%%%%%%%%%%%%%%%%%%%%%%%%%%%%%%%%%%%%%%%%%%%%%%%%%%%%
%%%%%%%%%%%%%%%%%%%%%%%%%%%%%%%% edf %%%%%%%%%%%%%%%%%%%%%%%%%%%%%%%%%
%%%%%%%%%%%%%%%%%%%%%%%%%%%%%%%%%%%%%%%%%%%%%%%%%%%%%%%%%%%%%%%%%%%%%%
\subsection{Proof for Proposition~\ref{prop:edf}}
\label{app:edf}
\begin{flushleft}
P{\scriptsize ROPOSITION}~\ref{prop:edf}. \emph{Consider a link shared by $n$ token bucket controlled flows, where flow $i, 1 \leq i \leq n$, has traffic profile $(r_i, b_i)$ and deadline $d_i$, with $d_1 > d_2 > ... > d_n$ and $d_1<\infty$. The earliest deadline first (EDF) scheduler realizes $\Gamma_{sc}$ under a link bandwidth of $R^*$.}
\end{flushleft}

\begin{proof}
We first show that EDF satisfies $\Gamma_{sc}$, and then shows that EDF requires a minimum required bandwidth of $R^*$.

We show that EDF satisfies $\Gamma_{sc}$ by contradiction. Suppose EDF cannot achieve $\Gamma_{sc}$. Then there exists $\hat{t} \geq d_i$ and $1 \leq i \leq n$, such that $\widehat{SC}_i(\hat{t}) < SC_i(\hat{t})$, where $\widehat{SC}_i$ is the service curve that EDF assigns to flow $i$. To satisfy flow $i$'s deadline, at $\hat{t} - d_i$ EDF should yield a virtual delay no large than $d_i$, \ie
$$\inf_{\tau \geq 0}\left\{b_i + r_i(\hat{t} - d_i) \leq \widehat{SC}_i(\hat{t} - d_i + \tau)\right\} \leq d_i,$$
which then gives $\widehat{SC}_i(\hat{t} - d_i + d_i) \geq b_i + r_i(\hat{t} - d_i)$. As $b_i + r_i(\hat{t} - d_i) = SC_i(\hat{t})$, this contradicts the assumption that $\widehat{SC}_i(\hat{t}) < SC_i(\hat{t})$.

Next, we show that EDF requires a minimum bandwidth of $R^*$.
Suppose flow $i$'s data sent at $t$ has a deadline of $(t + d_{i})$.
We show that EDF satisfies all flows' deadlines with a bandwidth of $R^* = \max_{1 \leq h \leq n}\left\{\sum_{i = 1}^n r_i, \frac{\sum_{i = h}^n b_i + r_i (d_h - d_i)}{d_h} \right\}$. Based on the utilization of the shared link, we consider two cases separately, where in both cases we prove the result by contradiction. Specifically, suppose that under EDF, $R^*$ cannot satisfy all flows' latency requirements. Then there exists $1 \leq \hat{h} \leq n$ and $\hat{t} \geq 0$, such that EDF processes at least one bit sent by flow $\hat{h}$ at time $\hat{t}$ after time $(\hat{t} + d_{\hat{h}})$. We consider first the case where the shared link uses up all its bandwidth during the period $[0,\hat{t}+ d_{\hat{h}}]$ to transmit data with absolute deadlines no larger than $(\hat{t} + d_{\hat{h}})$, and then consider the case where there exists $t_0 \in [0,\hat{t}+ d_{\hat{h}}]$, such that at $t_0$ the shared link is not busy with data whose absolute deadline is no larger than $(\hat{t} + d_{\hat{h}})$. Showing that $R^*$ is enough for EDF to meet all flows' deadlines, establishes the result. 

\begin{enumerate}
%%%%%%%%%%%%%%%% case 1 %%%%%%%%%%%%%%%%%%%%%
\item Consider the case where for all $t \in [0, \hat{t} + d_{\hat{h}}]$ the shared link uses up all its bandwidth to send bits with an absolute deadline no larger than $(\hat{t} + d_{\hat{t}})$. Then by $(\hat{t} + d_{\hat{h}})$ the shared link cumulatively has processed $R^*(\hat{t}+ d_{\hat{h}})$ amount of data that all have deadlines no larger than $(\hat{t} + d_{\hat{t}})$. From the fact that EDF violates an absolute deadline of $(\hat{t} + d_{\hat{h}})$, we know that there exists an arrival pattern consistent with the token bucket constraints, such that cumulatively flows send more than $R^* ( \hat{t} + d_{\hat{h}})$ amount of data with absolute deadlines no larger than $(\hat{t} + d_{\hat{h}})$. From the token bucket constraints, we know that by $(\hat{t} + d_{\hat{h}})$, flows can send at most $\sum_{i = 1}^n \left[b_i + r_i(\hat{t} + d_{\hat{h}}-d_i)\right]\bf{I}_{\hat{t} + d_{\hat{h}}-d_i \geq 0}$ amount of data whose absolute deadline is at most $(\hat{t} + d_{\hat{h}})$. Therefore, we have $R^* ( \hat{t} + d_{\hat{h}}) < \sum_{i = 1}^n \left[b_i + r_i(\hat{t} + d_{\hat{h}}-d_i)\right]\bf{I}_{\hat{t} + d_{\hat{h}}-d_i \geq 0}$. Define $d_0 = \infty$. There exists $1 \leq \hat{n} \leq n$ such that $\hat{t} + d_{\hat{h}} \in [d_{\hat{n}}, d_{\hat{n}-1})$ so that 
\bequn
R^* < \frac{\sum_{i = 1}^n \left[b_i + r_i(\hat{t} + d_{\hat{h}}-d_i)\right]\bf{I}_{\hat{t} + d_{\hat{h}}-d_i \geq 0}}{\hat{t} + d_{\hat{h}}} =  \sum_{i = \hat{n}}^n \left( r_i  + \frac{ b_i - r_i d_i}{\hat{t} + d_{\hat{h}}} \right):= R'.
\eequn

If $\sum_{i = \hat{n}}^n b_i - r_i d_i \leq $ $0$, we have $R^* < R' \leq \sum_{i = 1}^n r_i$, which contradicts to $R^* > \sum_{i = 1}^n r_i$. Hence we consider only $\sum_{i = \hat{n}}^n b_i - r_i d_i > 0$, where $R'$ decreases with $(\hat{t}+d_{\hat{h}})$. Define $\hat{R}(u) = \sum_{i = \hat{n}}^n \left( r_i  + \frac{ b_i - r_i d_i}{u} \right)$. We then have 
\bequn
R^* < R' \leq \hat{R}( d_{\hat{n}}) = \frac{\sum_{i = \hat{n}}^n \left[b_i + r_i(d_{\hat{n}}-d_i)\right]}{d_{\hat{n}}},
\eequn  
which contradicts to the definition of $R^*$.
%%%%%%%%%%%%%%%% case 2 %%%%%%%%%%%%%%%%%%%%%
\item Otherwise, the shared link uses less than all its bandwidth at $t_0 \in [0, \hat{t} + d_{\hat{h}}]$, and uses up all its bandwidth for all $t \in (t_0, \hat{t} + d_{\hat{h}}]$ to send bits with an absolute deadline no larger than $(\hat{t} + d_{\hat{t}})$. Then during $(t_0, \hat{t} + d_{\hat{h}}]$ the shared link processes $R^*(\hat{t} + d_{\hat{h}} - t_0)$ amount of data with absolute deadlines no larger than $(\hat{t} + d_{\hat{h}})$, and flows send strictly more than $R^*(\hat{t} + d_{\hat{h}} - t_0)$ amount of data with absolute deadlines no larger than $(\hat{t} + d_{\hat{h}})$. From the token bucket constraints, we know that during $(t_0, \hat{t} + d_{\hat{h}}]$ flows can send at most $\sum_{i = 1}^n \left[b_i + r_i(\hat{t} + d_{\hat{h}} - t_0 -d_i)\right]\bf{I}_{\hat{t} + d_{\hat{h}} - t_0 -d_i \geq 0}$ amount of data whose absolute deadlines are no larger than $(\hat{t} + d_{\hat{h}})$. Thus we have $R^* < \frac{\sum_{i = 1}^n \left[b_i + r_i(\hat{t} + d_{\hat{h}} - t_0 -d_i)\right]\bf{I}_{\hat{t} + d_{\hat{h}} - t_0 -d_i \geq 0}}{d_{\hat{h}}+ \hat{t} - t_0}$, which, as before, contradicts the definition of $R^*$.
\end{enumerate}
\end{proof}

\subsection{Proof for Proposition~\ref{prop:sc_noshaper}}
\label{app:sc_noshaper}

\begin{flushleft}
P{\scriptsize ROPOSITION}~\ref{prop:sc_noshaper}. \emph{Consider a link shared by $n$ token bucket controlled flows, where flow $i, 1 \leq i \leq n$, has traffic profile $(r_i, b_i)$ and deadline $d_i$, with $d_1 > d_2 > ... > d_n$ and $d_1<\infty$. Reprofiling flows will not decrease the minimum bandwidth required to meet the flows' deadlines.}
\end{flushleft}

\begin{proof}
We show that adding a reprofiler to any of the flow does not decrease the optimal minimum required bandwidth. 

Suppose we apply a reprofiler $b'_i$ to flow $i$, where $1 \leq i \leq n$ and $b'_i \geq b_i - d_i r_i$. Then flow $i$ incurs a reshaping delay of~$\frac{b_i - b'_i}{r_i}$.  This then leaves a maximum in-network delay of~$d_i - \frac{b_i - b'_i}{r_i}$, which is greater than~$0$ since $b'_i \geq b_i - d_i r_i$. Hence, to meet its deadline flow~$i$ requires a service curve of at least
$$
SC'_i(t) = \left\{\begin{aligned}
&0, &\text{when } t < d_i - \frac{b_i - b'_i}{r_i}\\
&b'_i + r_i \left(t - d_i +  \frac{b_i - b'_i}{r_i}\right) = b_i + r_i( t - d_i), &\text{otherwise}
\end{aligned} \right.
$$
which is greater than 
$SC_i(t) = \left\{
\begin{aligned}
&0, &\text{when } t < d_i \\
& b_i + r_i(t - d_i), &\text{otherwise}
\end{aligned}
\right.$. Consequently, according to Proposition~\ref{prop:shifted_ac} the system needs a bandwidth of at least 
\beq
R = \max \left\{\sum_{i = 1}^n r_i, \sup_{t \geq 0} \frac{\sum_{j \neq i} SC_j(t) + SC'_i(t)}{t} \right\} \geq \max \left\{\sum_{i = 1}^n r_i, \sup_{t \geq 0} \frac{ \sum_{1 \leq j \leq n} SC_j(t)}{t} \right\} = R^*_{sc},
\eeq
to meet each flow's deadline.  
\end{proof}

\section{Proofs for Static Priority Scheduler}
\label{app:one_static}

\subsection{Proof for Proposition~\ref{prop:order}}
\label{app:static_order}
We actually prove Proposition~\ref{prop:order} under the more general packet-based model. By assuming that all packets have a length of~$0$, the packet-based model defaults to the fluid model.

\begin{flushleft}
P{\scriptsize ROPOSITION}~\ref{prop:order}. \emph{Consider a link shared by $n$ token bucket controlled flows, where flow $i, 1 \leq i \leq n$, has traffic profile $(r_i, b_i)$ and deadline $d_i$, with $d_1 > d_2 > ... > d_n$ and $d_1<\infty$. Under a static-priority scheduler, there exists an assignment of flows to priorities that minimizes link bandwidth while meeting all flows deadlines such that flow~$i$ is assigned a priority strictly greater than that of flow~$j$ only if $d_i<d_j$.}
\end{flushleft}

\begin{proof}
For a mechanism $\Gamma$, denote flows with priority $h$ under $\Gamma$ as $G_h(\Gamma)$. Define $d^{(max)}_h(\Gamma) = \max_{i \in G_h(\Gamma)} d_i$ and $d^{(min)}_h(\Gamma) = \min_{i \in G_h(\Gamma)} d_i$. Assume priority class $h+1$ has a higher priority than priority class $h$. Then we will prove the proposition by induction on the number $k$ of priority classes. Our induction hypothesis $S(k)$ is expressed in the following statement:

\begin{flushleft}
$S(k)$: For link shared by any number of flows, there exists an optimal $k$-priority mechanism $\Gamma_k$ such that $\forall s < l \leq k$, $d^{(max)}_l(\Gamma_k) < d^{(min)}_s(\Gamma_k)$.
\end{flushleft}
%%%%%%%%%%%%%%%%%%%%%%%%%%%% base case %%%%%%%%%%%%%%%%%%%%%%%%%%%%%%
\begin{itemize}
\item \textbf{Base case:} consider the case when $k=2$. We show that for any mechanism $\Gamma_2$, if $d^{(min)}_1(\Gamma_2) < d^{(max)}_2(\Gamma_2)$, then there exists a 2-priority mechanism $\Gamma'_2$ such that $d^{(min)}_1(\Gamma'_2) > d^{(max)}_2(\Gamma'_2)$ and $R^*(\Gamma'_2) \leq R^*(\Gamma_2)$.

For mechanism $\Gamma_2$, denote $l_1^{(max)}(\Gamma_2)$ to be the maximum packet size for flows in $G_1(\Gamma_2)$. To satisfy each flow's deadline, it requires a bandwidth $R$ such that
\beq\label{order:base_case}
\left\{ \quad
\begin{aligned}
&\frac{\sum_{i \in G_2(\Gamma_2)} b_i + l_1^{(max)}(\Gamma_2)}{R} \leq d^{(min)}_2(\Gamma_2), \\
&\frac{\sum_i b_i}{R - \sum _{i \in G_2(\Gamma_2)} r_i} \leq d^{(min)}_1(\Gamma_2), \\
&\sum_{i = 1}^n r_i \leq R,
\end{aligned}
\right.
\eeq
which gives 
$$R^*(\Gamma_2) = \max\left\{\ \sum_{i = 1}^n r_i, \frac{\sum_{i \in G_2(\Gamma_2)} b_i + l_1^{(max)}(\Gamma_2)}{d^{(min)}_2(\Gamma_2)}, \frac{\sum_i b_i}{d^{(min)}_1(\Gamma_2)} + \sum _{i \in G_2(\Gamma_2)} r_i \ \right\}.$$

Define $G'_1(\Gamma_2) = \{ i \in G_2(\Gamma_2)\ | \ d_i > d^{(max)}_1(\Gamma_2) \}$ and $G'_2(\Gamma_2) = G_2(\Gamma_2) - G'_1(\Gamma_2)$. Consider the mechanism $\Gamma'_2$ such that $G_2(\Gamma'_2) = G'_2(\Gamma_2)$, and $G_1(\Gamma'_2) = G'_1(\Gamma_2) \cup G_1(\Gamma_2)$. Note that $d^{(min)}_1(\Gamma'_2) > d^{(max)}_2(\Gamma'_2)$, and note also that $d^{(min)}_i(\Gamma'_2) = d^{(min)}_i(\Gamma_2), i = 1,2$. Similar as \Eqref{order:base_case}, we have

\resizebox{.9 \textwidth}{!} 
{
$R^*(\Gamma'_2) = \max\left\{\ \sum_{i = 1}^n r_i, \frac{\sum_{i \in G_2(\Gamma'_2)} b_i + l_1^{(max)}(\Gamma'_2)}{d^{(min)}_2(\Gamma_2)}, \frac{\sum_i b_i}{d^{(min)}_1(\Gamma_2)} + \sum_{i \in G_2(\Gamma_2) - G'_1(\Gamma_2)}r_i \ \right\}$
}

Note that $\frac{\sum_i b_i}{d^{(min)}_1(\Gamma_2)} + \sum _{i \in G_2(\Gamma_2)} r_i \geq \frac{\sum_i b_i}{d^{(min)}_1(\Gamma_2)} + \sum_{i \in G_2(\Gamma_2) - G'_1(\Gamma_2)}r_i$.
Next we show $ \sum_{i \in G_2(\Gamma_2)} b_i + l_1^{(max)}(\Gamma_2) \geq \sum_{i \in G_2(\Gamma'_2)} b_i + l_1^{(max)}(\Gamma'_2)$, from which we then have $R^*(\Gamma_2) \geq R^*(\Gamma'_2)$, and therefore $S(2)$. Since $G_1(\Gamma_2) \varsubsetneqq G_1(\Gamma'_2)$, it has $l_1^{(max)}(\Gamma'_2) \geq l_1^{(max)}(\Gamma_2)$.
\begin{itemize}
\item When $l_1^{(max)}(\Gamma'_2) = l_1^{(max)}(\Gamma_2)$, from $G_2(\Gamma'_2) \varsubsetneqq G_2(\Gamma_2)$ we have $ \sum_{i \in G_2(\Gamma_2)} b_i \geq \sum_{i \in G_2(\Gamma'_2)} b_i $, and therefore have $ \sum_{i \in G_2(\Gamma_2)} b_i + l_1^{(max)}(\Gamma_2) \geq \sum_{i \in G_2(\Gamma'_2)} b_i + l_1^{(max)}(\Gamma'_2)$.
\item When $l_1^{(max)}(\Gamma'_2) > l_1^{(max)}(\Gamma_2)$, \ie there exists a flow $\hat{i} \in G'_1(\Gamma_2)$ such that $l_{\hat{i}} >  l_1^{(max)}(\Gamma_2)$. From $b_{\hat{i}} \geq l_{\hat{i}}$ and $ G_2(\Gamma'_2) \subseteq G_2(\Gamma_2) - \hat{i}$, we have
\bequn
\begin{aligned}
 \sum_{i \in G_2(\Gamma_2)} b_i + l_1^{(max)}(\Gamma_2) 
= &  \sum_{i \in G_2(\Gamma_2) - \hat{i}} b_i + b_{\hat{i}}+ l_1^{(max)}(\Gamma_2)  
\geq \sum_{i \in G_2(\Gamma'_2)} b_i + l_{\hat{i}} + l_1^{(max)}(\Gamma_2)\\
> & \sum_{i \in G_2(\Gamma'_2)} b_i + l_1^{(max)}(\Gamma'_2)
\end{aligned}
\eequn
\end{itemize}

%%%%%%%%%%%%%%%%%%%%%%% induction step %%%%%%%%%%%%%%%%%%%%%%%%%%%%%%%
\item \textbf{Induction Step:} Let $k \geq 2$ and suppose $S(k)$ holds. Next, we show that $S(k+1)$ holds.

Consider any (k+1)-priority mechanism $\Gamma_{k+1}$. For $1 \leq h \leq (1+k)$, denote $l_h^{(max)}(\Gamma_{k+1})$ to be the maximum packet size for all flows with priority strictly smaller than $h$, and define $l_h^{(max)}(\Gamma_{k+1})$ to be~$0$ if no flow has a priority strictly smaller than $h$. Define $B_h^{(k)}(\Gamma_{k+1})  = \sum_{j = h}^{k} \sum_{i \in G_j(\Gamma_{k+1})} b_i$, \ie the sum of bucket sizes for flows with priority $h \leq j \leq k$, and $R_h^{(k)}(\Gamma_{k+1}) = \sum_{j = h}^{k}\sum_{i \in G_j(\Gamma_{k+1})} r_i$, \ie the sum of rates for flows with priority in $h \leq j \leq k$. Then to satisfy each flow's deadline $\Gamma_{k+1}$ requires a bandwidth $R$ satisfying
\beq
\begin{aligned}
\left\{
\begin{array}{ll}
&\frac{B_h^{(k+1)}(\Gamma_{k+1}) +  l_h^{(max)}(\Gamma_{k+1})}{R - R_{h+1}^{(k+1)}(\Gamma_{k+1})} \leq  d^{(min)}_h(\Gamma_{k+1}), \quad  \forall h \in [1, k+1]; \\
\\
& \sum_{i = 1}^n r_i \leq R;
\end{array}
\right.
\end{aligned}
\eeq
which gives a minimum required bandwidth of 
\beq\label{order:min_R}
R^*(\Gamma_{k+1}) = \max_{1 \leq h \leq k+1}\left\{\sum_{i = 1}^n r_i, \ \frac{B_h^{(k+1)}(\Gamma_{k+1}) +  l_h^{(max)}(\Gamma_{k+1})}{d^{(min)}_h(\Gamma_{k+1})} + R_{h+1}^{(k+1)}(\Gamma_{k+1}) \right\}.
\eeq

Afterwards, we first show that there exists a $(k+1)$-priority mechanism $\Gamma'_{k+1}$ satisfying

\bigskip
\begin{itemize}
    \item \textbf{Condition 1}: $ G_{k+1}(\Gamma'_{k+1}) = \{  i \ | \ d_n \leq d_i < d_{\hat{n}}\} \text{ where } 1 \leq \hat{n} \leq n$, and $R^*(\Gamma'_{k+1}) \leq R^*(\Gamma_{k+1})$, where $G_{k+1}(\Gamma'_{k+1})=\emptyset$ if $\hat{n} = n$. 
\end{itemize}
\bigskip

After that, under a slight abuse of notation, we show that for any $\Gamma_{k+1}$ satisfying Condition 1, there exists a $(k+1)$-priority mechanism $\Gamma'_{k+1}$ satisfying

\bigskip
\begin{itemize}
    \item \textbf{Condition 2}: $ R^*(\Gamma'_{k+1}) \leq R^*(\Gamma_{k+1})$, and $d^{(max)}_i(\Gamma'_{k+1}) < d^{(min)}_j(\Gamma'_{k+1})$ for all $j < i \leq k+1$.
\end{itemize}    
\bigskip
    
Combining them gives $S(k+1)$.
\begin{enumerate}
%%%%%%%%%%%%%%%%%%%%%%%%%%%%%%%%%%%%%%%%%%%%%%%%%%%%%%%%%%%%%%%%%%
\item We first show the existence of a mechanism $\Gamma'_{k+1}$ satisfying condition 1. If $\Gamma_{k+1}$ satisfies Condition 1, then $\Gamma_{k+1} = \Gamma'_{k+1}$. Otherwise, for all $1 \leq \hat{n} < n$, $G_{k+1}(\Gamma_{k+1}) \neq \{ i \ | \ d_n \leq d_i < d_{\hat{n}}\}$. Define $\hat{i} = \max\{ 1\leq i \leq n \ | \ i \notin G_{k+1}(\Gamma_{k+1})\}$ and suppose $\hat{i} \in G_{\hat{h}}(\Gamma_{k+1})$. Further define $G'_{k+1} = \{ i \in G_{k+1}(\Gamma_{k+1})\ | \ d_i < d^{(min)}_{\hat{h}}(\Gamma_{k+1}) \}$ and $G'_{\hat{h}} = G_{k+1}(\Gamma_{k+1}) - G'_{k+1}$.

Consider the mechanism $\Gamma'_{k+1}$ such that 1) $G_{k+1}(\Gamma'_{k+1}) = G'_{k+1}$, 2)  $G_{\hat{h}}(\Gamma'_{k+1}) = G'_{\hat{h}} + G_{\hat{h}}(\Gamma_{k+1})$, and 3) $G_i(\Gamma'_{k+1}) = G_i(\Gamma_{k+1})$, when $1 \leq i \leq k$ and $i\neq \hat{h}$.   
Note that 
\begin{itemize}
\item when $h \leq \hat{h}$, $B_h^{(k+1)}(\Gamma'_{k+1}) = B_h^{(k+1)}(\Gamma_{k+1})$, $R_{h}^{(k+1)}(\Gamma'_{k+1}) = R_{h}^{(k+1)}(\Gamma_{k+1})$, and $l_h^{(max)}(\Gamma'_{k+1}) = l_h^{(max)}(\Gamma_{k+1})$;
\item when $h > \hat{h}$, $B_h^{(k+1)}(\Gamma'_{k+1}) = B_h^{(k+1)}(\Gamma_{k+1}) -\sum_{i \in G'_{\hat{h}}} b_i$, $R_{h}^{(k+1)}(\Gamma'_{k+1}) = R_{h}^{(k+1)}(\Gamma'_{k+1}) - \sum_{i \in G'_{\hat{h}}} r_i$, and $l_h^{(max)}(\Gamma'_{k+1}) = \max\left\{l_h^{(max)}(\Gamma_{k+1}),\  \max_{i \in G'_{\hat{h}}} l_i \right\}$.
\end{itemize}

Next we show that $R^*(\Gamma'_{k+1}) \leq R^*(\Gamma_{k+1})$. 
To satisfy each flow's deadline, $\Gamma'_{k+1}$ requires a bandwidth $R$ such that\footnote{When $G_{k+1}(\Gamma'_{k+1}) = \emptyset$, $\frac{B_{k+1}^{(k+1)}(\Gamma_{k+1}) - \sum_{i \in G'_{\hat{h}}} b_i + l^{(max)}_{k+1}(\Gamma'_{k+1})}{R} = \frac{\max_i l_i}{R}$, which is also no greater than $d_n$.}
\bequn
\left\{ \quad 
\begin{aligned}
&\sum_{i = 1}^n r_i \leq R, \\
&\frac{B_{k+1}^{(k+1)}(\Gamma_{k+1}) - \sum_{i \in G'_{\hat{h}}} b_i + l^{(max)}_{k+1}(\Gamma'_{k+1})}{R} \leq d_{n},  \\
& \frac{B_{\hat{h}}^{(k+1)}(\Gamma_{k+1}) +  l_{\hat{h}}^{(max)}(\Gamma_{k+1})}{R - R_{\hat{h}+1}^{(k+1)}(\Gamma_{k+1}) + \sum_{i \in G'_{\hat{h}}} r_i} \leq  d^{(min)}_{\hat{h}}(\Gamma_{k+1}) \\
&\frac{B_h^{(k+1)}(\Gamma_{k+1}) - \sum_{i \in G'_{\hat{h}}} b_i +  l_h^{(max)}(\Gamma'_{k+1})}{R - R_{h+1}^{(k+1)}(\Gamma_{k+1}) + \sum_{i \in G'_{\hat{h}}} r_i} \leq  d^{(min)}_h(\Gamma_{k+1}),  
\text{ when } \hat{h} < h \leq k  \\
&\frac{B_h^{(k+1)}(\Gamma_{k+1}) +  l_h^{(max)}(\Gamma_{k+1})}{R - R_{h+1}^{(k+1)}(\Gamma_{k+1})} \leq  d^{(min)}_h(\Gamma_{k+1}), \text{ when } h < \hat{h} \\
\end{aligned}
\right.
\eequn 
which gives a minimum required bandwidth $R^*(\Gamma')$ of 
\bequn
\max \left\{ \quad 
\begin{aligned}
& \sum_{i = 1}^n r_i, \\
& \frac{B_{k+1}^{(k+1)}(\Gamma_{k+1}) - \sum_{i \in G'_{\hat{h}}} b_i + l^{(max)}_{k+1}(\Gamma'_{k+1})}{d_n}, \\ &\frac{B_{\hat{h}}^{(k+1)}(\Gamma_{k+1}) +  l_{\hat{h}}^{(max)}(\Gamma_{k+1})}{d^{(min)}_{\hat{h}}(\Gamma_{k+1})} + R_{\hat{h}+1}^{(k+1)}(\Gamma_{k+1}) - \sum_{i \in G'_{\hat{h}}} r_i \\
& \frac{B_h^{(k+1)}(\Gamma_{k+1}) - \sum_{i \in G'_{\hat{h}}} b_i +  l_h^{(max)}(\Gamma'_{k+1})}{d^{(min)}_h(\Gamma_{k+1})} + R_{h+1}^{(k+1)}(\Gamma_{k+1}) - \sum_{i \in G'_{\hat{h}}} r_i,  
\text{ when } \hat{h} < h \leq k \\
& \frac{B_h^{(k+1)}(\Gamma_{k+1}) +  l_h^{(max)}(\Gamma_{k+1})}{d^{(min)}_h(\Gamma_{k+1})} + R_{h+1}^{(k+1)}(\Gamma_{k+1}), \text{ when } h < \hat{h}
\end{aligned}
\right.
\eequn
Note that if for all $\hat{h} < h \leq k+1$, $l_h^{(max)}(\Gamma'_{k+1}) - \sum_{i \in G'_{\hat{h}}} b_i \leq l_h^{(max)}(\Gamma_{k+1})$, we will then have $R^*(\Gamma'_{k+1}) \leq R^*(\Gamma_{k+1})$. In fact, as $\max_{i \in G'_{\hat{h}}} l_i - \sum_{i \in G'_{\hat{h}}} b_i \leq 0$ we have
\bequn
\begin{aligned}
l_h^{(max)}(\Gamma'_{k+1}) - \sum_{i \in G'_{\hat{h}}} b_i &= \max\left\{l_h^{(max)}(\Gamma_{k+1}),\  \max_{i \in G'_{\hat{h}}} l_i \right\} - \sum_{i \in G'_{\hat{h}}} b_i \\
&\leq \max\left\{l_h^{(max)}(\Gamma_{k+1}) - \sum_{i \in G'_{\hat{h}}} b_i, 0\right\} <  l_h^{(max)}(\Gamma_{k+1}).
\end{aligned}
\eequn
Thus, we show the existence of a mechanism $\Gamma'_{k+1}$ satisfying Condition 1.
%%%%%%%%%%%%%%%%%%%%%%%%%%%%%%%%%%%%%%%%%%%%%%%%%%%%%%%%%%%%%%%%%%
\item Next we show that for any $(k+1)$-priority mechanism $\Gamma_{k+1}$ satisfying Condition 1, there exists a $(k+1)$-priority mechanism $\Gamma'_{k+1}$ satisfying Condition 2.

For $\Gamma_{k+1}$, there to exist $1 \leq \hat{n} \leq n$ such that $G_{k+1}(\Gamma_{k+1}) = \{ i\ | \ d_n \leq d_i < d_{\hat{n}}\ \}$. If $G_{k+1}(\Gamma_{k+1}) = \emptyset$, by induction of hypothesis $S(k)$ we have $S(k+1)$. Afterwards we consider the case where $G_{k+1}(\Gamma_{k+1}) \neq \emptyset$.

Consider flows $\tilde{F} = \{ (r_1, b_1, l_1, d_1), ..., (r_{\hat{n}-1}, b_{\hat{n}-1}, l_{\hat{n}-1}, d_{\hat{n}-1}), (r_{\hat{n}}, \sum_{i \geq \hat{n}}b_i, l_{\hat{n}}, d_{\hat{n}}) \}$. According to $S(k)$, there exists a $k$-priority mechanism $\Gamma'_k$ for $\tilde{F}$ such that $\forall j < i \leq k$, $\tilde{d}^{(max)}_i (\Gamma'_k) < \tilde{d}^{(min)}_j(\Gamma'_k)$. $\Gamma_k$ gives a minimum required bandwidth of 
\bequn
R^*(\Gamma'_k) = \max_{1 \leq h \leq k}\left\{\sum_{i = 1}^{\hat{n}} r_i, \ \frac{\tilde{B}_h^{(k)}(\Gamma'_k) +  l_h^{(max)}(\Gamma'_k)}{d^{(min)}_h(\Gamma'_k)} + \tilde{R}_{h+1}^{(k)}(\Gamma'_k) \right\} 
\eequn

Consider the $k$-priority mechanism $\Gamma_k$, where $G_h(\Gamma_k) = G_h(\Gamma_{k+1})$ for all $1 \leq h \leq k$. Applying $\Gamma_k$ to $\tilde{F}$, we have 
\bequn
\begin{aligned}
R^*(\Gamma_k) &= \max_{1 \leq h \leq k} \left\{\sum_{i = 1}^{\hat{n}} r_i, \frac{\tilde{B}_h^{(k)}(\Gamma_k) +  l_h^{(max)}(\Gamma_k)}{d^{(min)}_h(\Gamma_k)} + \tilde{R}_{h+1}^{(k)}(\Gamma_k)\right\} \\
& = \max_{1 \leq h \leq k} \left\{\sum_{i = 1}^{\hat{n}} r_i, \frac{B_h^{(k+1)}(\Gamma_{k+1}) +  l_h^{(max)}(\Gamma_{k+1})}{d^{(min)}_h(\Gamma_{k+1})} + R_{h+1}^{(k+1)}(\Gamma_{k+1}) - \sum_{i = \hat{n}+1}^n r_i\right\} .
\end{aligned}
\eequn
From $R^*(\Gamma_k) \geq R^*(\Gamma'_k)$, we know that 
\bequn
\begin{aligned}
& \max_{1 \leq h \leq k}\left\{\frac{\tilde{B}_h^{(k)}(\Gamma'_k) +  l_h^{(max)}(\Gamma'_k)}{d^{(min)}_h(\Gamma'_k)} + \tilde{R}_{h+1}^{(k)}(\Gamma'_k) \right\} \\
& \qquad \leq \max\left\{\ 
\sum_{i = 1}^{\hat{n}} r_i,\
\max_{1 \leq h \leq k} \left\{\frac{B_h^{(k+1)}(\Gamma_{k+1}) +  l_h^{(max)}(\Gamma)}{d^{(min)}_h(\Gamma_{k+1})} + R_{h+1}^{(k+1)}(\Gamma_{k+1}) - \sum_{i = \hat{n}+1}^n r_i \right\}
\  \right\}
\end{aligned}
\eequn
which further gives
\beq\label{order:relation}
\begin{aligned}
& \max_{1 \leq h \leq k}\left\{\frac{\tilde{B}_h^{(k)}(\Gamma'_k) +  l_h^{(max)}(\Gamma'_k)}{d^{(min)}_h(\Gamma'_k)} + \tilde{R}_{h+1}^{(k)}(\Gamma'_k) + \sum_{i = \hat{n}+1}^n r_i \right\} \\
& \qquad 
\leq \max\left\{ \ 
\sum_{i = 1}^{n} r_i, 
\max_{1 \leq h \leq k} \left\{\frac{B_h^{(k+1)}(\Gamma_{k+1}) +  l_h^{(max)}(\Gamma_{k+1})}{d^{(min)}_h(\Gamma_{k+1})} + R_{h+1}^{(k+1)}(\Gamma_{k+1}) \right\}
\ \right\}
\end{aligned}
\eeq

Now consider the $(k+1)$-priority mechanism $\Gamma'_{k+1}$, where $G_h(\Gamma'_{k+1}) = G_h(\Gamma'_k)$ for all $1 \leq h \leq k$, and $G_{k+1}(\Gamma'_{k+1}) = G_{k+1}(\Gamma_{k+1})$. By the definition of $G_{k+1}(\Gamma_{k+1})$ and $\Gamma'_k$, we know that $d^{(max)}_i(\Gamma'_{k+1}) < d^{(min)}_j(\Gamma'_{k+1}), \forall j < i \leq k+1$. Next we show that $R^*(\Gamma'_{k+1}) \leq R^*(\Gamma_{k+1})$.

Applying $\Gamma'_{k+1}$ to flow $F = \{ (r_i, b_i, l_i, d_i) \ | \ 1 \leq i \leq n \}$, we have
\bequn
\begin{aligned}
& R^*(\Gamma'_{k+1}) = \max_{1 \leq h \leq k+1}\left\{\sum_{i = 1}^{n} r_i, \ \frac{B_h^{(k+1)}(\Gamma'_{k+1}) +  l_h^{(max)}(\Gamma'_{k+1})}{d^{(min)}_h(\Gamma'_{k+1})} + R_{h+1}^{(k+1)}(\Gamma'_{k+1}) \right\}  \\
& \quad = \max_{1 \leq h \leq k}\left\{\sum_{i = 1}^{n} r_i, \frac{B_{k+1}^{(k+1)}(\Gamma_{k+1}) +  l_{k+1}^{(max)}(\Gamma_{k+1})}{d^{(min)}_{k+1}(\Gamma_{k+1})}, \frac{B_h^{(k+1)}(\Gamma'_{k+1}) +  l_h^{(max)}(\Gamma'_{k+1})}{d^{(min)}_h(\Gamma'_{k+1})} + R_{h+1}^{(k+1)}(\Gamma'_{k+1}) \right\} \\
& \quad= \max_{1 \leq h \leq k}\left\{\sum_{i = 1}^{n} r_i, \frac{B_{k+1}^{(k+1)}(\Gamma_{k+1}) +  l_{k+1}^{(max)}(\Gamma_{k+1})}{d^{(min)}_{k+1}(\Gamma_{k+1})}, \frac{\tilde{B}_h^{(k)}(\Gamma'_k) +  l_h^{(max)}(\Gamma'_k)}{d^{(min)}_h(\Gamma'_k)} + \tilde{R}_{h+1}^{(k)}(\Gamma'_k) + \sum_{i = \hat{n}+1}^n r_i
\right\}
\end{aligned}
\eequn
Combining it with \Eqref{order:relation}, we have
\bequn
\begin{aligned}
R^*(\Gamma'_{k+1}) &\leq \max\left\{
\begin{array}{ll}
\sum_{i = 1}^{n} r_i, \\
\\
\frac{B_{k+1}^{(k+1)}(\Gamma_{k+1}) +  l_{k+1}^{(max)}(\Gamma_{k+1})}{d^{(min)}_{k+1}(\Gamma_{k+1})},\\
\\
\max_{1 \leq h \leq k} \left\{\frac{B_h^{(k+1)}(\Gamma_{k+1}) +  l_h^{(max)}(\Gamma_{k+1})}{d^{(min)}_h(\Gamma_{k+1})} + R_{h+1}^{(k+1)}(\Gamma_{k+1}) \right\}  
\end{array}
\right.\\
&= \max\left\{\sum_{i = 1}^{n} r_i, \max_{1 \leq h \leq k+1} \left\{\frac{B_h^{(k+1)}(\Gamma_{k+1}) +  l_h^{(max)}(\Gamma_{k+1})}{d^{(min)}_h(\Gamma_{k+1})} + R_{h+1}^{(k+1)}(\Gamma_{k+1}) \right\} \right\} \\
&= R^*(\Gamma_{k+1})
\end{aligned}
\eequn
Hence we show the existence of a mechanism $\Gamma'_{k+1}$ satisfying Condition 2.
\end{enumerate}
\end{itemize}
\end{proof}
\subsection{Proof for Proposition~\ref{nflow:delay}}
\label{app:static+reshape}
\begin{flushleft}
P{\scriptsize ROPOSITION}~\ref{nflow:delay}. \textit{
Consider a link shared by $n$ token bucket controlled flows, where flow $i, 1 \leq i \leq n$, has traffic profile $(r_i, b_i)$.
%%and a deadline $d_i$, with $d_1 > d_2 > ... > d_n$ and $d_1<\infty$. 
Assume a static priority scheduler that assigns flow $i$ a priority of $i$, where priority $n$ is the highest priority, and reprofiles flow $i$ to $(r_i, b'_i)$, where $0 \leq b'_i \leq b_i$. Given a link bandwidth of $R \geq \sum_{j = 1}^n r_j$, the worst-case delay for flow $i$ is 
\beq\tag{\ref{eq:worst_case_delay}}
D^*_i = \max\left\{\frac{b_i + B'_{i+1}}{R - R_{i+1}}, \ \frac{b_i - b'_i}{r_i} + \frac{B'_{i+1}}{R-R_{i+1}} \right\}.
\eeq 
}
\end{flushleft}
\begin{proof}
Flow $1 \leq i \leq n$ receives a service curve of $SC^{(i)}_1 = \left\{
\begin{aligned}
& b'_i + r_i t,\text{ when } t > 0 \\
& 0 \qquad \text{otherwise}
\end{aligned}
\right.$
inside the reprofiler, and a service curve of 
$SC^{(i)}_2(t) = \left[(R-R_{i+1})t - B'_{i+1} \right]^+$ at the shared link. Overall it receives a service curve of\footnote{See THEOREM 1.4.6 in~\cite{nc}, page 28} 
$$
SC^{(i)}(t) = SC^{(i)}_1 \otimes SC^{(i)}_2 = \min\left\{ \left[ b'_i + r_i \left( t - \frac{B'_{i+1}}{R-R_{i+1}}\right) \right]^+, [(R-R_{i+1})t - B'_{i+1}]^+ \right\}.
$$
Hence, we have
\bequn
D^*_i = \sup_{t\geq 0} \inf_{\tau \geq 0} \left\{ b_i + r_i t \leq SC^{(i)}(t + \tau) \right\} = \max\left\{\frac{b_i + B'_{i+1}}{R-R_{i+1}}, \ \frac{b_i - b'_i}{r_i} + \frac{B'_{i+1}}{R-R_{i+1}} \right\}.
\eequn
\end{proof}

%%%%%%%%%%%%%%%%%%%%%%%%%%%%%%%%%%%%%%%%%%%%%%%%%%%%%%%%%%%%%%%%%%%%%%
%%%%%%%%%%%%%%%%%%%%%%%%% Proof: optimization %%%%%%%%%%%%%%%%%%%%%%%%
%%%%%%%%%%%%%%%%%%%%%%%%%%%%%%%%%%%%%%%%%%%%%%%%%%%%%%%%%%%%%%%%%%%%%%
\subsection{Proofs for Proposition~\ref{nflow:R} and~\ref{nflow:b}}
\label{sec:opt-rsp}

This section provides the solution for \textbf{OPT\_RSP}, from which Propositions~\ref{nflow:R} and~\ref{nflow:b} derive. For the reader's convenience, we restate \textbf{OPT\_RSP}. Remember that we define $B'_i = \sum_{j = i}^n b'_j$ and $R_i = \sum_{j = i}^n r_j$, where $B'_i, R_i = 0$ when $i > n$. 
\bequn
\begin{aligned}
&\text{\bf{OPT\_RSP}} && \min_{\boldsymbol{b}'} 
R &\\
& \text{s.t}
& & \max\left\{\frac{b_i + B'_{i+1}}{R - R_{i+1}}, \ \frac{b_i - b'_i}{r_i} + \frac{B'_{i+1}}{R-R_{i+1}} \right\} \leq d_i, \ &\forall \ 1 \leq i \leq n, \\
&&& R_1 \leq R, \qquad b'_1 \leq b_1, \qquad 0 \leq b'_i \leq b_i, & \forall \ 2 \leq i \leq n. \\
\end{aligned}
\eequn
Instead of solving \textbf{OPT\_RSP}, for technical simplicity we consider \textbf{OPT\_RSP'}, whose solution directly gives that for \textbf{OPT\_RSP}. Next we first demonstrate the relationship between \textbf{OPT\_RSP} and \textbf{OPT\_RSP'} (Lemma~\ref{nflow:opt_n'}), and then proceed to solve \textbf{OPT\_RSP'} (Lemma~\ref{nflow:lemma_sol}). Combining Lemma~\ref{nflow:opt_n'} and~\ref{nflow:lemma_sol}, we then have Proposition~\ref{nflow:R} and~\ref{nflow:b}.
 
%%%%%%%%%%%%%%%%%%%%%%% OPT_N vs. OPT_N' %%%%%%%%%%%%%%%%%%%%%%%%%%%%
\begin{lemma}\label{nflow:opt_n'}
For $1 \leq i \leq n$, define $H_i = b_i - r_i d_i$, and $\boldsymbol{B'} = (B'_1, ..., B'_n)$. Consider the following optimization:
\beq
\begin{aligned}
&\text{\bf{OPT\_RSP'}} && \min_{\boldsymbol{B}'} 
R &\\
& \text{s.t}
& & R_1 \leq R. & \\
&&& B'_2 \leq d_1(R-R_2) - b_1, & \\
&&& B'_i \in \left[\max\left\{\frac{R-R_{i+1} + r_i}{R-R_{i+1}}B'_{i+1} + H_i,\ B'_{i+1}  \right\}, \ B'_{i+1} + \frac{b_i(R-R_i)}{R-R_{i+1}} \right], & \forall \ 2 \leq i \leq n. 
\end{aligned}
\eeq
Suppose the optimal solution for \textbf{OPT\_RSP'} is $(R^*, \boldsymbol{B}'^*)$. Then $(R^*, \boldsymbol{b}'^*)$, where $$\boldsymbol{b}'^* = (b_1, B'^*_2 - B'^*_3, ....,  B'^*_{n-1} - B'^*_n, B'^*_n),$$ is an optimal solution for \textbf{OPT\_RSP}.
\end{lemma}

%%%%%%%%%%%%%%%%%%%%%% proof: OPT_N vs. OPT_N' %%%%%%%%%%%%%%%%%%%%%%%
\begin{proof}
We first show that $\boldsymbol{b'}^* = (b_1, B'^*_2 - B'^*_3, ....,  B'^*_{n-1} - B'^*_n, B'^*_n)$ and $R^*$ satisfy all the constraints for \textbf{OPT\_RSP}, and then show that $(R^*, \boldsymbol{b'}^*)$ is an optimal solution for \textbf{OPT\_RSP}.

Substituting $b'_1 = b_1$ into $\max\left\{\frac{b_1 + B'_2}{R - R_2}, \ \frac{b_1 - b'_1}{r_1} + \frac{B'_2}{R-R_2} \right\} \leq d_1$ gives $\frac{b_1 + B'_2}{R - R_2} \leq d_1$, which is equivalent to $B'_2 \leq d_1(R-R_2)-b_1$. Thus, to show the feasibility of $(R^*, \boldsymbol{b'}^*)$, we only need to show that  $(R^*, \boldsymbol{b'}^*)$ satisfies $\max\left\{\frac{b_i + B'^*_{i+1}}{R^* - R_{i+1}}, \ \frac{b_i - b'^*_i}{r_i} + \frac{B'^*_{i+1}}{R^*-R_{i+1}} \right\} \leq d_i$ and $b'^*_i \in [0, b_i]$ for all $2 \leq i \leq n$. Below we consider each constraint separately. 
\begin{itemize}
\item Basic algebraic manipulation gives that 
\beq\label{nflow:max_relation}
\max\left\{\frac{b_i + B'^*_{i+1}}{R^* - R_{i+1}}, \ \frac{b_i - b'^*_i}{r_i} + \frac{B'^*_{i+1}}{R^*-R_{i+1}} \right\} = \left\{
\begin{aligned}
& \frac{b_i + B'^*_{i+1}}{R^* - R_{i+1}}, & \text{ when } b'^*_i \geq \frac{b_i(R^*-R_i)}{R^*-R_{i+1}}, \\
& \frac{b_i - b'^*_i}{r_i} + \frac{B'^*_{i+1}}{R^*-R_{i+1}}, &\text{ otherwise}.
\end{aligned}
\right. 
\eeq
From \textbf{OPT\_RSP'} we have $B'^*_i \leq B'^*_{i+1} + \frac{b_i(R^*-R_i)}{R^*-R_{i+1}}$, \ie $b'^*_i = B'^*_i - B'^*_{i+1} \leq \frac{b_i(R^*-R_i)}{R^*-R_{i+1}}$, and therefore $\max\left\{\frac{b_i + B'^*_{i+1}}{R^* - R_{i+1}}, \ \frac{b_i - b'^*_i}{r_i} + \frac{B'^*_{i+1}}{R^*-R_{i+1}} \right\} = \frac{b_i - b'^*_i}{r_i} + \frac{B'^*_{i+1}}{R^*-R_{i+1}}$. 
From \textbf{OPT\_RSP'} we also have $B'^*_i \geq \frac{R^*-R_{i+1} + r_i}{R^*-R_{i+1}}B'^*_{i+1} + H_i$, \ie $B'^*_i - B'^*_{i+1} - H_i  = b'^*_i - b_i + r_i d_i \geq \frac{r_i B'^*_{i+1} }{R-R_{i+1}}$, which is equivalent to $\frac{b_i - b'^*_i}{r_i} + \frac{B'^*_{i+1}}{R-R_{i+1}} \leq d_i$. 
Combining them gives $\max\left\{\frac{b_i + B'^*_{i+1}}{R^* - R_{i+1}}, \ \frac{b_i - b'^*_i}{r_i} + \frac{B'^*_{i+1}}{R^*-R_{i+1}} \right\} \leq d_i$.
\item From \textbf{OPT\_RSP'} we have $B'^*_i \geq B'^*_{i+1}$, and therefore $b'^*_i = B'^*_i - B'^*_{i+1}  \geq 0$. From \textbf{OPT\_RSP'} we also have $B'^*_i \leq B'^*_{i+1} + \frac{b_i(R^*-R_i)}{R^*-R_{i+1}}$, and therefore $b'^*_i \leq \frac{b_i(R^*-R_i)}{R^*-R_{i+1}} < b_i$. Thus, we have $b'^*_i \in [0, b_i]$.
\end{itemize}

%%%%%%%%%%%%%%%%%%%%%%%%%%%%%%%%%%%%%%%%%%%%%%%%%%%%%%%%%%%%%%%%%%%%%%%%
Next we show by contradiction that $(R^*, \boldsymbol{b'}^*)$ is optimal for \textbf{OPT\_RSP}. Suppose $(\tilde{R}, \boldsymbol{\tilde{b}'})$, where $\tilde{R} < R^*$ is an optimal solution for \textbf{OPT\_RSP}. Denote $\tilde{B}'_i = \sum_{j = i}^n \tilde{b}'_j$, and $\boldsymbol{\tilde{B}'} = (\tilde{B}'_1, ..., \tilde{B}'_n)$.

\begin{itemize}
\item If $(\tilde{R}, \boldsymbol{\tilde{B}'})$ satisfies all constraints for \textbf{OPT\_RSP'}, then by $(R^*, \boldsymbol{b'}^*)$'s optimality we know that $\tilde{R} \geq R^*$, which contradicts to the assumption that $\tilde{R} < R^*$.
\item Otherwise, there exists $2 \leq i \leq n$ such that $\max\left\{\frac{b_i + \tilde{B}'_{i+1}}{\tilde{R} - R_{i+1}}, \ \frac{b_i - \tilde{b}'_i}{r_i} + \frac{\tilde{B}'_{i+1}}{\tilde{R}-R_{i+1}} \right\} \leq d_i$ and $\tilde{b}'_i \in [0, b_i]$, whereas $\tilde{B}'_i \notin \left[\max\left\{\frac{\tilde{R}-R_{i+1} + r_i}{\tilde{R}-R_{i+1}} \tilde{B}'_{i+1} + H_i,\ \tilde{B}'_{i+1}  \right\}, \ \tilde{B}'_{i+1} + \frac{b_i(\tilde{R}-R_i)}{\tilde{R}-R_{i+1}} \right]$.
\begin{itemize}
\item When $\tilde{b}'_i \leq \frac{b_i(\tilde{R}-R_i)}{\tilde{R}-R_{i+1}}$, from \Eqref{nflow:max_relation} we have that $\max\left\{\frac{b_i + \tilde{B}'_{i+1}}{\tilde{R} - R_{i+1}}, \ \frac{b_i - \tilde{b}'_i}{r_i} + \frac{\tilde{B}'_{i+1}}{\tilde{R}-R_{i+1}} \right\}  = \frac{b_i - \tilde{b}'_i}{r_i} + \frac{\tilde{B}'_{i+1}}{\tilde{R}-R_{i+1}}$. Combining $\frac{b_i - \tilde{b}'_i}{r_i} + \frac{\tilde{B}'_{i+1}}{\tilde{R}-R_{i+1}} \leq d_i$ and $\tilde{b}'_i \in [0, \frac{b_i(\tilde{R}-R_i)}{\tilde{R}-R_{i+1}}]$ with $\tilde{b}'_i = \tilde{B}'_i - \tilde{B}'_{i-1}$, we have $$\tilde{B}'_i \in \left[\max\left\{\frac{\tilde{R}-R_{i+1} + r_i}{\tilde{R}-R_{i+1}} \tilde{B}'_{i+1} + H_i,\ \tilde{B}'_{i+1}  \right\}, \ \tilde{B}'_{i+1} + \frac{b_i(\tilde{R}-R_i)}{\tilde{R}-R_{i+1}} \right],$$ and therefore a  contradiction. 
\item When $\tilde{b}'_i > \frac{b_i(\tilde{R}-R_i)}{\tilde{R}-R_{i+1}}$, we show that there exists an optimal solution $(\tilde{R},\boldsymbol{\hat{b}'})$ for \textbf{OPT\_RSP} such that $\hat{b}'_i \leq \frac{b_i(\tilde{R}-R_i)}{\tilde{R}-R_{i+1}}$. Define $\boldsymbol{\hat{b}'}$ as 1) $\hat{b}'_i = \frac{b_i(\tilde{R}-R_i)}{\tilde{R}-R_{i+1}}$, and 2) $\hat{b}'_h = \tilde{b}'_h$ when $h \neq i$. Denote $\hat{B}'_h = \sum_{j = h}^n \hat{b}'_j \leq \tilde{B}'_h$, and $\boldsymbol{\hat{B}'} = (\hat{B}'_1, ..., \hat{B}'_n)$. Next we show that $(\tilde{R}, \boldsymbol{\hat{b}'})$ satisfies all the constraints in \textbf{OPT\_RSP}, and therefore is optimal. Observe first that from $\hat{B}'_2 < \tilde{B}'_2$, we have $\frac{b_1 + \hat{B}'_2}{\tilde{R} - R_2} < \frac{b_1 + \tilde{B}'_2}{\tilde{R} - R_2} \leq d_1$. Below we show that for all $2 \leq h \leq n$, $\max\left\{\frac{b_h + \hat{B}'_{h+1}}{\tilde{R} - R_{h+1}}, \ \frac{b_h - \hat{b}'_h}{r_h} + \frac{\hat{B}'_{h+1}}{\tilde{R}-R_{h+1}} \right\} \leq d_h$. 
\begin{itemize}
\item When $h > i$, by definition $\hat{B}'_{h+1} = \tilde{B}'_{h+1}$. Thus we have $$\max\left\{\frac{b_h + \hat{B}'_{h+1}}{\tilde{R} - R_{h+1}}, \ \frac{b_h - \hat{b}'_h}{r_h} + \frac{\hat{B}'_{h+1}}{\tilde{R}-R_{h+1}} \right\} = \max\left\{\frac{b_h + \tilde{B}'_{h+1}}{\tilde{R} - R_{h+1}}, \ \frac{b_h - \tilde{b}'_h}{r_h} + \frac{\tilde{B}'_{h+1}}{\tilde{R}-R_{h+1}} \right\} \leq d_h.$$
\item When $h = i$, $\max\left\{\frac{b_h + \hat{B}'_{h+1}}{\tilde{R} - R_{h+1}}, \ \frac{b_h - \hat{b}'_h}{r_h} + \frac{\hat{B}'_{h+1}}{\tilde{R}-R_{h+1}} \right\} = \frac{b_h + \hat{B}'_{h+1}}{\tilde{R} - R_{h+1}}$. Combining it with $\hat{B}'_{h+1} < \tilde{B}'_{h+1}$ and $\max\left\{\frac{b_h + \tilde{B}'_{h+1}}{\tilde{R} - R_{h+1}}, \ \frac{b_h - \tilde{b}'_h}{r_h} + \frac{\tilde{B}'_{h+1}}{\tilde{R}-R_{h+1}} \right\} = \frac{b_h + \tilde{B}'_{h+1}}{\tilde{R} - R_{h+1}} \leq d_h$ gives the result. 
\item When $h < i$, from $\hat{B}'_{h+1} < \tilde{B}'_{h+1}$ we have $\frac{b_h + \hat{B}'_{h+1}}{\tilde{R} - R_{h+1}} < \frac{b_h + \tilde{B}'_{h+1}}{\tilde{R} - R_{h+1}}$ and $\frac{b_h - \hat{b}'_h}{r_h} + \frac{\hat{B}'_{h+1}}{\tilde{R}-R_{h+1}} < \frac{b_h - \tilde{b}'_h}{r_h} + \frac{\tilde{B}'_{h+1}}{\tilde{R}-R_{h+1}}$, \ie $\max\left\{\frac{b_h + \hat{B}'_{h+1}}{\tilde{R} - R_{h+1}}, \ \frac{b_h - \hat{b}'_h}{r_h} + \frac{\hat{B}'_{h+1}}{\tilde{R}-R_{h+1}} \right\} < \max\left\{\frac{b_h + \tilde{B}'_{h+1}}{\tilde{R} - R_{h+1}}, \ \frac{b_h - \tilde{b}'_h}{r_h} + \frac{\tilde{B}'_{h+1}}{\tilde{R}-R_{h+1}} \right\} \leq d_h.$
\end{itemize}
If $(\tilde{R}, \boldsymbol{\hat{b}'})$ satisfies all the constraints for \textbf{OPT\_RSP}, it contradicts to the assumption that $\tilde{R} < R^*$. Otherwise, from the case for $\tilde{b}'_i \leq \frac{b_i(\tilde{R}-R_i)}{\tilde{R}-R_{i+1}}$, we know it again to produce a contradiction. 
\end{itemize}
\end{itemize}
\end{proof}
%%%%%%%%%%%%%%%%%%%%%%%%%%%%% proof ends %%%%%%%%%%%%%%%%%%%%%%%%%%%%%

Next, we proceed to solve \textbf{OPT\_RSP'}, which when combine with Lemma~\ref{nflow:opt_n'} then gives Propositions~\ref{nflow:R} and~\ref{nflow:b}. For the reader's convenience, we restate the Propositions.
\begin{flushleft}
P{\scriptsize ROPOSITION}~\ref{nflow:R}. \textit{For $1 \leq i \leq n$, denote $H_i = b_i - d_i r_i$, $\Pi_i(R) = \frac{r_i + R- R_{i+1}}{R-R_{i+1}}$ and $V_i(R) = d_i(R-R_{i+1}) - b_i$. Define $S_1(R) = \left\{ V_1(R) \right\}$, and $S_i(R) = S_{i-1}(R) \bigcup \left\{ V_i(R) \right\} \bigcup \left\{\frac{ s - H_i}{\Pi_i(R)} \ | \ s \in S_{i-1}(R)  \right\}$ for $2 \leq i \leq n$. Then we have $\tilde{R}^*_R = \max\left\{R_1, \inf \{ R \ | \ \forall s \in S_n(R), s \geq 0 \} \right\}$.}
\end{flushleft}

\begin{flushleft}
P{\scriptsize ROPOSITION}~\ref{nflow:b}. \textit{The optimal reprofiling solution $\boldsymbol{b}'^*$ satisfies}
\beq\tag{\ref{eq:static_opt_b}}
b'^*_i = \left\{
\begin{aligned}
& \max\{0, b_n - r_n d_n \}, & i = n; \\
& \max \left\{0,\  b_i - r_i d_i + \frac{r_i B'^*_{i+1}}{\tilde{R}^*_R-R_{i+1}} \right\}, & 2 \leq i \leq n-1.
\end{aligned}
\right.
\eeq
\end{flushleft}

Denote $\mathbb{B}_i := \left[\max\left\{\frac{R-R_{i+1} + r_i}{R-R_{i+1}}B'_{i+1} + H_i,\ B'_{i+1}  \right\}, \ B'_{i+1} + \frac{b_i(R-R_i)}{R-R_{i+1}} \right]$, where the interval overlaps with that in the third constraint of \textbf{OPT\_RSP'}.
We then show that the system meets each flow's deadline only if the shared link has a bandwidth no less than $\max \left\{R_1,  \inf \{ R \ | \ \forall s \in S_n(R), s \geq 0 \right\}$, which in turn gives the minimum required bandwidth $R^*$. As mentioned before, we achieve this by first solving \textbf{OPT\_RSP'}, from which we then get the solution for \textbf{OPT\_RSP} based on Lemma~\ref{nflow:opt_n'}.

%%%%%%%%%%%%%%%%%%%%%%% proof: solve OPT_N' %%%%%%%%%%%%%%%%%%%%%%%%%%%%
\begin{lemma}\label{nflow:lemma_sol}
Define $s^{(i)}_1 = \max\left\{\Pi_i(R)B'_{i+1} + H_i,\ B'_{i+1}  \right\}$, $s^{(i)}_2 = \min\left\{ S_{i-1}(R),  B'_{i+1} + \frac{b_i(R-R_i)}{R-R_{i+1}} \right\}$, and $\mathbb{S}_i = \left[s^{(i)}_1, s^{(i)}_2\right]$ for $2 \leq i \leq n$, where $\Pi_i(R) = \frac{r_i + R- R_{i+1}}{R-R_{i+1}}$, $H_i = b_i - d_i r_i$, $S_1(R) = \left\{ V_1(R) \right\}$, $V_i(R) = d_i(R-R_{i+1}) - b_i$, and $S_i(R) = S_{i-1}(R) \bigcup \left\{ V_i(R) \right\} \bigcup \left\{\frac{ s - H_i}{\Pi_i(R)} \ | \ s \in S_{i-1}(R)  \right\}$. Then $R$ and $\boldsymbol{b}'$ satisfies all the constraints in \textbf{OPT\_RSP'} iff $\mathbb{S}_n \neq \emptyset$ and $R \geq R_1$, \ie $R \geq \max\left\{R_1, \inf \{ R \ | \ \forall s \in S_n(R), s \geq 0 \} \right\}$.
\end{lemma}
\begin{proof}
We rely on the following statements to show the Lemma:

\bigskip
\textbf{Statement 1.} $\left\{ (R, \boldsymbol{b}') \ | \ B'_{i-1} \in \mathbb{S}_{i-1} \right\} \bigcap \left\{ (R, \boldsymbol{b}') \ | \ B'_i \in \mathbb{B}_i \right\} = \left\{ (R, \boldsymbol{b}') \ | \ B'_i \in \mathbb{S}_i \right\}$.
\bigskip

Given Statement 1, we then show that

\bigskip
\textbf{Statement 2.} $\left\{ (R, \boldsymbol{b}') \ | \ B'_2 \leq V_1(R), B'_i \in \mathbb{B}_i, \forall \ 2 \leq i \leq n \right\} = \left\{ (R, \boldsymbol{b}') \ | \ B'_n \in \mathbb{S}_n \right\}.$
\bigskip

Note that $B'_2 \leq V_1(R)$ corresponds to the second constraint in \textbf{OPT\_RSP'}, while $B'_i \in \mathbb{B}_i, \forall \ 2 \leq i \leq n $ corresponds to its third constraint. Therefore, from Statement 2 we have that $R$ and $\boldsymbol{b}'$ satisfies all the constraints in \textbf{OPT\_RSP'} iff $\mathbb{S}_n \neq \emptyset$ and $R \geq R_1$. Note that $\mathbb{S}_n \neq \emptyset$ iff $s^{(n)}_1 \leq s^{(n)}_2$, \ie $\max\{H_n, 0\} \leq \min\left\{ S_{n-1}(R), \frac{b_n(R-R_n)}{R} \right\}$. As $(R-R_n) \geq 0$, $\max\{H_n, 0\} \leq \frac{b_n(R-R_n)}{R}$ iff $V_n(R) \geq 0$, whereas $\max\{H_n, 0\} \leq \min\left\{ S_{n-1}(R)\right\}$ iff $s \geq \max\{0, H_n\}, \forall s \in S_{n-1}(R)$. Since $\Pi_n(R) > 0$, by the definition of $S_n$ we have $\mathbb{S}_n \neq \emptyset$ and $R \geq R_1$ iff $R \geq \max\left\{R_1, \inf \{ R \ | \ \forall s \in S_n(R), s \geq 0 \} \right\}$.

%%%%%%%%%%%%%%%%%%%%%%%%% proof: statement 1 %%%%%%%%%%%%%%%%%%%%%%%%%%
\paragraph{Proof for Statement 1.} Note that $\left\{ (R, \boldsymbol{b}') \ | \ B'_{i-1} \in \mathbb{S}_{i-1} \right\} = \left\{ (R, \boldsymbol{b}') \ | \ \mathbb{S}_{i-1} \neq \emptyset \right\}$. Below we prove that $\mathbb{S}_{i-1} \neq \emptyset$ iff $B'_i \leq \min\{S_{i-1}(R)\}$. As $s^{(i)}_2 = \min\left\{ S_{i-1}(R),  B'_{i+1} + \frac{b_i(R-R_i)}{R-R_{i+1}} \right\}$, combining $B'_i \leq \min\{S_{i-1}(R)\}$ with $B'_i \in \mathbb{B}_i = \left[s^{(1)}_i, \ B'_{i+1} + \frac{b_i(R-R_i)}{R-R_{i+1}} \right]$ directly gives $B'_i \in \mathbb{S}_i = \left[s^{(i)}_1 ,s^{(i)}_2  \right]$. 

From $\mathbb{S}_{i-1} \neq \emptyset \Leftrightarrow s^{(i-1)}_2 \geq s^{(i-1)}_1$, we have
\bequn
\left\{
\begin{aligned}
& B'_{i} \leq s^{(i-1)}_2 = \min\left\{ S_{i-2}(R),  B'_{i} + \frac{b_{i-1}(R-R_{i-1})}{R-R_i} \right\} \Leftrightarrow B'_{i} \leq \min \left\{ S_{i-2}(R) \right\} \\
& \Pi_{i-1}(R)B'_i + H_{i-1} \leq s^{(i-1)}_2 \Leftrightarrow  B'_{i} \leq \min \{V_{i-1}(R)\} \bigcup \left\{ \frac{s - H_{i-1}}{\Pi_{i-1}(R)}\ | \ s \in S_{i-2}(R) \right\}
\end{aligned}
\right.
\eequn
As $S_{i-1}(R) = S_{i-2}(R) \bigcup \{V_{i-1}(R)\} \bigcup \left\{ \frac{s - H_{i-1}}{\Pi_{i-1}(R)}\ | \ s \in S_{i-2}(R) \right\}$, we have $\mathbb{S}_{i-1} \neq \emptyset$ iff $B'_i \leq \min\{S_{i-1}(R)\}$.

%%%%%%%%%%%%%%%%%%%%%%%%% proof: statement 2 %%%%%%%%%%%%%%%%%%%%%%%%%%
\paragraph{Proof for Statement 2.} We show by induction on the value of $n$.
\begin{itemize}
\item Base case: when $n = 2$, basic algebraic manipulation gives that 
\bequn
\begin{aligned}
\left\{ (R, \boldsymbol{b}') \ | \ B'_2 \leq V_1(R), B'_2 \in \mathbb{B}_i \right\} &= \left\{ (R, \boldsymbol{b}') \ | \ B'_2 \in \left[s^{(2)}_1, \min\left\{V_1(R), B'_3 + \frac{b_2(R-R_2)}{R-R_3} \right\} \right]\right\}  \\
&= \left\{ (R, \boldsymbol{b}') \ | \ B'_2 \in \mathbb{S}_2 \right\}
\end{aligned}
\eequn
\item Induction step: suppose $\left\{ (R, \boldsymbol{b}') \ | \ B'_2 \leq V_1(R), B'_i \in \mathbb{B}_i, \forall \ 2 \leq i \leq k \right\} = \left\{ (R, \boldsymbol{b}') \ | \ B'_k \in \mathbb{S}_k \right\}$. Then we have
\bequn
\begin{aligned}
&\left\{ (R, \boldsymbol{b}') \ | \ B'_2 \leq V_1(R), B'_i \in \mathbb{B}_i, \forall \ 2 \leq i \leq k+1 \right\} \\
 = &\left\{ (R, \boldsymbol{b}') \ | \ B'_2 \leq V_1(R), B'_i \in \mathbb{B}_i, \forall \ 2 \leq i \leq k \right\} \bigcap \left\{ (R, \boldsymbol{b}') \ | \ B'_{k+1} \in \mathbb{B}_{k+1} \right\} \\
 = & \left\{ (R, \boldsymbol{b}') \ | \ B'_k \in \mathbb{S}_k \right\} \bigcap \left\{ (R, \boldsymbol{b}') \ | \ B'_{k+1} \in \mathbb{B}_{k+1} \right\} \\
 =& \left\{ (R, \boldsymbol{b}') \ | \ B'_{k+1} \in \mathbb{S}_{k+1} \right\}
\end{aligned}
\eequn
where the last equation comes from Statement~$1$.
\item Conclusion: by the principle of induction, we have 
$$\left\{ (R, \boldsymbol{b}') \ | \ B'_2 \leq V_1(R), B'_i \in \mathbb{B}_i, \forall \ 2 \leq i \leq n \right\} = \left\{ (R, \boldsymbol{b}') \ | \ B'_n \in \mathbb{S}_n \right\}.$$
\end{itemize}

\end{proof}

Observe from the proof of Lemma~\ref{nflow:lemma_sol} that for all $2 \leq i \leq n$, setting $B'_i = \max\left\{\Pi_i(R)B'_{i+1} + H_i,\ B'_{i+1}\right\}$ gives $\tilde{R}^*_R$. Combining it with Lemma~\ref{nflow:opt_n'}, we know that $R^*$ is the optimal solution for \textbf{OPT\_RSP}, and therefore we have Proposition~\ref{nflow:R}. Besides, since $\tilde{R}^*_R$ can be achieved by setting $b'_i = \max\left\{ \frac{r_i B'_{i+1}}{R-R_{i+1}} + H_i,\ 0\right\}$, we then have Proposition~\ref{nflow:b}.

\section{Proofs for FIFO Scheduler}
\label{app:one_fifo}

\subsection{Proof for Proposition~\ref{prop:delay_fifo}}
\label{app:fifo+reshape}
\begin{flushleft}
P{\scriptsize ROPOSITION}~\ref{prop:delay_fifo}. \textit{
Consider a system with $n$ token bucket controlled flows with traffic profiles $(r_i, b_i), 1 \leq i\leq n$, sharing a fifo link with bandwidth $R \geq R_1 = \sum_{j=1}^n r_j$. Assume that the system reprofiles flow $i$ to $(r_i, b'_i)$. The worst-case delay for flow $i$ is then
\beq
\tag{\ref{eq:fifo_delay}}
\widehat{D}^*_i = \max\left\{\frac{b_i - b'_i}{r_i}+ \frac{\sum_{j \neq i} b'_j}{R}, \frac{\sum_{j = 1}^n b'_j}{R} + \frac{(b_i - b'_i)R_1}{r_i R} \right\}.
\eeq
}
\end{flushleft}
%Remark: we will hardly rely on network calculus to show Proposition~\ref{prop:delay_fifo}. This is because current work regarding aggregate multiplexing is not very rich, and none of the existing work provides a tight bound for our system. 
\begin{proof}
W.l.o.g we consider the worst-case delay for flow~$1$. First we show that there always exists a traffic pattern such that flow $1$'s worst-case delay can be achieved by the last bit inside a burst of size $b_1$, and then we characterize the worst-case delay for that $b_1^{th}$ bit. 
\begin{itemize}
\item Consider the traffic pattern $\boldsymbol{T}(t) = \left\{T_1(t), ..., T_n(t) \right\}$ that realizes flow $1$'s worst-case delay, where $T_i(t)$ is right continuous and specifies the cumulative amount of data sent by flow $i$ during time $[0,t]$. Suppose the worst-case delay is achieved at $t_0$,  and at $t_0$ flow~$1$ sends a burst of $b \leq b_1$.
%, \ie $\lim_{\epsilon \rightarrow 0 }T_1(t_0) - T_1(t_0 - \epsilon) = b \leq b_1$.
As under FIFO the last bit gets a strictly larger delay compared with all the other bits inside the burst, the $b^{th}$ bit sent at $t_0$ achieves the worst-case delay. 

If $b = b_1$, flow~$1$'s worst-case delay is achieved by the $b_1^{th}$ bit inside a burst, \ie $\boldsymbol{T}(t)$ is the traffic pattern we want. Afterwards we consider the case $b < b_1$. 

First note that $b$ is the maximum amount of data flow~$1$ can send at $t_0$ without violating its arrival-curve constraint, \ie there exists $t_s \in [0, t_0)$ such that $T(t_0) - T(t_s) = b_1 + r_1(t_0 - t_s)$. Otherwise, we can produce a worse delay by increasing $b$ to the maximum value that remains conformant with the arrival-curve constraint. Next we show that if $b < b_1$, there exists $T'_1(t)$ sending a burst of $b_1$ at $t_0$, such that under $\boldsymbol{T}'(t) = \left\{T'_1(t), ..., T_n(t) \right\}$ the last bit flow~$1$ sends at $t_0$ also achieves flow~$1$'s worst-case delay.

Define $\hat{t} = \sup\{ t \ | \  T_1(t_0) - T(t) \geq b_1 \}$. As $b < b_1$, $\hat{t} \in (t_s, t_0)$. Note that $T_1(t_0) - T_1(\hat{t}) \leq b_1$ as $T_1(t)$ is right continuous. Define $$T'_1(t) = \left\{
\begin{aligned}
&T_1(t), & \text{ when } t < \hat{t} \\
&T_1(t_0) - b_1, &\text{ when } \hat{t} \leq t < t_0 \\
&T_1(t_0), &\text{ otherwise }
\end{aligned}
\right.$$
which sends a burst of $b_1$ at $t_0$. Note that $T'_1(t)$ satisfies flow~$1$'s arrival curve. Specifically, consider any $0 \leq t^{(1)} < t^{(2)}$. 
\begin{itemize}
\item When $t^{(2)} < \hat{t}$, it has $T'_1\left(t^{(2)}\right) - T'_1\left(t^{(1)}\right) = T_1\left(t^{(2)}\right) - T_1\left(t^{(1)}\right)$. 
\item When $\hat{t} \leq t^{(2)} < t_0$, 
\begin{itemize}
\item If $t^{(1)} < \hat{t}$, $T'_1\left(t^{(2)}\right) - T'_1\left(t^{(1)}\right) = T_1(t_0) - b_1 - T_1\left(t^{(1)}\right) \leq T_1(\hat{t}) - T_1\left(t^{(1)}\right)
\leq
T_1\left(t^{(2)}\right) - T_1\left(t^{(1)}\right) $
\item If $\hat{t} \leq t^{(1)} < t_0$, $T'_1\left(t^{(2)}\right) - T'_1\left(t^{(1)}\right) = T_1(t_0) - b_1 - [T_1(t_0) - b_1] = 0$
\end{itemize}
\item When $t^{(2)} \geq t_0$,
\begin{itemize}
\item If $t^{(1)} \leq \hat{t}$, $T'_1\left(t^{(2)}\right) - T'_1\left(t^{(1)}\right) = T_1(t_0) -  T_1\left(t^{(1)}\right) \leq T_1\left(t^{(2)}\right) - T_1\left(t^{(1)}\right)$
\item If $\hat{t} < t^{(1)} < t_0$, $T'_1\left(t^{(2)}\right) - T'_1\left(t^{(1)}\right) = T_1(t_0) - [T_1(t_0) - b_1] = b_1 \leq b_1 + r_1\left(t^{(2)} - t^{(1)} \right)$
\item If $t^{(1)} \geq t_0$, $T'_1\left(t^{(2)}\right) - T'_1\left(t^{(1)}\right) = T_1(t_0) - T_1(t_0) = 0$
\end{itemize}

\end{itemize}

We then show that under $\boldsymbol{T}'(t)$ the last bit sent at $t_0$ also achieves flow~$1$'s worst-case delay. First observe that under $\boldsymbol{T}'(t)$ the last bit sent at $t_0$ arrives at the shared link no earlier than that under $\boldsymbol{T}(t)$. This is because it arrives at the reprofiler later than under $T(t)$ and experiences a no smaller reshaping delay upon its arrival. Particularly, given the reprofiler's service curve of $\left\{
\begin{aligned}
& b'_1 + r_1 t, &\text{ when } t > 0 \\
& 0  &\text{otherwise}
\end{aligned}
\right.$, 
the $b_1^{th}$ bit of a burst gets a delay of $\frac{b_1 - b'_1}{r_1}$, which equals the worst-case delay for flow~$1$ inside the reprofiler. Next we show that under $\boldsymbol{T}'(t)$ the last bit leaves the shared link no earlier than that under $\boldsymbol{T}(t)$, \ie overall it gets a no smaller delay under $\boldsymbol{T}'(t)$. Since under $\boldsymbol{T}(t)$ the last bit achieves the worst-case delay, so does under $\boldsymbol{T}(t)$.

Suppose the last bit arrives at the shared link at $\hat{t}_0$ under $\boldsymbol{T}(t)$, and at $\hat{t}'_0 \geq \hat{t}_0$ under $\boldsymbol{T}'(t)$. If under $\boldsymbol{T}(t)$ the last bit arrives to find the shared link with an empty queue, \ie it has no delay at the shared link, then combining it with $\hat{t}'_0 \geq \hat{t}_0$ gives what we want. Afterwards, we consider the case where under $\boldsymbol{T}(t)$ the last bit arrives at the shared link with a non-empty queue, \ie there exists $\hat{t}_s < \hat{t}_0$ and $\delta > 0$ such that the shared link processes data at full speed $R$ during $[\hat{t}_s, \hat{t}_0]$, and at a speed strictly less than $R$ during $[\hat{t}_s - \delta, \hat{t}_s)$. 

Under $\boldsymbol{T}(t)$, $T_1(\hat{t}_s, \hat{t}_0)$ data from flow~$1$ arrives at the shared link during $[\hat{t}_s, \hat{t}_0]$. Then the last bit leaves the shared link at time $\hat{t}_e = \hat{t}_s + \frac{\sum_{j \geq 2}[ T_j(\hat{t}_0) - T_j( \hat{t}_s) ]+ T_1(\hat{t}_s, \hat{t}_0)}{R}$. 
Whereas under $\boldsymbol{T}'(t)$, suppose there is $M \geq 0$ amount of data in the buffer at $\hat{t}_s$, and $T'_1(\hat{t}_s, \hat{t}'_0)$ amount of data from flow~$1$ arrives at the shared link during $[\hat{t}_s, \hat{t}'_0]$. As $T'_1(t)$ delays some data to $t_0$, and by $\hat{t}'_0$ all of the delayed data arrives at the shared link, it has  $T'_1(\hat{t}_s, \hat{t}'_0) \geq T_1(\hat{t}_s, \hat{t}_0)$. Then the $b_1^{th}$ bit leaves the shared link at a time no less than $\hat{t}_s + \frac{M + \sum_{j \geq 2}[ T_j(\hat{t}'_0) - T_j( \hat{t}_s) ]+ T'_1(\hat{t}_s, \hat{t}'_0)}{R}$, which is no less than $\hat{t}_e$.

%%%%%%%%%%%%%%%%%%%%%%%%%%%%%%%%%%%%%%%%%%%%%%%%%%%%%%%%%%%
\item Next we characterize the worst-case delay. Given that there always exists a traffic pattern that the $b_1^{th}$ bit of a burst gives flow~$1$'s worst-case delay, w.l.o.g we assume the worst-case delay to be achieved by the $b_1^{th}$ bit sent at $t_0$.

Remember that the $b_1^{th}$ bit of a burst gets a delay of $\frac{b_1 - b'_1}{r_1}$ inside the reprofiler. Next we consider the delay at the shared link. Suppose the $b_1^{th}$ bit arrives at the shared link at $\hat{t}_0$. If at $\hat{t}_0$ the shared link processes at a speed strictly less than $R$, then the $b_1^{th}$ bit gets no delay at the shared link, and therefore gets a overall worst-case delay of $\frac{b_1 - b'_1}{r_1}$. Otherwise, suppose the last busy period at the shared link starts at $0 \leq \hat{t}_s \leq \hat{t}_0$. 
\begin{enumerate}
\item When $\hat{t}_0 - \hat{t}_s \geq \frac{b_1 - b'_1}{r_1}$, during $[\hat{t}_s, \hat{t}_0]$ at most $\sum_{j= 1}^n b'_j + \sum_{j= 1}^n r_j \left( \hat{t}_0 - \hat{t}_s \right)$ amount of data arrives at the shared link. Thus, the delay for the $b_1^{th}$ bit at the shared link is $\frac{\sum_{j= 1}^n b'_j + \sum_{j= 1}^n r_j \left( \hat{t}_0 - \hat{t}_s \right)}{R} - \hat{t}_0$, which decreases with $\hat{t}_0$ since $R \geq \sum_{j= 1}^n r_j$. Given $\hat{t}_0 \geq \frac{b_1 - b'_1}{r_1} + \hat{t}_s$, the worst-case delay at the shared link is achieved at $\hat{t}_0 = \frac{b_1 - b'_1}{r_1} + \hat{t}_s$, and has a value of 
$\frac{\sum_{j= 1}^n b'_j + \frac{b_1 - b'_1}{r_1} \sum_{j= 1}^n r_j }{R} - \frac{b_1 - b'_1}{r_1} - \hat{t}_s.$
Thus the overall worst-case delay is achieved at $\hat{t}_s = 0$, with a value of 
$$ \frac{\sum_{j= 1}^n b'_j + \frac{b_1 - b'_1}{r_1} \sum_{j= 1}^n r_j }{R} - \frac{b_1 - b'_1}{r_1} + \frac{b_1 - b'_1}{r_1} = \frac{\sum_{j = 1}^n b'_j}{R} + \frac{(b_1 - b'_1)R_1}{r_1 R}.$$

\item When $\hat{t}_0 - \hat{t}_s < \frac{b_1 - b'_1}{r_1}$, as flow~$1$'s burst of $b'_1$ arrived at the shared link before $\hat{t}_0 - \frac{b_1 - b'_1}{r_1}$, it should has been cleared before $\hat{t}_s$. Hence, during $[\hat{t}_s, \hat{t}_0]$ at most $\sum_{j \neq 1}^n b'_j + \sum_{j= 1}^n r_j \left( \hat{t}_0 - \hat{t}_s \right)$ data arrive at the shared link. Therefore, the $b_1^{th}$ bit gets a delay of $\frac{\sum_{j \neq 1}^n b'_j + \sum_{j= 1}^n r_j \left( \hat{t}_0 - \hat{t}_s \right)}{R} - \hat{t}_0$ at the shared link, which decreases with $\hat{t}_0$ under $R \geq \sum_{j= 1}^n r_j$. Consequently, its worst-case delay is achieved at $\hat{t}_0 = \hat{t}_s$, with a value of  
$\frac{\sum_{j \neq 1}^n b'_j}{R} - \hat{t}_s$, which is maximized at $\hat{t}_s = 0$. Thus the overall worst-case delay is 
$\frac{b_1 - b'_1}{r_1} + \frac{\sum_{j \neq 1}^n b'_j}{R}.$
\end{enumerate}
Combining case 1 and 2 gives flow~$1$'s worst-case delay, \ie
$$\widehat{D}^*_1 = \max\left\{\frac{b_1 - b'_1}{r_1}+ \frac{\sum_{j \neq 1} b'_j}{R}, \frac{\sum_{j = 1}^n b'_j}{R} + \frac{(b_1 - b'_1)R_1}{r_1 R} \right\}.$$
\end{itemize}
\end{proof}

%%%%%%%%%%%%%%%%%%%%%%%%%%%% optimization %%%%%%%%%%%%%%%%%%%%%%%%%%%%%%
\subsection{Proofs for Proposition~\ref{fifo:R} and~\ref{fifo:B}}
\label{sec:opt-rf}

This section provides the solution for \textbf{OPT\_RF}, from which we then have Propositions~\ref{fifo:R} and~\ref{fifo:B}. For the reader's convenience, we restate \textbf{OPT\_RF}, where $R_1 = \sum_{i = 1}^n r_i$.
\bequn
\begin{aligned}
&\text{\textbf{OPT\_RF}} && \min_{\boldsymbol{b}'} 
R &\\
& \text{s.t}
& & \max\left\{\frac{b_i - b'_i}{r_i}+ \frac{\sum_{j \neq i} b'_j}{R}, \frac{\sum_{j = 1}^n b'_j}{R} + \frac{(b_i - b'_i)R_1}{r_i R} \right\} \leq d_i, \ &\forall \ 1 \leq i \leq n, \\
&&& R_1 \leq R, \quad 0 \leq b'_i \leq b_i, & \forall \ 1 \leq i \leq n. \\
\end{aligned}
\eequn

%%%%%%%%%%%%%%%%%%%%%%% lemma 1: OPT_F' %%%%%%%%%%%%%%%%%%%%%%%%%%%%%%%
\begin{lemma}\label{lemma:fifo_opt'}
For $1 \leq i \leq n$ define $T_i^{(1)} = \frac{R}{R+r_i}\left(H_i + \frac{r_i}{R}\widehat{B}'_n\right)$ and $T_i^{(2)} = b_i + \frac{r_i(\widehat{B}'_n - R d_i)}{R_1}$. Denote $\widehat{B}'_0 = 0$ and $\boldsymbol{\widehat{B}'} = (\widehat{B}'_1, ... \widehat{B}'_n)$. Consider the following optimization: 
\bequn
\begin{aligned}
&\text{\textbf{OPT\_RF'}} && \min_{\boldsymbol{\widehat{B}}'} 
R &\\
& \text{s.t}
& & \max\left\{0, T_i^{(1)}, T_i^{(2)} \right\} \leq \widehat{B}'_i -  \widehat{B}'_{i-1} \leq b_i, \ &\forall \ 1 \leq i \leq n, \\
&&& R_1 \leq R. \\
\end{aligned}
\eequn
Suppose the optimal solution for \textbf{OPT\_RF'} is 
$(R^*, \boldsymbol{\widehat{B}}'^*)$. 
Then $(R^*, \boldsymbol{b}'^*)$, where
$$\boldsymbol{b}'^* = (\widehat{B}'^*_1, \widehat{B}'^*_2 - \widehat{B}'^*_1, ..., \widehat{B}'^*_n - \widehat{B}'^*_{n-1})$$
is an optimal solution for \textbf{OPT\_RF}.
\end{lemma}

\begin{proof}
Define $\widehat{B}_i = \sum_{j = 1}^i b_j$. From basic algebraic manipulation \textbf{OPT\_RF} is equivalent to \textbf{OPT\_RF"}:
\bequn
\begin{aligned}
&\text{\textbf{OPT\_RF''}} && \min_{\boldsymbol{b'}} 
R &\\
& \text{s.t}
& & \max\left\{ 0, T_i^{(1)}, T_i^{(2)} \right\} \leq \widehat{B}'_i -  \widehat{B}'_{i-1} \leq b_i \ &\forall \ 1 \leq i \leq n, \\
&&& R_1 \leq R. \\
\end{aligned}
\eequn
As the constraints of \textbf{OPT\_RF'} and \textbf{OPT\_RF''} are the same, the two optimizations share the same optimal value $R^*$. From $\widehat{B}_i = \sum_{j = 1}^i b_j$, $\boldsymbol{b}'^* = (\widehat{B}'^*_1, \widehat{B}'^*_2 - \widehat{B}'^*_1, ..., \widehat{B}'^*_n - \widehat{B}'^*_{n-1})$ is then the optimal variable for \textbf{OPT\_RF''}. Hence, $(R^*, \boldsymbol{b}'^*)$ is an optimal solution for \textbf{OPT\_RF''}, and therefore an optimal solution for \textbf{OPT\_RF}.
\end{proof}

%%%%%%%%%%%%%%%%%%%%%%%%%%% solution  %%%%%%%%%%%%%%%%%%%%%%%%%%%%%%%

Next we proceed to solve \textbf{OPT\_RF'}, which when combined with Lemma~\ref{lemma:fifo_opt'} then gives Propositions~\ref{fifo:R} and~\ref{fifo:B}. 
Define $\mathbb{Z}_i = \{ 1 \leq j \leq i \mid j \in \mathbb{Z}\}$ for $1 \leq i \leq n$, 
$$
X_F(R) = \max_{P_1, P_2 \subseteq \mathbb{Z}_n, P_2 \neq \mathbb{Z}_n, P_1 \bigcap P_2 = \emptyset} \frac{\sum_{i \in P_1} \frac{R H_i}{R+r_i} + \sum_{i \in P_2} \left(b_i - \frac{r_i d_i R}{R_1} \right)}{1 - \sum_{i \in P_1}\frac{r_i}{R+r_i} - \sum_{i \in P_2} \frac{r_i}{R_1}},
$$ 
and 
$$
Y_F(R) = \min_{1 \leq i \leq n-1}\left\{\widehat{B}_n, Rd_n, 
\min_{P_1, P_2 \subseteq \mathbb{Z}_i, P_1 \bigcap P_2 = \emptyset, P_1 \bigcup P_2 \neq \emptyset} \left\{\frac{\widehat{B}_i - \sum_{j \in P_1} \frac{R H_j}{R+r_j} -  \sum_{j \in P_2} \left( b_j - \frac{r_j d_j R}{R_1}\right)}{\sum_{j \in P_1}\frac{r_j}{R+r_j} +  \sum_{j \in P_2} \frac{r_j}{R_1}} \right\} \right\},
$$ 
where $\widehat{B}_i = \sum_{j = 1}^i b_j$ for $1 \leq i \leq n$. Then from Lemma~\ref{lemma:fifo_region} we know that $R$ forms a feasible solution for \textbf{OPT\_RF'} iff $R \geq \max\left\{R_1,  \frac{\widehat{B}_n R_1}{\sum_{i = 1}^n r_i d_i}\right\}$ and $X_F(R) \leq Y_F(R)$. Therefore, $R^*$ is the minimum value satisfying these conditions.  

Given $R^*$, from Lemma~\ref{lemma:fifo_region} we know that there exists an optimal solution with $\widehat{B}'^*_n = X_F(R^*)$. From Statement 2 in Lemma~\ref{lemma:fifo1}, we know that suppose there exists an optimal solution with $\widehat{B}'_n = \widehat{B}'^*_n$ and $\widehat{B}'_i = \widehat{B}'^*_i$ where $2 \leq i \leq n$, then there exists an optimal solution with $\widehat{B}'_n = \widehat{B}'^*_n$, $\widehat{B}'_i = \widehat{B}'^*_i$, and $\widehat{B}'_{i-1} = \max\left\{\sum_{j = 1}^{i-1} T_j, \widehat{B}'^*_i - b_i \right\}$. Thus, based on $\widehat{B}'^*_n$ we can sequentially compute $\widehat{B}'^*_i$ from $i = n-1$ to $i = 1$. 

\begin{lemma}\label{lemma:fifo_region}
Define $T_i = \max\left\{0, T_i^{(1)}, T_i^{(2)} \right\}$. Then 
$\left\{\boldsymbol{\widehat{B}}' \ | \ T_h \leq \widehat{B}'_h -  \widehat{B}'_{h-1} \leq b_h, \ \forall 1 \leq h \leq n \right\} \neq \emptyset$ iff all of the following conditions hold: 
\begin{itemize}
\item $R \geq \frac{\sum_{i = 1}^n b_i R_1}{\sum_{i = 1}^n r_i d_i} = \frac{\widehat{B}_n R_1}{\sum_{i = 1}^n r_i d_i}$ 
\item $\widehat{B}'_n \geq \max_{P_1, P_2 \subseteq \mathbb{Z}_n, P_2 \neq \mathbb{Z}_n, P_1 \bigcap P_2 = \emptyset} \frac{\sum_{i \in P_1} \frac{R H_i}{R+r_i} + \sum_{p \in P_2} \left(b_i - \frac{r_i d_i R}{R_1} \right)}{1 - \sum_{i \in P_1}\frac{r_i}{R+r_i} - \sum_{i \in P_2} \frac{r_i}{R_1}}$, 
\item $\widehat{B}'_n \leq\min_{1 \leq i \leq n-1}\left\{\widehat{B}_n, Rd_n, 
\min_{P_1, P_2 \subseteq \mathbb{Z}_i, P_1 \bigcap P_2 = \emptyset, P_1 \bigcup P_2 \neq \emptyset} \left\{\frac{\widehat{B}_i - \sum_{j \in P_1} \frac{R H_j}{R+r_j} -  \sum_{j \in P_2} \left( b_j - \frac{r_j d_j R}{R_1}\right)}{\sum_{j \in P_1}\frac{r_j}{R+r_j} +  \sum_{j \in P_2} \frac{r_j}{R_1}} \right\} \right\}$
\end{itemize}
\end{lemma}
\begin{proof}
Define $\widehat{B}_i = \sum_{j = 1}^i b_j$ for $1 \leq i \leq n$, and define\\
$\widehat{\mathbb{B}}_i = \left[\max\left\{ T_i, \widehat{B}'_{i+1} - b_{i+1} \right\}, \ \min\left\{\widehat{B}_i, \widehat{B}'_{i+1}- T_{i+1} \right\} \right]$ for $1 \leq i \leq n-1$.
As $\widehat{B}'_0 = 0$, from basic algebraic manipulation $T_i = \max\left\{0, T_i^{(1)}, T_i^{(2)} \right\} \leq \widehat{B}'_i -  \widehat{B}'_{i-1} \leq b_i, \ \forall 1 \leq i \leq n$ is equivalent to 
\bequn
\left\{
\begin{aligned}
& \widehat{B}'_1 \in \left[\max\left\{ T_1, \widehat{B}'_2 - b_2 \right\}, \ \min\left\{b_1, \widehat{B}'_2- T_2 \right\} \right] = \widehat{\mathbb{B}}_1, \\
& T_i \leq \widehat{B}'_i -  \widehat{B}'_{i-1} \leq b_i, \ \forall \  2 < i \leq n
\end{aligned}
\right.
\eequn
Define $\boldsymbol{\widehat{B}}'_i = \left\{\widehat{B}_i, ..., \widehat{B}_n \right\}$, and $\widehat{\mathbb{B}}^{(i)} = \left\{\boldsymbol{\widehat{B}}'_i \ | \ T_h \leq \widehat{B}'_h -  \widehat{B}'_{h-1} \leq b_h, \ \forall i < h \leq n \right\}$. Then we have 
$$\left\{\boldsymbol{B}' \ | \ T_i \leq \widehat{B}'_i -  \widehat{B}'_{i-1} \leq b_i, \ \forall 1 \leq i \leq n \right\} \neq \emptyset \iff \widehat{\mathbb{B}}_1 \neq \emptyset \text{ and } \widehat{\mathbb{B}}^{(2)} \neq \emptyset,$$
which from Lemma~\ref{lemma:fifo1} is equivalent to 
\beq\label{eq:fifo_set}
\left\{ \sum_{j = 1}^n T_j \leq \widehat{B}'_n \leq \min \left\{\widehat{B}_n, Rd_n \right\} \mid \sum_{j = 1}^{i} T_j \leq \widehat{B}_i, \ \forall 1 \leq i \leq n-1 \right\} \neq \emptyset.
\eeq
Denote $\mathbb{Z}_i = \{ 1 \leq j \leq i \mid j \in \mathbb{Z}\}$ for $1 \leq i \leq n$, then 
\begin{itemize}
\item $\sum_{j = 1}^n T_j \leq \widehat{B}'_n$ implies that for all $P_1, P_2 \subseteq \mathbb{Z}_n$ and $P_1 \bigcap P_2 = \emptyset$, $\sum_{i \in P_1}T_i^{(1)} + \sum_{i \in P_2}T_i^{(2)} \leq B'_n$, \ie $\sum_{i \in P_1} \frac{R H_i}{R+r_i} + \sum_{p \in P_2}\left(b_i - \frac{r_i d_i R}{R_1} \right) \leq \left(1 - \sum_{i \in P_1}\frac{r_i}{R+r_i} -  \sum_{i \in P_2} \frac{r_i}{R_1} \right)B'_n$, which is equivalent to $ R \geq \frac{\sum_{i = 1}^n b_i R_1}{\sum_{i = 1} r_i d_i}$ when $P_2 = \mathbb{Z}_n$, and $\widehat{B}'_n \geq \frac{\sum_{i \in P_1} \frac{R H_i}{R+r_i} + \sum_{p \in P_2} \left(b_i - \frac{r_i d_i R}{R_1} \right)}{1 - \sum_{i \in P_1}\frac{r_i}{R+r_i} - \sum_{i \in P_2} \frac{r_i}{R_1}}$ otherwise.
\item For $1 \leq i \leq n-1$, $\sum_{j = 1}^{i} T_j \leq \widehat{B}_i$ implies that for all $P_1, P_2 \subseteq \mathbb{Z}_i$ and $P_1 \bigcap P_2 = \emptyset$, $\sum_{i \in P_1}T_i^{(1)} + \sum_{i \in P_2}T_i^{(2)} \leq \widehat{B}_i$, \ie $\sum_{i \in P_1} \frac{R H_i}{R+r_i} + \sum_{p \in P_2} \left( b_i - \frac{r_i d_i R}{R_1}\right) + \left( \sum_{i \in P_1}\frac{r_i}{R+r_i} +  \sum_{i \in P_2} \frac{r_i}{R_1} \right)B'_n \leq \widehat{B}_i$, which is equivalent to $\widehat{B}'_n \leq \frac{\widehat{B}_i - \sum_{j \in P_1} \frac{R H_j}{R+r_j} -  \sum_{j \in P_2} \left( b_j - \frac{r_j d_j R}{R_1}\right)}{\sum_{j \in P_1}\frac{r_j}{R+r_j} +  \sum_{j \in P_2} \frac{r_j}{R_1}}$.
\end{itemize}
Therefore, \Eqref{eq:fifo_set} holds iff 
\bequn
\left\{
\begin{aligned}
& R \geq \frac{\sum_{i = 1}^n b_i R_1}{\sum_{i = 1} r_i d_i} \\
& \widehat{B}'_n \geq \max_{P_1, P_2 \subseteq \mathbb{Z}_n, P_2 \neq \mathbb{Z}_n, P_1 \bigcap P_2 = \emptyset} \frac{\sum_{i \in P_1} \frac{R H_i}{R+r_i} + \sum_{i \in P_2} \left(b_i - \frac{r_i d_i R}{R_1} \right)}{1 - \sum_{i \in P_1}\frac{r_i}{R+r_i} - \sum_{i \in P_2} \frac{r_i}{R_1}}\\
& \widehat{B}'_n \leq \min_{1 \leq i \leq n-1}\left\{\widehat{B}_n, Rd_n, 
\min_{P_1, P_2 \subseteq \mathbb{Z}_i, P_1 \bigcap P_2 = \emptyset, P_1 \bigcup P_2 \neq \emptyset} \left\{\frac{\widehat{B}_i - \sum_{j \in P_1} \frac{R H_j}{R+r_j} -  \sum_{j \in P_2} \left( b_j - \frac{r_j d_j R}{R_1}\right)}{\sum_{j \in P_1}\frac{r_j}{R+r_j} +  \sum_{j \in P_2} \frac{r_j}{R_1}} \right\} \right\}
\end{aligned}
\right.
\eequn
\end{proof}

%%%%%%%%%%%%%%%%%%%%%%%%%%%%%%%%%%%%%%%%%%%%%%%%%%%%%%%%%%%%%%%%%%%%%

Next we establish Lemma~\ref{lemma:fifo1}.
For $\widehat{\mathbb{B}}_1 = \left[\max\left\{ T_1, \widehat{B}'_2 - b_2 \right\}, \ \min\left\{b_1, \widehat{B}'_2- T_2 \right\} \right]$, $\boldsymbol{\widehat{B}}'_2 = \left\{\widehat{B}'_2, ..., \widehat{B}'_n \right\}$, and $\widehat{\mathbb{B}}^{(2)} = \left\{\boldsymbol{\widehat{B}}'_2  \ | \ T_h \leq \widehat{B}'_h -  \widehat{B}'_{h-1} \leq b_h, \ \forall 2 < h \leq n \right\}$, we have

\begin{lemma}\label{lemma:fifo1}
$\widehat{\mathbb{B}}_1 \neq \emptyset$ and $\widehat{\mathbb{B}}^{(2)}\neq \emptyset$ iff 
$$\left\{ \sum_{j = 1}^n T_j \leq \widehat{B}'_n \leq \min \left\{\widehat{B}_n, Rd_n \right\} \mid \sum_{j = 1}^{i} T_j \leq \widehat{B}_i, \ \forall 1 \leq i \leq n-1 \right\} \neq \emptyset.$$
\end{lemma}

\begin{proof}
For $1 \leq i \leq n$ define $\widehat{B}_i = \sum_{j = 1}^i b_j$ and $\boldsymbol{\widehat{B}}'_i = \left\{\widehat{B}_i, ..., \widehat{B}_n \right\}$.
For $1 \leq i \leq n-1$ further define $\widehat{\mathbb{B}}_i = \left[\max\left\{ \sum_{j = 1}^i T_j, \widehat{B}'_{i+1} - b_{i+1} \right\}, \ \min\left\{\widehat{B}_i, \widehat{B}'_{i+1}- T_{i+1} \right\} \right]$ and\\
$\widehat{\mathbb{B}}^{(i)} = \left\{\boldsymbol{\widehat{B}}'_i \ | \ T_h \leq \widehat{B}'_h -  \widehat{B}'_{h-1} \leq b_h, \ \forall i < h \leq n \right\}$. Suppose we have 
\begin{flushleft}
\textbf{Statement 1.} $\widehat{\mathbb{B}}_1 \neq \emptyset$ and $\widehat{\mathbb{B}}^{(2)}\neq \emptyset$ iff 1) $\widehat{\mathbb{B}}_{n-1} \neq \emptyset$, 2) $\sum_{j = 1}^h T_h \leq \widehat{B}_h$, for $1 \leq h \leq n-2$, and 3) $\widehat{B}'_n \leq Rd_{n-1}$.
\end{flushleft}
Basic algebraic manipulations give that $\widehat{\mathbb{B}}_{n-1} \neq \emptyset$, \ie $\max\left\{ \sum_{j = 1}^{n-1} T_j, \widehat{B}'_n - b_n \right\} \leq \min\left\{\widehat{B}_{n-1}, \widehat{B}'_n- T_n \right\}$, iff 1) $\sum_{j = 1}^n T_j \leq \widehat{B}'_n \leq \widehat{B}_n$, 2) $\sum_{j = 1}^{n-1} T_j \leq \widehat{B}_{n-1}$, and 3) $ \widehat{B}'_n \leq R d_n$. Combining them with $\sum_{j = 1}^h T_h \leq \widehat{B}_h$, for $1 \leq h \leq n-2$ and $\widehat{B}'_n \leq Rd_{n-1}$ gives $\sum_{j = 1}^n T_j \leq \widehat{B}'_n \leq \min\left\{\widehat{B}_n, Rd_n\right\}$ and $\sum_{j = 1}^h T_h \leq \widehat{B}_h$, for $1 \leq h \leq n-1$. Therefore, we have Lemma~\ref{lemma:fifo1}.

Next, we show Statement 1 based on Statement 2. For convenience, define $\widehat{\mathbb{B}}^{(n)} = \{1\}$. Then we have

\begin{flushleft}
\textbf{Statement 2.} For $1 \leq i \leq n-2$, $\widehat{\mathbb{B}}_i \neq \emptyset$ and $\widehat{\mathbb{B}}^{(i+1)}\neq \emptyset$ iff 1) $\widehat{\mathbb{B}}_{i+1} \neq \emptyset$ and $\widehat{\mathbb{B}}^{(i+2)} \neq \emptyset$, 2) $\sum_{j = 1}^i T_j \leq \widehat{B}_i$, and 3) $\widehat{B}'_n \leq Rd_{i+1}$.
\end{flushleft}

\paragraph{Proof for Statement 1} we show Statement 1 by induction. For $1 \leq i \leq n-1$, define 
$$S_i: 
\widehat{\mathbb{B}}_i \neq \emptyset, \
\widehat{\mathbb{B}}^{(i+1)}\neq \emptyset,  \
\widehat{B}'_n \leq Rd_i,  
\text{ and }  \sum_{j = 1}^h T_h \leq \widehat{B}_h, \forall 1 \leq h \leq i-1.$$
\begin{itemize}
\item When $i = 1$, Statement 2 directly gives that $\widehat{\mathbb{B}}_1 \neq \emptyset$ and $\widehat{\mathbb{B}}^{(2)}\neq \emptyset$ iff $S_2$ holds.
\item When $i \geq 1$, suppose $\widehat{\mathbb{B}}_1 \neq \emptyset$ and $\widehat{\mathbb{B}}^{(2)}\neq \emptyset$ iff $S_i$ holds, \ie 1) $\widehat{\mathbb{B}}_i \neq \emptyset$ and $\widehat{\mathbb{B}}^{(i+1)}\neq \emptyset$, 2) $\widehat{B}'_n \leq Rd_i$, and 3)$\sum_{j = 1}^h T_h \leq \widehat{B}_h, \forall 1 \leq h \leq i-1$. Note that by Statement 2 we have $\widehat{\mathbb{B}}_i \neq \emptyset$ and $\widehat{\mathbb{B}}^{(i+1)}\neq \emptyset$ iff i) $\widehat{\mathbb{B}}_{i+1} \neq \emptyset$ and $\widehat{\mathbb{B}}^{(i+2)} \neq \emptyset$, ii) $\sum_{j = 1}^i T_i \leq \widehat{B}_i$, and iii) $\widehat{B}'_n \leq Rd_{i+1}$. Thus, $\widehat{\mathbb{B}}_1 \neq \emptyset$ and $\widehat{\mathbb{B}}^{(2)}\neq \emptyset$ iff 1) $\widehat{\mathbb{B}}_{i+1} \neq \emptyset$ and $\widehat{\mathbb{B}}^{(i+2)} \neq \emptyset$, 2) $\widehat{B}'_n \leq \min\left\{ Rd_i, Rd_{i+1}\right\} = Rd_{i+1}$, and 3) $\sum_{j = 1}^h T_h \leq \widehat{B}_h, \forall 1 \leq h \leq i$, \ie $S_{i+1}$ holds.
\end{itemize}
Thus, we have $\widehat{\mathbb{B}}_1 \neq \emptyset$ and $\widehat{\mathbb{B}}^{(2)}\neq \emptyset$ iff $S_{n-1}$ holds: $\widehat{\mathbb{B}}_{n-1} \neq \emptyset, \
\widehat{\mathbb{B}}^{(n)} = \{ 1\} \neq \emptyset,  \
\widehat{B}'_n \leq Rd_{n-1}$,  
and $\sum_{j = 1}^h T_h \leq \widehat{B}_h, \forall 1 \leq h \leq n-2$, which then gives Statement 1.

\paragraph{Proof for Statement 2} Consider $\widehat{\mathbb{B}}_i \neq \emptyset$ and $\widehat{\mathbb{B}}^{(i+1)} \neq \emptyset$.
\begin{itemize}
\item $\widehat{\mathbb{B}}_i \neq \emptyset$ iff $\max\left\{ \sum_{j = 1}^i T_j, \widehat{B}'_{i+1} - b_{i+1} \right\} \leq \min\left\{\widehat{B}_i, \widehat{B}'_{i+1}- T_{i+1} \right\}$, which from basic algebraic manipulation is equivalent to 1) $\sum_{j = 1}^{i+1} T_j \leq \widehat{B}'_{i+1} \leq \widehat{B}_{i+1}$, 2) $\sum_{j = 1}^i T_j \leq \widehat{B}_i$, and 3) $b_{i+1} \geq T_{i+1} \iff \widehat{B}'_n \leq R d_{i+1}$.  
\item Consider $\widehat{\mathbb{B}}^{(i+1)} \neq \emptyset$. When $i < n-2$, from basic algebraic manipulation it is equivalent to 1) $T_{n+2} \leq \widehat{B}'_{i+2} -  \widehat{B}'_{i+1} \leq b_{i+2}$ and 2) $\widehat{\mathbb{B}}^{(i+2)} \neq \emptyset$. When $i = n-2$, $\widehat{\mathbb{B}}^{(i+1)} = \left\{\boldsymbol{\widehat{B}}'_{n-1} \mid T_n \leq \widehat{B}'_n - \widehat{B}'_{n-1} \leq b_n \right\}$, which is non-empty iff $ T_n \leq \widehat{B}'_n - \widehat{B}'_{n-1} \leq b_n$. Since $\widehat{\mathbb{B}}^{n} = \{1\}$, it also has $\widehat{\mathbb{B}}^{(i+2)} \neq \emptyset$.
\end{itemize}
Note that $\sum_{j = 1}^{i+1} T_j \leq \widehat{B}'_{i+1} \leq \widehat{B}_{i+1}$ and $T_{n+2} \leq  \widehat{B}'_{i+2} -  \widehat{B}'_{i+1} \leq b_{i+2}$ iff $\widehat{\mathbb{B}}_{i+1} \neq \emptyset$. Thus we have Statement 2.
\end{proof}

\section{On the Benefit of Grouping Flows With Different Deadlines}
\label{app:one_static_merge}

In this section, we explore scenarios that consist of two flows sharing a common link whose access is arbitrated by a static priority scheduler.  The goal is to identify configurations that minimize the link bandwidth required to meet the flows' deadlines.  Of particular interest is assessing when the two flows should be assigned different priorities or instead merged into the same priority class.

Recalling the discussion of Section~\ref{sec:perf_2flow}, specializing  Propositions~\ref{nflow:R} and~\ref{fifo:R} to two flows, we find that the minimum required bandwidth for the two-flow scenario under static priority+shaping is
\beq
\tag{\ref{eq:2flow_stat+s}}
\widetilde{R}^*_R = \left\{
\begin{aligned}%\label{eq:2flow_prio}
 & max\left\{ r_1 + r_2, \frac{b_2}{d_2}, \frac{b_1 + b_2 - r_2 d_2}{d_1} + r_2 \right\}, & \text{ when } \frac{b_2}{r_2} \geq \frac{b_1}{r_1} \\
 & max\left\{ r_1 + r_2, \frac{b_2}{d_2}, \frac{b_1 + \max\{ b_2 - r_2 d_2, 0\}}{d_1} + r_2 \right\},  & \text{ otherwise}
\end{aligned}
\right.
\eeq
and that under fifo+shaping it is
\beq %\label{eq:2flow_fifo}
\tag{\ref{eq:2flow_fifo+s}}
\widehat{R}^*_R = \max \left\{ r_1 + r_2, \frac{b_2}{d_2}, \frac{(b_1 + b_2)(r_2 + r_2)}{d_1 r_1 + d_2 r_2}, \frac{b_1 + b_2 - d_1 r_1 + \sqrt{(b_1 + b_2 - d_1 r_1 )^2 + 4 r_1 d_2 b_2}}{2d_2} \right\}.
\eeq
Comparing them gives
\begin{proposition}
For the two-flow scenario, $\widetilde{R}^*_R > \widehat{R}^*_R$ iff 
$$d_1 \in \left(\frac{b_2}{r_2}, \frac{b_1}{r_1} \right) \text{ and } d_2 \in \left( \frac{(b_1 + b_2)(r_1 + r_2)}{r_2(b_1/d_1 + r_2)} - \frac{d_1 r_1}{r_2},d_1\right).$$
\end{proposition}
\begin{proof}
 When $\widetilde{R}^*_R = max\left\{ r_1 + r_2, \frac{b_2}{d_2} \right\}$, from \Eqref{eq:2flow_fifo} $\widetilde{R}^*_R \leq \widehat{R}^*_R$. Below we consider 1) $\frac{b_2}{d_2} \geq \frac{b_1}{d_1}$ and 2) $\frac{b_2}{d_2} < \frac{b_1}{d_1}$ separately under $\widetilde{R}^*_R > max\left\{ r_1 + r_2, \frac{b_2}{d_2} \right\}$.
\begin{enumerate}
    \item When $\frac{b_2}{d_2} \geq \frac{b_1}{d_1}$, we show that $\widetilde{R}^*_R \leq \widehat{R}^*_R$. Specifically, $\widetilde{R}^*_R > max\left\{ r_1 + r_2, \frac{b_2}{d_2} \right\}$ iff $\frac{b_1 + b_2 - r_2 d_2}{d_1} + r_2 > \max \left\{ r_1 + r_2, \frac{b_2}{d_2} \right\}$, which from basic algebraic manipulation equivalents to $b_1 + b_2 > r_1 d_1 + r_2 d_2$ and $d_2 < \frac{b_1 + b_2 - \sqrt{(b_1 + b_2)^2 - 4 r_2 b_2 d_1}}{2r_2}$. Next we show that under $b_1 + b_2 > r_1 d_1 + r_2 d_2$, it has $\frac{(b_1 + b_2)(r_2 + r_2)}{d_1 r_1 + d_2 r_2} \geq \frac{b_1 + b_2 - r_2 d_2}{d_1} + r_2$, and therefore $\widetilde{R}^*_R \leq \widehat{R}^*_R$.
    
    Consider $f(d_2) = \frac{(b_1 + b_2)(r_2 + r_2)}{d_1 r_1 + d_2 r_2} - \frac{b_1 + b_2 - r_2 d_2}{d_1} - r_2$, which equals~$0$ when $d_2 = d_1$. Basic algebraic gives that 
    \bequn
    \begin{aligned}
    \frac{d f(d_2)}{d d_2} &= \frac{r_2}{d_1} - \frac{(b_1 + b_2)(r_2 + r_2) r_2}{(d_1 r_1 + d_2 r_2)^2} \\
    & = \frac{r_2}{(d_1 r_1 + d_2 r_2)^2} \left(\frac{(d_1 r_1 + d_2 r_2)^2}{d_1} -(b_1 + b_2)(r_2 + r_2) \right) \\
    & \leq \frac{r_2}{(d_1 r_1 + d_2 r_2)^2} \left(\frac{(d_1 r_1 + d_2 r_2)^2}{d_1} -(d_1 r_1 + d_2 r_2)(r_2 + r_2) \right) \\
    & = \frac{r_2^2(d_2 - d_1)}{d_1(d_1 r_1 + d_2 r_2)} \leq 0
    \end{aligned}
    \eequn
    Thus, for all $d_2 \leq d_1$, it has $f(d_2) \geq f(d_1) = 0$, \ie $\frac{(b_1 + b_2)(r_2 + r_2)}{d_1 r_1 + d_2 r_2} \geq \frac{b_1 + b_2 - r_2 d_2}{d_1} + r_2$.
    
    \item When $\frac{b_2}{d_2} < \frac{b_1}{d_1}$, we show that $\widetilde{R}^*_R > \widehat{R}^*_R$ iff $d_1 \in \left(\frac{b_2}{r_2}, \frac{b_1}{r_1} \right)$ and $d_2 \in \left( \frac{(b_1 + b_2)(r_1 + r_2)}{r_2(b_1/d_1 + r_2)} - \frac{d_1 r_1}{r_2},d_1\right).$
    Specifically, $\widetilde{R}^*_R > max\left\{ r_1 + r_2, \frac{b_2}{d_2} \right\}$ iff $\frac{b_1 + \max\{ b_2 - r_2 d_2, 0\}}{d_1} + r_2 > \max \left\{ r_1 + r_2, \frac{b_2}{d_2} \right\}$, which from basic algebraic manipulation equivalents to 
    \bequn
    \left\{
    \begin{aligned}
    & \frac{b_1 + b_2 - r_2 d_2}{d_1} + r_2 > \max \left\{ r_1 + r_2, \frac{b_2}{d_2} \right\}, &\text{ when } d_2 \leq \frac{b_2}{r_2} \\
    & \frac{b_1}{d_1} + r_2, &\text{ otherwise }.
    \end{aligned}
    \right.
    \eequn
    When $d_2 \leq \frac{b_2}{r_2}$, similar as before we have $\widetilde{R}^*_R \leq \widehat{R}^*_R$. When $d_2 > \frac{b_2}{r_2}$, basic algebraic manipulation gives that $ \frac{b_1}{d_1} + r_2 > max\left\{ r_1 + r_2, \frac{b_2}{d_2} \right\}$ iff $d_1 < \frac{b_1}{r_1}$. Combining them gives that $\widetilde{R}^*_R \leq \widehat{R}^*_R$ when $(d_1, d_2) \notin \left\{ (d_1, d_2) \ | \ \frac{b_2}{r_2} < d_2 < d_1 < \frac{b_1}{r_1} \right\}$. 
     When $(d_1, d_2) \in \left\{ (d_1, d_2) \ | \ \frac{b_2}{r_2} < d_2 < d_1 < \frac{b_1}{r_1} \right\}$, basic algebraic manipulation gives that $\frac{(b_1 + b_2)(r_2 + r_2)}{d_1 r_1 + d_2 r_2} > \frac{b_1 + b_2 - d_1 r_1 + \sqrt{(b_1 + b_2 - d_1 r_1 )^2 + 4 r_1 d_2 b_2}}{2d_2}$ and $r_1 + r_2 > \frac{b_2}{d_2}$, \ie 
     $$\widehat{R}^*_R = \max\left\{r_1 + r_2, \frac{(b_1 + b_2)(r_2 + r_2)}{d_1 r_1 + d_2 r_2} \right\} = \left\{ 
     \begin{aligned}
     & r_1 + r_2, &\text{ if } b_1 + b_2 \leq r_1d_1 + r_2 d_2 \\
     & \frac{(b_1 + b_2)(r_2 + r_2)}{d_1 r_1 + d_2 r_2}, & \text{otherwise}
     \end{aligned}
     \right.$$
     When $b_1 + b_2 \leq r_1d_1 + r_2 d_2$, it has $\widetilde{R}^*_R > \widehat{R}^*_R$. Otherwise, basic algebraic manipulation gives that $\widetilde{R}^*_R > \widehat{R}^*_R$ iff $d_2 > \frac{(b_1 + b_2)(r_1 + r_2)}{r_2(b_1/d_1 + r_2)} - \frac{d_1 r_1}{r_2} := g(d_1)$. Note that when $\frac{b_2}{r_2} < d_1 < \frac{b_1}{r_1}$, $g(d_1) > \frac{b_2}{r_2}$; and when $d_1 = \frac{b_2}{r_2}$ or $d_1 = \frac{b_1}{r_1}$, $g(d_1) = d_1$. Therefore, $\widetilde{R}^*_R > \widehat{R}^*_R$ iff $d_2 \in \left( \frac{(b_1 + b_2)(r_1 + r_2)}{r_2(b_1/d_1 + r_2)} - \frac{d_1 r_1}{r_2},d_1\right).$

\end{enumerate}

\end{proof}

\section{ Extensions to Packet-Based Models in the Two-Flow Static Priority Case}\label{app:one_2flow}
In this section, we consider a more general packet-based model, where flow $i$ has maximum packet size of $l_i$ and the scheduler relies on static priorities. For ease of exposition, we only consider scenarios that consist of $n=2$ flows, and consequently two priority classes (low and high).  

We first characterize in Proposition~\ref{delay-h} the worst-case delay of high-priority packets, and use the result to identify a condition for when adding a reprofiler can help lower the required bandwidth (Corollary~\ref{improve}).  We also confirm (Proposition~\ref{delay-l0}) the intuitive property that reshaping the low-priority flow does not contribute to lowering the required bandwidth.  We then proceed to characterize in Proposition~\ref{delay-l} the worst-case delay of low-priority packets.  The results of Propositions~\ref{delay-h} and~\ref{delay-l} are used to formulate an optimization, {\bf OPT\_2}, that seeks to identify the optimum reshaping parameters for the high-priority flow that minimizes the link bandwidth required to meet the flows' deadlines.  The bulk the section is devoted to solving this optimization, while also establishing the intermediate result that the optimal reshaping can be realized simply by reducing the flow's burst size, \ie keeping its rate constant.

Returning to our two-flow, packet-based scenario, consider two token-bucket controlled flows $(r_1, b_1)$ and $(r_2, b_2)$ sharing a link with a bandwidth of $R$ whose access is controlled by a static priority scheduler. Assume that flow $(r_i, b_i)$ has a deadline of $d_i$ ($d_1 > d_2 > 0$), and $(r_2, b_2)$ has non-preemptive priority over $(r_1, b_1)$ at the shared link. Denote the maximum packet length of $(r_i, b_i)$ as $l_i < b_i$. Note that by setting $l_i = 0$, the packet-based model defaults to the fluid model.
To guarantee that none of the packets in $(r_1, b_1)$ or $(r_2, b_2)$ misses the deadline, $R$ needs to satisfy~\cite{nc}: 
\begin{subequations}\label{delay0}
\begin{equation} \label{delay0-r} 
r_1 + r_2 \leq R,
\end{equation}
\begin{equation}\label{delay0-h} 
\frac{b_2 + l_1 }{R} \leq d_2,
\end{equation}
\begin{equation} \label{delay0-l} 
\frac{b_2 + b_1}{R-r_2} \leq d_1.
\end{equation}
\end{subequations}

Therefore, the minimum bandwidth for the link satisfies: 
\beq
\label{eq:minR_noshap}
\tilde{R}^{(2)*} = \max\left\{r_1 + r_2, \frac{l_1 +b_2}{d_2}, \frac{b_1 + b_2}{d_1} + r_2\right\}.
\eeq

Now consider adding a lossless packet-based greedy leaky-bucket (re)profiler for each flow before the shared link, as shown in the above figure. Denote $(r_i, b_i)$'s reprofiler as $(r'_i, b'_i)$, where $b'_i \geq l_i$. To guarantee a finite delay inside the reprofiler, we also need $r'_i \geq r_i$. Under this assumption, if $b'_i \geq b_i$, the reprofiler has no effect. Hence, we further require that $b'_i < b_i$. 

Next, we proceed to characterize the optimal minimum required bandwidth under static priority and (re)shaping. Denote it as $\tilde{R}^{(2)*}_R$.

Under a non-preemptive static priority discipline, the worst-case delay of the high-priority flow is unaffected by the low-priority flow's arrival curve $(r_1, b_1)$, and only depends on its maximum packet size $l_1$. This is because high-priority packets arriving to an empty high-priority queue wait for at most the transmission time of one low-priority packet. This property holds whether reprofilers are present or not. Specifically,

\begin{proposition}\label{delay-h}
For a high priority token bucket-controlled flow $(r_2, b_2)$ traversing a lossless packet-based greedy token bucket reprofiler $(r'_2, b'_2)$, where $r_2 \leq r'_2$ and $b_2 > b'_2$, before going through a shared link with bandwidth $R>r_2$, the worst-case delay is 
$$D^*_2 = \max\left\{\frac{b_2 + l_1}{R}, \frac{b_2 - b'_2}{r'_2} + \frac{l_1+l_2}{R} \right\},$$
where $l_1$ is the maximum packet length of the low-priority flow with which it shares the link, and $l_2$ is its own maximum packet length. 
\end{proposition} 

\begin{proof}
Denote the virtual delay at $t$ inside the reprofiler as $D_1(t)$, and that at the shared link as $D_2(t)$. Then for a packet arriving the system at $t$, its virtual delay inside the system is $D_1(t) + D_2(t + D_1(t))$. 
For $D_1(t)$, we have 
$$D_1(t) = \inf_{0 \leq \tau} \left\{b_2 + r_2 t \leq b'_2 + t'_2(t + \tau) \right\} = \left[\frac{b_2 - b'_2 + r_2 t}{r'_2} - t \right]^+.$$
For $D_2(t + D_1(t))$, we have
$$D_2(t + D_1(t)) = \frac{l_1 + l_2}{R} + \inf_{0 \leq \tau}\left\{b_2 + r_2 t - l_2 \leq R(t + \tau + D_1(t)) \right\} = \frac{l_1 + l_2}{R} + \left[\frac{b_2 + r_2 t - l_2}{R} - t - D_1(t) \right]^+.$$ 
Then we have 
\beq
\begin{aligned}
D^*_2 &= \sup_{0 \leq t}\left\{D_1(t) + D_2(t + D_1(t)) \right\} \\
& \leq \sup_{0 \leq t} \left\{\max \left\{ \frac{l_1 + l_2}{R} + \frac{b_2 - b'_2 + r_2 t}{r'_2} - t, \frac{l_1 + l_2}{R} + \frac{b_2 + r_2 t - l_2}{R} - t\right\}\right\} \\
& \leq \max \left\{ \frac{l_1 + l_2}{R} + \frac{b_2 - b'_2}{r'_2}, \frac{l_1 + b_2}{R}\right\}
\end{aligned}
\eeq
\end{proof}

Note that after adding the reprofiler, we have $D^*_2 \geq \frac{b_2 + l_1}{R}$. Comparing this expression to \Eqref{delay0-h}, we know that adding reprofilers will never decrease the high-priority flow's worst-case delay. Furthermore, since ensuring stability of the shared queue mandates $R \geq r_1 + r_2$ irrespective of whether reprofilers are used, \Eqref{eq:minR_noshap} then gives

\begin{corollary}\label{improve}
Adding reprofilers decreases the minimum required bandwidth only when 
\beq
\tilde{R}^{(2)*} = \frac{b_1 + b_2}{d_1} + r_2 > \max\left\{ r_1 + r_2, \frac{b_2+ l_1}{d_2}\right\}. 
\eeq 
\end{corollary}

Conversely, we note that for the low-priority flow $(r_1, b_1)$, its service curve is determined by both the high-priority flow's arrival curve at the shared link and the shared link's bandwidth, and does not depend on the presence or absence of its own reprofiler. As a result, adding a reprofiler to the low-priority flow cannot decrease its worst-case delay (though it can increase it). Consequently, reprofiling the low-priority flow cannot contribute to reducing the bandwidth of the shared link while meeting the delay bounds of both the high and low-priority flows. This is formally stated in the next proposition that simplifies the investigation of the worst-case delay experienced by the low-priority flow by allowing us to omit the use of a reprofiler for it.

\begin{proposition}\label{delay-l0}
Given the service curve assigned to a low-priority flow, adding a packet-based greedy reprofiler cannot decrease its worst-case delay, and consequently cannot reduce the minimum link bandwidth required to meet the worst-case delay guarantees of both the high and low-priority flows.
\end{proposition} 

\begin{proof}
Denote the service curve of the low-priority flow as $\beta(t)$ and its arrival curve as $\alpha(t)$. Without reprofiler, the system's virtual delay at $t$ is 
$$D'(t) = \inf_{\tau \geq 0} \{\tau: \alpha(t) \leq \beta(t + \tau)\}.$$

Denote the reprofiler's maximum service curve as $\sigma(t)$, which is also the arrival curve for the shared link. Due to packetization, the system provides the flow a service curve no greater than
\bequn
\sigma \otimes \beta(t) = \inf_{0 \leq s \leq t} \left\{\sigma(s) + \beta(t-s) \right\} \leq \sigma(0) + \beta(t) = \beta(t),
\eequn
Hence, the virtual delay is
\bequn
D(t) \geq \inf_{\tau \geq 0} \left\{\tau: \alpha(t) \leq \inf_{0 \leq s \leq t} \left\{\sigma(s)  + \beta(t + \tau -s) \right\}\right\} 
\geq \inf_{\tau \geq 0} \{\tau: \alpha(t) \leq \beta(t + \tau)\} = D'(t).
\eequn

As $D(t) \geq D'(t)$, $\forall t \geq 0$, we have $\sup_{t \geq 0} \{D(t)\} \geq \sup_{t \geq 0} \{D'(t)\}$, \ie adding a reprofiler cannot decrease the system's worst-case delay.  

\end{proof}

Next, we characterize the worst-case delay $D_1^*$ of the low-priority flow. 
\begin{proposition}\label{delay-l}
Given a token bucket-controlled high priority flow $(r_2, b_2)$ with a packet-based greedy token bucket reprofiler $(r'_2, b'_2)$ going through a shared link with bandwidth $R$, the low-priority flow $(r_1, b_1)$'s worst-case delay $d^*_1$ is 
\begin{enumerate}[nosep]
\item when $r_2 = r'_2$, $D^*_1 = \frac{b_1 + b'_2}{R-r_2}$; 
\item when $r_2 < r'_2$ and $\frac{(R-r'_2)(b_2 - b'_2)}{r'_2 - r_2} - (b_1 + b'_2) < 0$, $D^*_1 = \frac{b_1 + b_2}{R-r_2}$; 
\item otherwise, \ie $r_2 < r'_2$ and $\frac{(R-r'_2)(b_2 - b'_2)}{r'_2 - r_2} - (b_1 + b'_2) \geq 0$ (recall that $r_2\leq r_2^{\prime}$)\\
$D^*_1 = \max\left\{\frac{b_1 + b'_2}{R-r'_2}, \frac{b_1 + b_2}{r_1} - \frac{(R- r_1 - r_2)(b_2 - b'_2)}{r_1(r'_2 - r_2)}\right\}$.
\end{enumerate}
\end{proposition}

\begin{proof}
The high-priority flow's arrival curve at the shared link is 
\beq
\alpha_2(t) = \min \left\{\gamma_{r,b}(t), \gamma_{r',b'}(t) \right\},
\eeq
and the low-priority flow's service curve at the shared link is 
\beq\label{ac_2}
\beta_1(t) = \left[Rt - \alpha_2(t) \right]^+ = \left[Rt - \min \left\{\gamma_{r_2,b_2}(t), \gamma_{r'_2,b'_2}(t) \right\} \right]^+.
\eeq
Hence, the virtual delay at $t > 0$ is
\beq
D(t) = \inf_{\tau \geq 0} \left\{ \tau: r_1t + b_1 \leq \left[ R(t + \tau) - \min\{r_2(t+\tau)+b_2, r'_2(t+\tau)+b'_2\}\right]^+\right\} 
%%& = \inf_{\tau \geq 0} \left\{ \tau: r_1t + b_1 \leq \max\left\{R(t + \tau) - r_2(t+\tau) - b_2, R(t + \tau) - r'_2(t+\tau) - b'_2\right\}\right\}
\eeq
When $r_2 = r'_2$, we have
\beq
\begin{aligned}
D(t) &= \inf_{\tau \geq 0} \left\{ \tau: r_1t + b_1 \leq \left[ R(t + \tau) - r_2(t+\tau)-b'_2\right]^+ \right\} \\
&= \left[ \frac{r_1t + b_1 + b'_2}{R-r_2}-t\right]^+ \leq \frac{b_1 + b'_2}{R-r_2}.
\end{aligned}
\eeq
When $r_2 < r'_2$, we can rewrite \Eqref{ac_2} as
\beq
\beta_1(t) = \left[ Rt - \min \left\{\gamma_{r,b}(t), \gamma_{r',b'}(t) \right\} \right]^+ = \left\{
\begin{array}{ll}
0 &\text{when } t = 0 \\
\left[(R - r'_2)t - b'_2\right]^+ &\text{when } t < \frac{b_2 - b'_2}{r'_2 - r_2}\\
\left[(R - r_2) t - b_2\right]^+  &\text{otherwise.
}  
\end{array}
\right.
\eeq
\begin{itemize}
\item When $\beta_1(\frac{b_2 - b'_2}{r'_2 - r_2}) \leq b_1 + r_1 t$, \ie $t \geq \frac{(R-r'_2)(b_2 - b'_2)}{r_1(r'_2 - r_2)} - \frac{b_1 + b'_2}{r_1}$ we have 
\beq\label{1st-half}
\begin{aligned}
D(t) &= \inf_{\tau \geq 0} \left\{ \tau: r_1t + b_1 \leq \left[ (R-r_2)(t + \tau) -b_2\right]^+ \right\} \\ 
&= \left[ \frac{b_1 + b_2}{R-r_2} - \frac{t(R-r_1-r_2)}{R-r_2}\right]^+ \\
& \leq \left[ \frac{b_1 + b_2}{R-r_2} - \frac{R-r_1-r_2}{R-r_2} \left[ \frac{(R-r'_2)(b_2 - b'_2)}{r_1(r'_2 - r_2)} - \frac{b_1 + b'_2}{r_1} \right]^+ \right]^+ \\
& =  \left\{
\begin{array}{ll}
\frac{b_1 + b_2}{R-r_2} &\text{when }\frac{(R-r'_2)(b_2 - b'_2)}{r_1(r'_2 - r_2)} - \frac{b_1 + b'_2}{r_1} < 0 \\
\left[ \frac{b_1 + b_2}{r_1} - \frac{(b_2 - b'_2)(R-r_1 - r_2)}{r_1 (r'_2 - r_2)}\right]^+ &\text{otherwise.}
\end{array}
\right.
\end{aligned}
\eeq

Note that when $\frac{(R-r'_2)(b_2 - b'_2)}{r_1(r'_2 - r_2)} - \frac{b_1 + b'_2}{r_1} < 0$, we have $t \geq \frac{(R-r'_2)(b_2 - b'_2)}{r_1(r'_2 - r_2)} - \frac{b_1 + b'_2}{r_1}, \forall t$. Therefore, $d^*_1 = \frac{b_1 + b_2}{R-r_2}$. Next we consider the case when $\frac{(R-r'_2)(b_2 - b'_2)}{r_1(r'_2 - r_2)} - \frac{b_1 + b'_2}{r_1} \geq 0$.

\item When $\beta_1(\frac{b_2 - b'_2}{r'_2 - r_2}) > b_1 + r_1 t$, \ie $t < \frac{(R-r'_2)(b_2 - b'_2)}{r_1(r'_2 - r_2)} - \frac{b_1 + b'_2}{r_1}$. As when $\frac{(R-r'_2)(b_2 - b'_2)}{r_1(r'_2 - r_2)} - \frac{b_1 + b'_2}{r_1} > 0$ implies $R-r'_2 > 0$, we have
\beq
\begin{aligned}
D(t) &= \inf_{\tau \geq 0} \left\{ \tau: r_1t + b_1 \leq \left[ (R-r'_2)(t + \tau) -b'_2\right]^+ \right\}\\
&= \left[ \frac{b_1 + b'_2}{R-r'_2}+\frac{t(r_1 + r'_2 - R)}{R-r'_2}\right]^+, 
\end{aligned}
\eeq
As $\left[ \frac{b_1 + b'_2}{R-r'_2}+\frac{t(r_1 + r'_2 - R)}{R-r'_2}\right]^+$ is a linear function with $t$, we know $D(t)$ achieves its maximum at either $t = 0$ or $t = \frac{(R-r'_2)(b_2 - b'_2)}{r_1(r'_2 - r_2)} - \frac{b_1 + b'_2}{r_1}$, which gives $d^*_1 = \max\left\{\frac{b_1 + b'_2}{R-r'_2}, \frac{b_1 + b_2}{r_1} - \frac{(R- r_1 - r_2)(b_2 - b'_2)}{r_1(r'_2 - r_2)}\right\}$.
Combine it with \Eqref{1st-half}, when $r'_2 > r_2$ we have:
\beq
D^*_1 = \left\{
\begin{array}{ll}
\frac{b_1 + b_2}{R-r_2}, &\text{when } \frac{(R-r'_2)(b_2 - b'_2)}{r_1(r'_2 - r_2)} - \frac{b_1 + b'_2}{r_1} < 0 \\
\\
\max\left\{\frac{b_1 + b'_2}{R-r'_2}, \frac{b_1 + b_2}{r_1} - \frac{(R- r_1 - r_2)(b_2 - b'_2)}{r_1(r'_2 - r_2)}\right\}, &\text{otherwise}
\end{array}
\right.
\eeq
\end{itemize}
\end{proof}

Note that when $r_2 < r'_2$ and $\frac{(R-r'_2)(b_2 - b'_2)}{r_1(r'_2 - r_2)} - \frac{b_1 + b'_2}{r_1} < 0$ (case~$2$ of
Proposition~\ref{delay-l}), $\tilde{R}^{(2)*}_R$ ensures $d_1 \geq d_1^* = \frac{b_1 + b_2}{R^*-r_2}$, \ie $R^* \geq \frac{b_1 + b_2}{d_1} + r_2$. 
Combined with Corollary~\ref{improve}, we then know that in this case $\tilde{R}^{(2)*}_R$ is no smaller than $\tilde{R}^{(2)*}$, \ie the optimal system needs no reprofiling. Therefore, we only need to
focus on cases~$1$ and~$3$
when seeking to characterize $\tilde{R}^{(2)*}_R$ in the presence of reprofilers.

Next, we establish that these two cases can be combined.  Specifically, note that $\frac{(R-r'_2)(b_2 - b'_2)}{r_1(r'_2 - r_2)} - \frac{b_1 + b'_2}{r_1}\geq 0$ implies $R-r'_2 > 0$.  This means that as $r'_2 \to r_2^+$, $\frac{(R-r'_2)(b_2 - b'_2)}{r_1(r'_2 - r_2)} - \frac{b_1 + b'_2}{r_1} \to +\infty >0$, so that we are always in case~$3$ as $r'_2 \to r_2^+$.  Furthermore, 
$\lim_{r'_2 \to r_2^+}\max\left\{\frac{b_1 + b'_2}{R-r'_2}, \frac{b_1 + b_2}{r_1} - \frac{(R- r_1 - r_2)(b_2 - b'_2)}{r_1(r'_2 - r_2)}\right\} = \max\left\{\frac{b_1 + b'_2}{R-r'_2}, -\infty \right\} = \frac{b_1 + b'_2}{R-r_2}$, or in other words the value of $d_1^*$ of case~$3$ is the same as that of case~$1$ as $r'_2 \to r_2^+$.  This therefore allows us to write that 
when $\frac{(R-r'_2)(b_2 - b'_2)}{r'_2 - r_2} - (b_1 + b'_2) \geq 0$, $d^*_1 = \max\left\{\frac{b_1 + b'_2}{R-r'_2}, \frac{b_1 + b_2}{r_1} - \frac{(R- r_1 - r_2)(b_2 - b'_2)}{r_1(r'_2 - r_2)}\right\}$.

Together with Proposition~\ref{delay-h}, this yields the following optimization for $\tilde{R}^{(2)*}_R$:
\beq\label{optimization}
\begin{aligned}
&\text{\bf OPT\_2} && \min_{r'_2,b'_2}%% {\text{minimize}}_{\substack{r'_2, b'_2}} 
R\\
& \text{subject to}
& & \max\left\{\frac{b_1 + b'_2}{R-r'_2}, \frac{b_1 + b_2}{r_1} - \frac{(R- r_1 - r_2)(b_2 - b'_2)}{r_1(r'_2 - r_2)} \right\} \leq d_1, \\
&&&  \max\left\{\frac{b_2 + l_1}{R}, \frac{b_2 - b'_2 }{r'_2} + \frac{l_1 + l_2}{R} \right\} \leq d_2, \\
&&& \frac{(R-r'_2)(b_2 - b'_2)}{r_1(r'_2 - r_2)} - \frac{b_1 + b'_2}{r_1} \geq 0, \\
&&& r_1 + r_2 \leq R, \qquad r_2 \leq r'_2, \qquad l_2 \leq b'_2 \leq b_2 \\
\end{aligned}
\eeq
Solving {\bf OPT\_2} gives the following combination of five cases, four of which yield values $\tilde{R}^{(2)*}_R<\tilde{R}^{(2)*}$, \ie the introduction of reprofilers helps reduce the link bandwidth required to meet the flows delay targets, where $r'^*_2$ and $b'^*_2$ defines the optimal reprofiler:
\begin{enumerate}[nosep,label=(\roman*)]
\item $\tilde{R}^{(2)*}_R = r_1 + r_2 < \tilde{R}^{(2)*}$, $r'^*_2 = r_2$, and $b'^*_2$ can be any values inside $[b_2 + r_2 \left(\frac{l_1 + l_2}{r_1 + r_2} - d_2 \right), d_1 r_1 - b_1] \cap [l_2, b_2)$, when $ d_1 \in [\frac{l_2 + b_1}{r_1}, \frac{b_1 + b_2}{r_1}) \text{ and }
d_2 \geq \max \left\{\frac{b_2 + l_1}{r_1 + r_2}, \frac{b_1 + b_2 - d_1 r_1}{r_2} + \frac{l_1 + l_2}{r_1 + r_2} \right\};$

\item $\tilde{R}^{(2)*}_R = \frac{b_2 + l_1}{d_2} < \tilde{R}^{(2)*}$,  $r'^*_2$ can be any values inside $[r_2, \min\left\{\tilde{R}^{(2)*}_R - r_1, \tilde{R}^{(2)*}_R - \frac{b_1 + l_2}{d_1 - (b_2 - l_2)/ \tilde{R}^{(2)*}_R}\right\}],$ and $b'^*_2$ can be any values inside $[b_2 - \frac{r'^*_2(b_2 - l_2)}{\tilde{R}^{(2)*}_R }, d_1 (\tilde{R}^{(2)*}_R  - r'^*_2) - b_1] \cap [l_2, b_2)$, when $d_2 < \frac{b_2 + l_1}{r_1 + r_2} \text{ and } d_1 \in [\frac{d_2(b_1 + l_2)}{b_2 + l_1 - d_2 r_2} + \frac{d_2(b_2 - l_2)}{b_2 + l_1}, \frac{d_2(b_1 + b_2)}{b_2 + l_1 - d_2 r_2}).$

\item $\tilde{R}^{(2)*}_R = \frac{l_2 + b_1}{d_1}+r_2< \tilde{R}^{(2)*}$, $r'^*_2 = r_2$, and $b'_2 = l_2$, when $d_1 < \min \left\{\frac{b_1 + l_2}{r_1}, \frac{(b_1 + l_2)\left( d_2 - \frac{b_2 - l_2}{r_2}\right)}{l_1 + b_2 - r_2 d_2} \right\}$ and $d_2 < \frac{l_1 + b_2}{r_2}$; 
and when $d_2 \geq \frac{l_1 + b_2}{r_2}$ and $d_1 < \frac{b_1 + l_2}{r_1}$.

\item $\tilde{R}^{(2)*}_R = \frac{(d_1 - d_2)r_2 + (b_1 + b_2) + \sqrt{((d_1 - d_2)r_2 + b_1 + b_2)^2 + 4 d_1 r_2 (l_1 + l_2)}}{2 d_1}< R^*_0$, $r'^*_2 = r_2$, and \newline $b'_2 = \frac{b_2 - b_1 - (d_1 + d_2)r_2 + \sqrt{((d_1 - d_2)r_2 + b_1 + b_2)^2 + 4 d_1 r_2 (l_1 + l_2)}}{2}$
\subitem $\circ$ when $d_2 < \frac{b_2 + l_1}{r_1 + r_2}, $ and $d_1 \in [\frac{(b_1 + l_2)\left( d_2 - \frac{b_2 - l_2}{r_2} \right)}{b_2 + l_1 - r_2 d_2},\frac{d_2(b_1 + l_2)}{b_2 + l_1 - d_2 r_2} + \frac{d_2(b_2 - l_2)}{b_2 + l_1});$ and
\subitem $\circ$ when $\frac{b_2 + l_1}{r_1 + r_2} \leq d_2 \leq \frac{b_2 + l_1}{r_2}$, and $d_1 \in [\frac{(b_1 + l_2)\left( d_2 - \frac{b_2 - l_2}{r_2} \right)}{b_2 + l_1 - r_2 d_2}, \frac{r_2 (l_1 + l_2)}{r_1 (r_1 + r_2)} + \frac{b_1 + b_2 - d_2 r_2}{r_1}).$

\item otherwise, $\tilde{R}^{(2)*}_R = \tilde{R}^{(2)*}$.
\end{enumerate}

From the solution of \textbf{OPT\_2}, we directly get
\begin{corollary}
\label{prop:onerate}
We can achieve the optimality of \textbf{OPT\_2} through setting $r'_2 = r_2$.
\end{corollary}

\subsubsection{Solving {\bf OPT\_2}}

We can divide the optimization into two sub-optimizations:

\begin{flushleft}
\textbf{ Sub-optimization~$1$:}
 \end{flushleft} 
\beq\label{op1}
\begin{aligned}
&{\text{minimize}}_{\substack{r'_2, b'_2}} 
& & R\\
& \text{subject to}
& & r_1 + r'_2 -R \leq 0, \qquad  \frac{b_1 + b'_2}{R-r'_2} - d_1 \leq 0, \\
&&& \frac{b_2 + l_1}{R} - d_2 \leq 0, \qquad \frac{b_2 - b'_2 + l_2}{r'_2} + \frac{l_1}{R} - d_2 \leq 0, \\
&&& \frac{b_1 + b'_2}{r_1} - \frac{(R-r'_2)(b_2 - b'_2)}{r_1(r'_2 - r_2)} \leq 0, \\
&&& r_1 + r_2 -R \leq 0, \qquad r_2 -r'_2 \leq 0, \\
&&& l_2-b'_2 \leq 0,  \qquad  b'_2 - b_2 \leq 0  
\end{aligned}
\eeq

\begin{flushleft}
\textbf{Sub-optimization~$2$:} 
\end{flushleft}
\beq\label{op2}
\begin{aligned}
&{\text{minimize}}_{\substack{r'_2, b'_2}} 
& & R\\
& \text{subject to}
& & R - r_1 - r'_2 < 0, \qquad \frac{b_1 + b_2}{r_1} - \frac{(R- r_1 - r_2)(b_2 - b'_2)}{r_1(r'_2 - r_2)} -d_1 \leq 0,\\
&&& \frac{b_2 + l_1}{R} - d_2 \leq 0, \qquad \frac{b_2 - b'_2 + l_2}{r'_2} + \frac{l_1}{R} - d_2 \leq 0, \\
&&& \frac{b_1 + b'_2}{r_1} - \frac{(R-r'_2)(b_2 - b'_2)}{r_1(r'_2 - r_2)} \leq 0, \\
&&& r_1 + r_2 -R \leq 0, \qquad r_2 -r'_2 \leq 0, \\
&&& l_2-b'_2 \leq 0,  \qquad  b'_2 - b_2 \leq 0  
\end{aligned}
\eeq

Denote the solution of sub-optimizations~$1$ and~$2$ as $R^*_1$ and $R^*_2$, respectively. Then we have $R^* = \min \{R^*_1, R^*_2\}$. Next, we solve sub-optimizations~$1$ and~$2$. Note than when $R = r_1 + r'_2$, the two sub-optimizations are the same. Therefore, when solving sub-optimization~$2$, we only consider the case where $R - r_1 - r'_2 < 0$.  Then we have:

\begin{lemma}\label{opt1-solution}
The solution for Sub-optimiaztion~$1$ is
\begin{itemize}
\item $R^*_1 = r_1 + r_2, \text{ when } d_1 \in [\frac{l_2 + b_1}{r_1}, \frac{b_1 + b_2}{r_1}) \text{ and }
d_2 \geq \max \left\{\frac{b_2 + l_1}{r_1 + r_2}, \frac{b_1 + b_2 - d_1 r_1}{r_2} + \frac{l_1 + l_2}{r_1 + r_2} \right\};$
\item $R^*_1 = \frac{b_2 + l_1}{d_2}, \text{ when } d_2 < \frac{b_2 + l_1}{r_1 + r_2} \text{ and } d_1 \in [\frac{d_2(b_1 + l_2)}{b_2 + l_1 - d_2 r_2} + \frac{d_2(b_2 - l_2)}{b_2 + l_1}, \frac{d_2(b_1 + b_2)}{b_2 + l_1 - d_2 r_2}); $
\item $R^*_1 = \frac{l_2 + b_1}{d_1}+r_2$, when $d_2 < \frac{l_1 + b_2}{r_2}$ and $d_1 < \min \left\{\frac{b_1 + l_2}{r_1}, \frac{(b_1 + l_2)\left( d_2 - \frac{b_2 - l_2}{r_2}\right)}{l_1 + b_2 - r_2 d_2} \right\}$; and when $d_2 \geq \frac{l_1 + b_2}{r_2}$ and $d_1 < \frac{b_1 + l_2}{r_1}$;
\item $R^*_1 = \frac{(d_1 - d_2)r_2 + (b_1 + b_2) + \sqrt{((d_1 - d_2)r_2 + b_1 + b_2)^2 + 4 d_1 r_2 (l_1 + l_2)}}{2 d_1}$, 
\subitem $\circ$ when $d_2 < \frac{b_2 + l_1}{r_1 + r_2}, $ and $d_1 \in [\frac{(b_1 + l_2)\left( d_2 - \frac{b_2 - l_2}{r_2} \right)}{b_2 + l_1 - r_2 d_2},\frac{d_2(b_1 + l_2)}{b_2 + l_1 - d_2 r_2} + \frac{d_2(b_2 - l_2)}{b_2 + l_1});$ and
\subitem $\circ$ when $\frac{b_2 + l_1}{r_1 + r_2} \leq d_2 \leq \frac{b_2 + l_1}{r_2}$, and $d_1 \in [\frac{(b_1 + l_2)\left( d_2 - \frac{b_2 - l_2}{r_2} \right)}{b_2 + l_1 - r_2 d_2}, \frac{r_2 (l_1 + l_2)}{r_1 (r_1 + r_2)} + \frac{b_1 + b_2 - d_2 r_2}{r_1}).$
\end{itemize}
\end{lemma}

\begin{lemma}\label{opt2-solution}
The solution for Sub-optimization~$2$ is
\begin{itemize}
\item when $d_2 < \frac{b_2 + l_1}{r_1 + r_2}$, and $d_1 \in [\frac{d_2(b_2 - l_2)}{b_2 + l_1} + \frac{b_1 + l_2}{r_1}, \frac{d_2(b_1 + b_2)}{b_2 + l_1 - d_2 r_2})$, $R^*_2 = \frac{b_2 + l_1}{d_2} < R^*_0 $;
\item otherwise, $R^*_2 = R^*_0$.
\end{itemize}
\end{lemma}

Basic algebraic manipulation gives that when $d_2 < \frac{b_2 + l_1}{r_1 + r_2}$, $\frac{d_2(b_2 - l_2)}{b_2 + l_1} + \frac{b_1 + l_2}{r_1} \geq \frac{d_2(b_1 + l_2)}{b_2 + l_1 - d_2 r_2} + \frac{d_2(b_2 - l_2)}{b_2 + l_1}$. Therefore, we have $R^*_2 \geq R^*_1$. 

%%%%%%%%%%%%%%%%%%%%%%%%%%%%%%%%%%%
%%%%%%%% sub-optimization 1 %%%%%%%%
%%%%%%%%%%%%%%%%%%%%%%%%%%%%%%%%%%%
\subsubsection*{Solution for Sub-optimization~$\pmb{1}$}
The Lagrangian function for sub-optimization~$1$ is 
\beq
\begin{aligned}
L_1(R, r'_2, b'_2, \pmb{\lambda}) &= R + \lambda_1 (r_1 + r'_2 -R) + \lambda_2 \left(\frac{b_1 + b'_2}{R-r'_2} - d_1\right) + \lambda_3 \left(\frac{b_2 + l_1}{R} - d_2\right) \\
&+ \lambda_4 \left(\frac{b_2 - b'_2}{r'_2} + \frac{l_1 + l_2}{R} - d_2 \right) + \lambda_5 \left(\frac{b_1 + b'_2}{r_1} - \frac{(R-r'_2)(b_2 - b'_2)}{r_1(r'_2 - r_2)} \right)\\
& + \lambda_6 (r_1 + r_2 -R) + \lambda_7 (r_2 -r'_2) + \lambda_8 (l_2-b'_2) + \lambda_9 (b'_2 - b_2)
\end{aligned}
\eeq 

\beq\label{dev1}
\nabla_{R,r'_2, b'_2}L_1 = \begin{bmatrix} 
1-\lambda_1 - \frac{\lambda_2(b_1 + b'_2)}{(R-r'_2)^2} - \frac{\lambda_3(b_2 + l_1)}{R^2} - \frac{\lambda_4 (l_1+l_2)}{R^2} - \frac{\lambda_5(b_2 - b'_2)}{r_1 (r'_2 - r_2)} - \lambda_6 \\
\\
\lambda_1 + \frac{\lambda_2(b_1 + b'_2)}{(R-r'_2)^2} - \frac{\lambda_4(b_2 - b'_2 )}{r^{'2}_2} + \frac{\lambda_5(b_2 - b'_2)(R-r_2)}{r_1(r'_2 - r_2)^2} - \lambda_7 \\
\\
\frac{\lambda_2}{R-r'_2} - \frac{\lambda_4}{r'_2} + \lambda_5\left(\frac{1}{r_1} + \frac{R-r'_2}{r_1(r'_2 - r_2)} \right) - \lambda_8 + \lambda_9
\end{bmatrix}
\eeq

\beq\label{diag1}
\text{diag}(\triangle_{R,r'_2, b'_2}L_1) = \begin{bmatrix}
\frac{2\lambda_2(b_1 + b'_2)}{(R-r'_2)^3} + \frac{2\lambda_3(b_2 + l_1)}{R^3} + \frac{2 \lambda_4(l_1 + l_2)}{R^3} \\
\\
\frac{2 \lambda_2(b_1 + b'_2)}{(R-r'_2)^3} + \frac{2 \lambda_4(b_2 - b'_2)}{r^{'3}_2} - \frac{2\lambda_5(b_2 - b'_2)(R - r_2)}{r_1 (r'_2 - r_2)^3} \\
\\
0
\end{bmatrix}
\eeq

From \Eqref{1st-half}, we know that when $\frac{b_1 + b'_2}{r_1} - \frac{(R-r'_2)(b_2 - b'_2)}{r_1(r'_2 - r_2)} = 0$, $d_1^* = \frac{b_1 + b_2}{R-r_2}$. Combine it with Corollary~\ref{improve}, we have $R^*_1 = R^*_0$. 

Next we consider the case where $\frac{b_1 + b'_2}{r_1} - \frac{(R-r'_2)(b_2 - b'_2)}{r_1(r'_2 - r_2)} > 0$. Then from KKT conditions' complementary slackness requirement, we have $\lambda_5 = 0$. Substitute $\lambda_5 = 0$ into \Eqref{diag1}, we have
\beq
\text{diag}(\triangle_{R,r'_2, b'_2}L_1) = \begin{bmatrix}
\frac{2\lambda_2(b_1 + b'_2)}{(R-r'_2)^3} + \frac{2\lambda_3(b_2 + l_1)}{R^3} + \frac{2 \lambda_4 (l_1+l_2)}{R^3} \\
\\
\frac{2 \lambda_2(b_1 + b'_2)}{(R-r'_2)^3} + \frac{2 \lambda_4(b_2 - b'_2)}{r^{'3}_2} \\
\\
0
\end{bmatrix}
\geq 0
\eeq 
Therefore, for any $(r'_2, b'_2, \pmb{\lambda})$ satisfying KKT's necessary conditions, it is a local optimum. The necessary conditions for the optimization under $\lambda_5 = 0$ is:
\beq\label{cons}
\begin{aligned}
& 1-\lambda_1 - \frac{\lambda_2(b_1 + b'_2)}{(R-r'_2)^2} - \frac{\lambda_3(b_2 + l_1)}{R^2} - \frac{\lambda_4 (l_1+l_2)}{R^2}  - \lambda_6 = 0,
& & \frac{\lambda_2}{R-r'_2} - \frac{\lambda_4}{r'_2} - \lambda_8 + \lambda_9 = 0,\\
& \lambda_1 + \frac{\lambda_2(b_1 + b'_2)}{(R-r'_2)^2} - \frac{\lambda_4(b_2 - b'_2)}{r^{'2}_2} - \lambda_7 = 0, 
& & \lambda_i \geq 0, \text{ for }i = 1, ...9, \\
& r_1 + r'_2 -R \leq 0, 
& & \lambda_1 (r_1 + r'_2 -R) = 0, \\
& \frac{b_1 + b'_2}{R-r'_2} - d_1 \leq 0, 
& &\lambda_2 \left(\frac{b_1 + b'_2}{R-r'_2} - d_1\right) = 0, \\
& \frac{b_2 + l_1}{R} - d_2 \leq 0, 
& &\lambda_3\left( \frac{b_2 + l_1}{R} - d_2\right) = 0, \\
& \frac{b_2 - b'_2}{r'_2} + \frac{l_1 + l_2}{R} - d_2 \leq 0,
& &\lambda_4 \left(\frac{b_2 - b'_2}{r'_2} + \frac{l_1 + l_2}{R} - d_2  \right) = 0, \\
& \frac{b_1 + b'_2}{r_1} - \frac{(R-r'_2)(b_2 - b'_2)}{r_1(r'_2 - r_2)} < 0, 
& &\lambda_5 =0, \\
& r_1 + r_2 -R \leq 0,
& &\lambda_6 (r_1 + r_2 -R) = 0,\\
& r_2 -r'_2 \leq 0, 
& &\lambda_7 (r_2 -r'_2) = 0, \\
& l_2-b'_2 \leq 0,  
& &\lambda_8 (l_2-b'_2) = 0,\\
& b'_2 - b_2 \leq 0 
& &\lambda_9 (b'_2 - b_2) = 0
\end{aligned}
\eeq
\begin{flushleft}
For the conditions in \Eqref{cons}, we have:
\begin{itemize}
\item $R^*_1 = r_1 + r_2, \text{ when } d_1 \in [\frac{l_2 + b_1}{r_1}, \frac{b_1 + b_2}{r_1}) \text{ and }
d_2 \geq \max \left\{\frac{b_2 + l_1}{r_1 + r_2}, \frac{b_1 + b_2 - d_1 r_1}{r_2} + \frac{l_1 + l_2}{r_1 + r_2} \right\};$
\item $R^*_1 = \frac{b_2 + l_1}{d_2} \geq r_1 + r_2, \text{ when } d_2 < \frac{b_2 + l_1}{r_1 + r_2} \text{ and } d_1 \in [\frac{d_2(b_1 + l_2)}{b_2 + l_1 - d_2 r_2} + \frac{d_2(b_2 - l_2)}{b_2 + l_1}, \frac{d_2(b_1 + b_2)}{b_2 + l_1 - d_2 r_2}); $
\item $R^*_1 = \frac{l_2 + b_1}{d_1}+r_2$, when $d_2 < \frac{l_1 + b_2}{r_2}$ and $d_1 < \min \left\{\frac{b_1 + l_2}{r_1}, \frac{(b_1 + l_2)\left( d_2 - \frac{b_2 - l_2}{r_2}\right)}{l_1 + b_2 - r_2 d_2} \right\}$; and when $d_2 \geq \frac{l_1 + b_2}{r_2}$ and $d_1 < \frac{b_1 + l_2}{r_1}$;
\item $R^*_1 = \frac{(d_1 - d_2)r_2 + (b_1 + b_2) + \sqrt{((d_1 - d_2)r_2 + b_1 + b_2)^2 + 4 d_1 r_2 (l_1 + l_2)}}{2 d_1}$, 
\subitem $\circ$ when $d_2 < \frac{b_2 + l_1}{r_1 + r_2}, $ and $d_1 \in [\frac{(b_1 + l_2)\left( d_2 - \frac{b_2 - l_2}{r_2} \right)}{b_2 + l_1 - r_2 d_2},\frac{d_2(b_1 + l_2)}{b_2 + l_1 - d_2 r_2} + \frac{d_2(b_2 - l_2)}{b_2 + l_1});$ and
\subitem $\circ$ when $\frac{b_2 + l_1}{r_1 + r_2} \leq d_2 \leq \frac{b_2 + l_1}{r_2}$, and $d_1 \in [\frac{(b_1 + l_2)\left( d_2 - \frac{b_2 - l_2}{r_2} \right)}{b_2 + l_1 - r_2 d_2}, \frac{r_2 (l_1 + l_2)}{r_1 (r_1 + r_2)} + \frac{b_1 + b_2 - d_2 r_2}{r_1}).$
\end{itemize}
\end{flushleft}
\begin{proof}
Note that when $b'_2 = b_2$, the reprofiler has no effect, \ie $R^* = R^*_0$. Therefore, we consider only $\lambda_9 = 0$ and $b'_2 < b_2$. Then from $\frac{\lambda_2}{R-r'_2} - \frac{\lambda_4}{r'_2} - \lambda_8 + \lambda_9 = 0$ we have 1) if $\lambda_2 = 0$, then $\lambda_4 = \lambda_8 = 0$; and 2) if $\lambda_2 > 0$, then $\lambda_4 + \lambda_8 > 0$.
\begin{itemize}
\item When $\lambda_2 = 0$, $\lambda_4 = \lambda_8 = 0$, from $\lambda_1 + \frac{\lambda_2(b_1 + b'_2)}{(R-r'_2)^2} - \frac{\lambda_4(b_2 - b'_2)}{r^{'2}_2} - \lambda_7 = 0$, we have $\lambda_1 = \lambda_7$.

$\bullet$ When $\lambda_1 = \lambda_7 > 0$, it has $R = r_1 + r'_2$ and $r_2 = r'_2$. Therefore, $R = r_1 + r_2$. Then the constraints become:
\beq\label{cons1-0}
\left\{
\begin{array}{ll}
\frac{b_1 + b'_2}{r_1} - d_1 \leq 0 \implies b'_2 \leq d_1 r_1 - b_1,  \\
\\
\frac{b_2 + l_1}{r_1 + r_2} - d_2 \leq 0 \implies d_2 \geq \frac{b_2 + l_1}{r_1 + r_2}, \\
\\
\frac{b_2 - b'_2}{r'_2} + \frac{l_1+l_2}{r_1 + r_2} - d_2 \leq 0, \implies b'_2 \geq b_2 + r_2 \left( \frac{l_1 + l_2}{r_1 + r_2} - d_2 \right) \\
\\
l_2 \leq b'_2 < b_2
\end{array}
\right.
\eeq

Hence we have $b'_2 \in [b_2 + r_2 \left( \frac{l_1 + l_2}{r_1 + r_2} - d_2 \right), d_1 r_1 - b_1] \cap [l_2, b_2)$. As $d_2 \geq \frac{b_2 + l_1}{r_1 + r_2} > \frac{l_1 + l_2}{r_1 + r_2}$, we have $b_2 - r_2 \left( \frac{l_1 + l_2}{r_1 + r_2} - d_2 \right) < b_2$. Therefore, to guarantee $[b_2 + r_2 \left( \frac{l_1 + l_2}{r_1 + r_2} - d_2 \right), d_1 r_1 - b_1] \cap [l_2, b_2) \neq \emptyset$:
\beq
\left\{
\begin{array}{ll}
d_1 r_1 - b_1 \geq l_2 \implies d_1 \geq \frac{l_2 + b_1}{r_1} \\
\\
b_2 + r_2 \left( \frac{l_1 + l_2}{r_1 + r_2} - d_2 \right) \leq d_1 r_1 - b_1
\implies d_2 \geq \frac{b_1 + b_2 - d_1 r_1}{r_2} + \frac{l_1 + l_2}{r_1 + r_2}
\end{array}
\right.
\eeq

Remember that adding a reprofiler is beneficial only when $R^*_0 = \frac{b_1 + b_2}{d_1} + r_2 > \max \left\{r_1 + r_2, \frac{b_2 + l_1}{d_2} \right\}$, \ie $d_1 < \frac{b_1 + b_2}{r_1}$ and $d_2 > \frac{b_2 + l_1}{\frac{b_1 + b_2}{d_1} + r_2}$, which gives $\frac{b_2 + l_1}{r_1 + r_2} >  \frac{b_2 + l_1}{\frac{b_1 + b_2}{d_1}}$.
Therefore, we have
\beq
\circ \ d_1 \in [\frac{l_2 + b_1}{r_1}, \frac{b_1 + b_2}{r_1}) \text{ and } d_2 \geq \max \left\{\frac{b_2 + l_1}{r_1 + r_2}, \ \frac{b_1 + b_2 - d_1 r_1}{r_2} + \frac{l_1 + l_2}{r_1 + r_2} \right\}.
\eeq

%%%%%%%%%%%%%%%%%%%%%%%%%%%%%%%
$\bullet$ When $\lambda_1 = \lambda_7= 0$, from $1-\lambda_1 - \frac{\lambda_2(b_1 + b'_2)}{(R-r'_2)^2} - \frac{\lambda_3(b_2 + l_1)}{R^2} - \frac{\lambda_4 (l_1+l_2)}{R^2}  - \lambda_6 = 0$ we have $\lambda_3 + \lambda_6 > 0$. If $\lambda_6 > 0$, it has $r_1 + r_2 -R = 0$. As $r_1 + r'_2 -R \leq 0$, it has $r'_2 = r_2$. Therefore, it produces the same optimization as \Eqref{cons1-0}. Therefore, we only consider $\lambda_6 = 0$ in this case. When $\lambda_6 = 0$, $\lambda_3 > 0$, then we have $R = \frac{b_2 + l_1}{d_2}$. Note that $R = \frac{b_2 + l_1}{d_2}$ implies $\frac{b_2 + l_1}{R} \geq \frac{b_2 - b'_2}{r'_2} + \frac{l_1 + l_2}{R}$, \ie $b'_2 \geq b_2 - \frac{r'_2 (b_2 - l_2)}{R}$. Then the constraints become
\beq\label{cons1-3}
\left\{
\begin{array}{ll}
r'_2 \in [r_2, R-r_1], \qquad b'_2 \in [l_2, b_2), \qquad r_1 + r_2 \leq R,  \\
\\
\frac{b_1 + b'_2}{R-r'_2} \leq d_1 \implies b'_2 \leq d_1 (R-r'_2) - b_1,  \\
\\
b'_2 \geq b_2 - \frac{r'_2 (b_2 - l_2)}{R}, \\
\\
b_1 + b'_2 - \frac{(R-r'_2)(b_2 - b'_2)}{r'_2 - r_2} < 0 \implies b'_2 < b_2 - \frac{(b_1 + b_2)(r'_2 - r_2)}{R-r_2}. \\
\end{array}
\right.
\eeq  
Note that $b_2 - \frac{(b_1 + b_2)(r'_2 - r_2)}{R-r_2} - d_1 (R-r'_2) + b_1 = (R-r'_2)\left(\frac{b_1 + b_2}{R-r_2} - d_1\right) = \frac{d_1 (R- r'_2)(R^*_0 - R)}{R-r_2}$. As $r'_2 < R$, under $R < R^*_0$, it has $ b_2 - \frac{(b_1 + b_2)(r'_2 - r_2)}{R-r_2} > d_1 (R-r'_2) - b_1$. Hence, we have $b'_2 \in [b_2 - \frac{r'_2 (b_2 - l_2)}{R}, d_1 (R-r'_2) - b_1]$. Next we configure the conditions where $\exists r'_2 \in [r_2, R-r_1]$, such that $[b_2 - \frac{r'_2 (b_2 - l_2)}{R}, d_1 (R-r'_2) - b_1] \cap [l_2, b_2) \neq \emptyset$. 

For $[b_2 - \frac{r'_2 (b_2 - l_2)}{R}, d_1 (R-r'_2) - b_1] \cap [l_2, b_2) \neq \emptyset$, it requires
\beq\label{opt1:case1-2}
\left\{
\begin{array}{ll}
b_2 - \frac{r'_2 (b_2 - l_2)}{R} < b_2 \implies b_2 > l_2, \\
\\
d_1 (R-r'_2) - b_1 \geq l_2 \implies r'_2 \leq R - \frac{b_1 + l_2}{d_1} , \\
\\
b_2 - \frac{r'_2 (b_2 - l_2)}{R} - d_1 (R-r'_2) + b_1 \leq 0 \implies r'_2 \leq R - \frac{b_1 + l_2}{d_1 - (b_2-l_2)/R}
\end{array}
\right.
\eeq
Basic algebraic manipulation gives $R - \frac{b_1 + l_2}{d_1} > R - \frac{b_1 + l_2}{d_1 -(b_2-l_2)/R}$. Hence, \Eqref{opt1:case1-2} gives $r'_2 \leq R - \frac{b_1 + l_2}{d_1 - (b_2-l_2)/R}$. Combine it with $r'_2 \in [r_2, R-r_1]$ and $R = \frac{b_2 + l_1}{d_2}$, we have
\beq
r_2 \leq R - \frac{b_1 + l_2}{d_1 - (b_2-l_2)/R} \implies d_1 \geq \frac{d_2(b_1 + l_2)}{b_2 + l_1 - d_2r_2} + \frac{(b_2-l_2)d_2}{b_2 + l_1},  
\eeq

Also, from $r_1 + r_2 < R < \frac{b_1 + b_2}{d_1} + r_2$, we have $d_2 < \frac{b_2 + l_1}{r_1 + r_2}$ and $d_1 < \frac{d_2(b_1 + b_2)}{b_2 + l_1 - d_2r_2}$. Hence, we have 
\beq
\circ \ d_2 < \frac{b_2 + l_1}{r_1 + r_2}, \text{ and } d_1 \in [\frac{d_2(b_1 + l_2)}{b_2 + l_1 - d_2r_2} + \frac{(b_2-l_2)d_2}{b_2 + l_1}, \frac{d_2(b_1 + b_2)}{b_2 + l_1 - d_2r_2}).
\eeq
Basic algebraic manipulation shows that the interval is always valid. 

\item When $\lambda_2 > 0$, as $\lambda_9 = 0$, from $\frac{\lambda_2}{R-r'_2}  - \frac{\lambda_4}{r'_2} - \lambda_8 + \lambda_9 = 0$ we have $\frac{\lambda_2}{R-r'_2}  = \frac{\lambda_4}{r'_2} + \lambda_8$ and $\lambda_4 + \lambda_8 > 0$. Combining it with $\lambda_1 + \frac{\lambda_2(b_1 + b'_2)}{(R-r'_2)^2} - \frac{\lambda_4(b_2 - b'_2)}{r^{'2}_2} - \lambda_7$ and $\frac{b_1 + b'_2}{R-r'_2} = d_1$. We have $\lambda_8 d_1 + \frac{\lambda_4}{r'_2}\left( d_1 - \frac{b_2 - b'_2}{r'_2}\right) + \lambda_1 - \lambda_7 = 0$. As $\frac{b_2 - b'_2}{r'_2} + \frac{l_1 + l_2}{R} \leq d_2$, $\lambda_8 d_1 + \frac{\lambda_4}{r'_2}\left( d_1 - \frac{b_2 - b'_2}{r'_2}\right) + \lambda_1 - \lambda_7 \geq  d_1 \lambda_8 + \lambda_1 - \lambda_7 + \frac{\lambda_4}{r'_2}\left(d_1 - d_2 + \frac{l_1 + l_2}{R} \right) = 0$. Therefore, given $\lambda_4 + \lambda_8 > 0$, we have $\lambda_7 > 0$, \ie $r'_2 = r_2$. 

$\bullet$ When $\lambda_8 = 0$, $\lambda_4 > 0$, \ie $\frac{b_2 - b'_2}{r_2} + \frac{l_1 + l_2}{R} = d_2$. Then the constraints become
\beq
\left\{
\begin{array}{ll}
r_1 + r_2 \leq R, \qquad b'_2 \in [l_2, b_2), \\
\\
\frac{b_1 + b'_2}{R-r_2} = d_1 \implies R = \frac{b_1 + b'_2}{d_1} + r_2, \\
\\
\frac{b_2 - b'_2}{r_2} + \frac{l_1 + l_2}{R} = d_2,  \\ 
\\
R \geq \frac{b_2 + l_1}{d_2}. 
\end{array}
\right.
\eeq
Substituting $R = \frac{b_1 + b'_2}{d_1} + r_2$ into $\frac{b_2 - b'_2}{r_2} + \frac{l_1 + l_2}{R} = d_2$ gives
$$\frac{d_1}{r_2}R^2 - \left[d_1 - d_2 + \frac{b_1 + b_2}{r_2} \right]R - (l_1+l_2) = 0,$$
which gives
$$R = \frac{(d_1 - d_2)r_2 + (b_1 + b_2) + \sqrt{\left((d_1 - d_2)r_2 + b_1 + b_2 \right)^2 + 4d_1 r_2(l_1 + l_2)}}{2d_1}, $$
and 
$$b'_2 = \frac{(b_2 - b_1)-(d_1 + d_2)r_2 + \sqrt{\left((d_1 - d_2)r_2 + b_1 + b_2 \right)^2 + 4d_1 r_2(l_1 + l_2)}}{2}.$$
$b'_2 \in [l_2, b_2)$ gives
$$d_2 \in (\ \frac{d_1(l_1+l_2)}{b_1 + b_2 + d_1r_2},\  \frac{d_1 (l_1+l_2)}{b_1 + l_2 + d_1r_2} + \frac{b_2-l_2}{r_2}\ ]. $$

Note that when $R^*_0 = \frac{b_1 + b_2}{d_1} + r_2$, we have $d_2 > \frac{(b_2 + l_1)d_1}{b_1 + b_2 + r_2 d_1} \in (\frac{d_1(l_1+l_2)}{b_1 + b_2 + d_1r_2}, \frac{d_1 (l_1+l_2)}{b_1 + l_2 + d_1r_2} + \frac{b_2-l_2}{r_2})$. Hence we have 
$d_2 \in (\ \frac{(b_2 + l_1)d_1}{b_1 + b_2 + r_2 d_1},\  \frac{d_1 (l_1+l_2)}{b_1 + l_2 + d_1r_2} + \frac{b_2-l_2}{r_2}\ ]$, \ie
\beq\label{opt-1-sqrt}
d_2 < \frac{b_2 + l_1}{r_2} \text{, and } d_1 \in [\frac{(b_1 + l_2)\left(d_2 -\frac{b_2 - l_2}{r_2}\right)}{b_2 + l_1 - r_2 d_2}, \frac{d_2 (b_1 + b_2)}{b_2 + l_1 - r_2 d_2}). 
\eeq

From $R \geq r_1 + r_2$ and $R \geq \frac{b_2 + l_1}{d_2}$, we have:
 when $d_2 < \frac{b_2 + l_1}{r_1 + r_2}$, $d_1 \leq \frac{r_2(l_1 + l_2)}{\frac{b_2 + l_1}{d_2} \left( \frac{b_2 + l_1}{d_2} - r_2\right)} + \frac{b_1 + b_2 - d_2 r_2}{\frac{b_2 + l_1}{d_2} - r_2} = \frac{d_2(b_1 + l_2)}{b_2 + l_1 - d_2r_2} + \frac{d_2(b_2 - l_2)}{b_2 + l_1}$, which is greater than $\frac{d_2 (b_1 + b_2)}{b_2 + l_1 - r_2 d_2}$;
and when $d_2 \geq \frac{b_2 + l_1}{r_1 + r_2}$, $d_1 \leq \frac{r_2(l_1 + l_2)}{r_1(r_1 + r_2)} + \frac{b_1 + b_2 - d_2 r_2}{r_1} < \frac{d_2 (b_1 + b_2)}{b_2 + l_1 - r_2 d_2}$. Combining it with \Eqref{opt-1-sqrt} gives: 
\subitem $\circ$ when $d_2 < \frac{l_1 + b_2}{r_2+r_1}$, $d_1 \in [\frac{(b_1 + l_2)\left(d_2 -\frac{b_2 - l_2}{r_2}\right)}{b_2 + l_1 - r_2 d_2},\frac{d_2 (b_1 + b_2)}{b_2 + l_1 - r_2 d_2})$;
\subitem $\circ$ when $\frac{l_1 + b_2}{r_1 + r_2} \leq d_2 \leq \frac{b_2 + l_1}{r_2}$, $d_1 \in [\frac{(b_1 + l_2)\left(d_2 -\frac{b_2 - l_2}{r_2}\right)}{b_2 + l_1 - r_2 d_2},  \frac{r_2(l_1 + l_2)}{r_1(r_1 + r_2)} + \frac{b_1 + b_2 - d_2 r_2}{r_1})$.

$\bullet$ When $\lambda_8 > 0$, \ie $b'_2 = l_2$,  we have $\frac{b_2 - b'_2}{r_2} + \frac{l_1+l_2}{R} = \frac{l_1 + b_2}{R} + \left(\frac{1}{r_2} - \frac{1}{R} \right)(b_2 - l_2) > \frac{b_2 + l_1}{R}$. Then the constraints become
\beq
\left\{
\begin{array}{ll}
\frac{b_1 + l_2}{R-r_2} = d_1 \implies R = \frac{b_1 + l_2}{d_1} + r_2 < \frac{b_2 + b_1}{d_1} + r_2 = R^*_0, \\
\\
R > r_1 + r_2 \implies d_1 < \frac{b_1 + l_2}{r_1}\\
\\
\frac{b_2 - l_2}{r_2} + \frac{l_1 + l_2}{R} \leq d_2 \implies d_2 \geq \frac{b_2-l_2}{r_2} + \frac{(l_1+l_2) d_1}{b_1 + l_2 + r_2 d_1} > \frac{(b_2 + l_1)d_1}{b_1 + b_2+r_2d_1s} \\
\end{array}
\right.
\eeq
Hence, we have $d_1 < \frac{b_1 + l_2}{r_1}$, and $d_2 \geq \frac{b_2-l_2}{r_2} + \frac{(l_1+l_2) d_1}{b_1 + l_2 + r_2 d_1}$. Basic algebraic manipulation gives that it is equivalent to:
\subitem $\circ$ when $d_2 \geq \frac{l_1 + b_2}{r_2}$, $d_1 < \frac{b_1 + l_2}{r_1}$;
\subitem $\circ$ when $d_2 < \frac{l_1 + b_2}{r_2}$, $d_1 \leq \min\left\{\frac{b_1 + l_2}{r_1}, \frac{(b_1 + l_2)\left(d_2 - \frac{b_2 - l_2}{r_2} \right)}{l_1 + b_2 - r_2 d_2} \right\}$.
\end{itemize}

In summary, the local optimums are:
\begin{itemize}
\item $R = r_1 + r_2, \text{ when } d_1 \in [\frac{l_2 + b_1}{r_1}, \frac{b_1 + b_2}{r_1}) \text{ and }
d_2 \in [\max \left\{\frac{b_2 + l_1}{r_1 + r_2}, \frac{b_1 + b_2 - d_1 r_1}{r_2} + \frac{l_1 + l_2}{r_1 + r_2} \right\}, d_1);$
\item $R = \frac{b_2 + l_1}{d_2} \geq r_1 + r_2, \text{ when } d_2 < \frac{b_2 + l_1}{r_1 + r_2} \text{ and } d_1 \in [\frac{d_2(b_1 + l_2)}{b_2 + l_1 - d_2 r_2} + \frac{d_2(b_2 - l_2)}{b_2 + l_1}, \frac{d_2(b_1 + b_2)}{b_2 + l_1 - d_2 r_2}); $
\item $R = \frac{(d_1 - d_2)r_2 + (b_1 + b_2) + \sqrt{((d_1 - d_2)r_2 + b_1 + b_2)^2 + 4 d_1 r_2 (l_1 + l_2)}}{2 d_1} \geq \max \left\{r_1 + r_2, \frac{b_2 + l_1}{d_2}, \frac{b_1 + l_2}{d_1} + r_2 \right\}$, 
\subitem $\circ$ when $d_2 < \frac{b_2 + l_1}{r_1 + r_2}, $ and $d_1 \in [\frac{(b_1 + l_2)\left( d_2 - \frac{b_2 - l_2}{r_2} \right)}{b_2 + l_1 - r_2 d_2}, \frac{d_2(b_1 + b_2)}{b_2 + l_1 - r_2 d_2});$
\subitem $\circ$ when $\frac{b_2 + l_1}{r_1 + r_2} \leq d_2 \leq \frac{b_2 + l_1}{r_2}$, and $d_1 \in [\frac{(b_1 + l_2)\left( d_2 - \frac{b_2 - l_2}{r_2} \right)}{b_2 + l_1 - r_2 d_2},  \frac{r_2 (l_1 + l_2)}{r_1 (r_1 + r_2)} + \frac{b_1 + b_2 - d_2 r_2}{r_1});$
\item $R = \frac{b_1 + l_2}{d_1} + r_2 > \max \left\{r_1 + r_2, \frac{b_2 + l_1}{d_2}\right\}$, 
\subitem $\circ$ when $d_2 < \frac{l_1 + b_2}{r_2}$ and $d_1 < \min \left\{\frac{b_1 + l_2}{r_1}, \frac{(b_1 + l_2)\left( d_2 - \frac{b_2 - l_2}{r_2}\right)}{l_1 + b_2 - r_2 d_2} \right\}$;
\subitem $\circ$ when $d_2 \geq \frac{l_1 + b_2}{r_2}$ and $d_1 < \frac{b_1 + l_2}{r_1}.$
\item $R = R^*_0$, otherwise.
\end{itemize}

Next we characterize the global optimum $R^*_1$. We considering three cases: $d_2 \geq \frac{b_2 + l_1}{r_2}$, $\frac{b_2 + l_1}{r_1 + r_2} < d_2 < \frac{b_2 + l_1}{r_2}$, and $d_2 \leq \frac{b_2 + l_1}{r_1 + r_2}$.
\begin{itemize}
\item When $d_2 \geq \frac{b_2 + l_1}{r_2}$, we consider whether $\frac{b_1 + b_2}{r_1 + r_2} + \frac{(l_1 + l_2)r_2}{(r_1 + r_2)^2} - \frac{l_2 + b_1}{r_1} \geq 0$ or not. 

$\circ$ When $\frac{b_1 + b_2}{r_1 + r_2} + \frac{r_2(l_1 + l_2)}{(r_1 + r_2)^2} - \frac{l_2 + b_1}{r_1} \geq 0$, \ie $r_1 r_2 (r_1 + r_2)\left(\frac{b_2 - l_2}{r_2} - \frac{b_1 - l_1}{r_1} \right) \geq r^2_2(l_1 + l_2)$, basic algebraic manipulation gives that for all $d_2 \geq \frac{b_1 + b_2}{r_1 + r_2} + \frac{r_2(l_1 + l_2)}{(r_1 + r_2)^2}$, $R = r_1 + r_2$. 
As $\frac{b_1 + b_2}{r_1 + r_2} + \frac{r_2(l_1 + l_2)}{(r_1 + r_2)^2} - \frac{l_1 + b_2}{r_2} = \frac{r_1 r_2}{r_2(r_2 + r_2)} \left(\frac{b_1 - l_1}{r_1} - \frac{b_2 - l_2}{r_2} \right) + \frac{(r^2_2 - r^2_1 - r_1 r_2)(l_1 + l_2)}{r_2(r_1 + r_2)^2} \leq$ 
%\frac{-r_2^2(l_1 + l_2) +(r^2_2 - r^2_1 - r_1 r_2)(l_1 + l_2)}{r_2(r_1 + r_2)^2} = 
$-r_1(r_1 + r_2)(l_1 + l_2) < 0$, 
we have $\frac{l_2 + b_1}{r_2} > \frac{b_1 + b_2}{r_1 + r_2} + \frac{r_2(l_1 + l_2)}{(r_1 + r_2)^2}$. Therefore, we have $R^*_1 = r_1 + r_2$.

$\circ$ When $\frac{b_1 + b_2}{r_1 + r_2} + \frac{(l_1 + l_2)r_2}{(r_1 + r_2)^2} - \frac{l_2 + b_1}{r_1} < 0$, from the local optimums we have: 
\subitem $\circ$ when $\frac{l_2 + b_1}{r_1} \leq \frac{l_1 + b_2}{r_2}$, $R^*_1 = r_1 + r_2$;
\subitem $\circ$  when $\frac{l_2 + b_1}{r_1} > \frac{l_1 + b_2}{r_2}$, $R^*_1 = r_1 + r_2$ when $d_1 \geq \frac{l_2 + b_1}{r_1}$, while $R^*_1 = \frac{b_1 + l_2}{d_1} + r_2$ when $d_1 < \frac{l_2 + b_1}{r_1}$.

\item When $\frac{b_2 + l_1}{r_1 + r_2} < d_2 < \frac{b_2 + l_1}{r_2}$, we separate it into three conditions: $d_1 \geq \frac{b_1 + b_2}{r_1 + r_2} + \frac{r_2(l_1 + l_2)}{(r_1 + r_2)^2}$, $d_1 \in (\frac{b_1 + b_2 - d_2r_2}{r_1} + \frac{r_2(l_1 + l_2)}{r_1(r_1 + r_2)}, \frac{b_1 + b_2}{r_1 + r_2} + \frac{r_2(l_1 + l_2)}{(r_1 + r_2)^2})$, and $d_1 < \frac{b_1 + b_2 - d_2r_2}{r_1} + \frac{r_2(l_1 + l_2)}{r_1(r_1 + r_2)}$.

$\circ$ When $d_1 \geq \frac{b_1 + b_2}{r_1 + r_2} + \frac{r_2(l_1 + l_2)}{(r_1 + r_2)^2}$, we have $\frac{b_2 + l_1}{r_1 + r_2} > \frac{b_1 + b_2 - d_1 r_1}{r_2} + \frac{l_1 + l_2}{r_1 + r_2} < \frac{b_2 + l_1}{r_1 + r_2} < d_2$. Therefore, we have $R^*_1 = r_1 + r_2$.

$\circ$ When $d_1 \in (\frac{b_1 + b_2 - d_2r_2}{r_1} + \frac{r_2(l_1 + l_2)}{r_1(r_1 + r_2)}, \frac{b_1 + b_2}{r_1 + r_2} + \frac{r_2(l_1 + l_2)}{(r_1 + r_2)^2})$, basic algebraic manipulation gives $d_1 > \frac{b_1 + b_2 - d_2 r_2}{r_1} + \frac{r_2 (l_1 + l_2)}{r_1(r_1 + r_2)} > \frac{b_2 + l_1}{r_1 + r_2}$. Therefore, we have $R^*_1 = r_1 + r_2$.

$\circ$ When $d_1 < \frac{b_1 + b_2 - d_2r_2}{r_1} + \frac{r_2(l_1 + l_2)}{r_1(r_1 + r_2)}$, from the local optimums we directly have:
\subitem $\circ$ when $d_2 \geq \frac{l_1 + b_2}{r_1 + r_2} + \frac{r_1 (b_2 - l_2)}{r_2 (r_1 + r_2)}$, $R^*_1 = \frac{b_1 + l_2}{d_1} + r_2$.
\subitem $\circ$ when $d_2 < \frac{l_1 + b_2}{r_1 + r_2} + \frac{r_1 (b_2 - l_2)}{r_2 (r_1 + r_2)}$, $R^*_1 = \frac{(d_1 - d_2)r_2 + (b_1 + b_2) + \sqrt{((d_1 - d_2)r_2 + b_1 + b_2)^2 + 4 d_1 r_2 (l_1 + l_2)}}{2 d_1}$ when 
$\quad d_1 \in [\frac{(b_1 + l_2)\left(d_2 - \frac{b_2 - l_2}{r_2} \right)}{b_2 + l_1 - r_2 d_2}, \frac{b_1 + b_2 - d_2r_2}{r_1} + \frac{r_2(l_1 + l_2)}{r_1(r_1 + r_2)})$, while $R^*_1 = \frac{b_1 + l_2}{d_1} + r_2$ when $d_1 < \frac{(b_1 + l_2)\left(d_2 - \frac{b_2 - l_2}{r_2} \right)}{b_2 + l_1 - r_2 d_2}$.

\item When $d_2 \leq \frac{b_2 + l_1}{r_1 + r_2}$, we have $\frac{b_1 + l_2}{r_1}> \frac{(b_2 + l_2) \left(d_2 - \frac{b_2 - l_2}{r_2} \right)}{l_1 + b_2 - r_2 d_2}$. Therefore, from local optimums we directly have:

$\circ$ when $d_1 \in (d_2, \frac{(b_2 + l_2) \left(d_2 - \frac{b_2 - l_2}{r_2} \right)}{l_1 + b_2 - r_2 d_2}]$, $R^*_1 = \frac{b_1 + l_2}{d_1} + r_2$;

$\circ$ when $d_1 \in [\frac{(b_2 + l_2) \left(d_2 - \frac{b_2 - l_2}{r_2} \right)}{l_1 + b_2 - r_2 d_2}, \frac{d_2(b_1 + l_2)}{b_2 + l_1 - d_2 r_2} + \frac{d_2(b_2 - l_2)}{b_2 + l_1})$, 

$ \quad \quad \quad R^*_1 = \frac{(d_1 - d_2)r_2 + (b_1 + b_2) + \sqrt{((d_1 - d_2)r_2 + b_1 + b_2)^2 + 4 d_1 r_2 (l_1 + l_2)}}{2 d_1}$;

$\circ$ when $d_1 \in ( \frac{d_2(b_1 + l_2)}{b_2 + l_1 - d_2 r_2} + \frac{d_2(b_2 - l_2)}{b_2 + l_1}), \frac{d_2(b_1 + b_2)}{b_2 + l_1 - d_2 r_2})$, $R^*_1 = \frac{b_2 + l_1}{d_2}$.

\end{itemize}

\textbf{\textit{Remark:}} The above analysis also shows:  $R^*_1 > R^*_0$ as long as $R^*_0 = \frac{b_1 + b_2}{d_1} + r_2 > \max\left\{\frac{b_2 + l_1}{d_2}, r_1 + r_2 \right\}$. 

In summary, the global optimum is:
\begin{itemize}
\item $R^*_1 = r_1 + r_2, \text{ when } d_1 \in [\frac{l_2 + b_1}{r_1}, \frac{b_1 + b_2}{r_1}) \text{ and }
d_2 \geq \max \left\{\frac{b_2 + l_1}{r_1 + r_2}, \frac{b_1 + b_2 - d_1 r_1}{r_2} + \frac{l_1 + l_2}{r_1 + r_2} \right\};$
\item $R^*_1 = \frac{b_2 + l_1}{d_2} \geq r_1 + r_2, \text{ when } d_2 < \frac{b_2 + l_1}{r_1 + r_2} \text{ and } d_1 \in [\frac{d_2(b_1 + l_2)}{b_2 + l_1 - d_2 r_2} + \frac{d_2(b_2 - l_2)}{b_2 + l_1}, \frac{d_2(b_1 + b_2)}{b_2 + l_1 - d_2 r_2}); $
\item $R^*_1 = \frac{l_2 + b_1}{d_1}+r_2$, when $d_2 < \frac{l_1 + b_2}{r_2}$ and $d_1 < \min \left\{\frac{b_1 + l_2}{r_1}, \frac{(b_1 + l_2)\left( d_2 - \frac{b_2 - l_2}{r_2}\right)}{l_1 + b_2 - r_2 d_2} \right\}$; and when $d_2 \geq \frac{l_1 + b_2}{r_2}$ and $d_1 < \frac{b_1 + l_2}{r_1}$;
\item $R^*_1 = \frac{(d_1 - d_2)r_2 + (b_1 + b_2) + \sqrt{((d_1 - d_2)r_2 + b_1 + b_2)^2 + 4 d_1 r_2 (l_1 + l_2)}}{2 d_1}$, 
\subitem $\circ$ when $d_2 < \frac{b_2 + l_1}{r_1 + r_2}, $ and $d_1 \in [\frac{(b_1 + l_2)\left( d_2 - \frac{b_2 - l_2}{r_2} \right)}{b_2 + l_1 - r_2 d_2},\frac{d_2(b_1 + l_2)}{b_2 + l_1 - d_2 r_2} + \frac{d_2(b_2 - l_2)}{b_2 + l_1});$ and
\subitem $\circ$ when $\frac{b_2 + l_1}{r_1 + r_2} \leq d_2 \leq \frac{b_2 + l_1}{r_2}$, and $d_1 \in [\frac{(b_1 + l_2)\left( d_2 - \frac{b_2 - l_2}{r_2} \right)}{b_2 + l_1 - r_2 d_2}, \frac{r_2 (l_1 + l_2)}{r_1 (r_1 + r_2)} + \frac{b_1 + b_2 - d_2 r_2}{r_1}).$
\end{itemize}
\end{proof}

%%%%%%%%%%%%%%%%%%%%%%%%%%%%%%%%%%%
%%%%%%%% sub-optimization 2 %%%%%%%%
%%%%%%%%%%%%%%%%%%%%%%%%%%%%%%%%%%%
\subsubsection*{Solution for Sub-optimization~$\pmb{2}$}
The Lagrangian function for sub-optimization~$2$ is 
\beq
\begin{aligned}
L_2(R, r'_2, b'_2, \pmb{\lambda}) &= R + \lambda_1 (R - r_1 - r'_2) + \lambda_2 \left(\frac{b_1 + b_2}{r_1} - \frac{(R- r_1 - r_2)(b_2 - b'_2)}{r_1(r'_2 - r_2)} -d_1 \right) \\ 
&+ \lambda_3 \left(\frac{b_2 + l_1}{R} - d_2 \right)+ \lambda_4 \left(\frac{b_2 - b'_2}{r'_2}+ \frac{l_1 + l_2}{R} - d_2 \right) \\
& + \lambda_5 \left(\frac{b_1 + b'_2}{r_1} - \frac{(R-r'_2)(b_2 - b'_2)}{r_1(r'_2 - r_2)} \right)+ \lambda_6 (r_1 + r_2 -R)\\
& + \lambda_7 (r_2 -r'_2) + \lambda_8 (l_2-b'_2) + \lambda_9 (b'_2 - b_2)
\end{aligned}
\eeq

\beq\label{dev2}
\nabla_{R,r'_2, b'_2}L_2 = \begin{bmatrix} 
1+\lambda_1 - \frac{\lambda_2(b_2- b'_2)}{r_1(r'_2-r_2)^2} - \frac{\lambda_3(b_2 + l_1)}{R^2} - \frac{\lambda_4 (l_1+l_2)}{R^2} - \frac{\lambda_5(b_2 - b'_2)}{r_1 (r'_2 - r_2)} - \lambda_6 \\
\\
-\lambda_1 + \frac{\lambda_2(R-r_1-r_2)(b_2 - b'_2)}{r_1(r'_2-r_2)^2} - \frac{\lambda_4(b_2 - b'_2)}{r^{'2}_2} + \frac{\lambda_5(b_2 - b'_2)(R-r_2)}{r_1(r'_2 - r_2)^2} - \lambda_7 \\
\\
\frac{\lambda_2(R-r_1-r_2)}{r_1(r'_2-r_2)} - \frac{\lambda_4}{r'_2} + \lambda_5\left(\frac{1}{r_1} + \frac{R-r'_2}{r_1(r'_2 - r_2)} \right) - \lambda_8 + \lambda_9
\end{bmatrix}
\eeq

\beq\label{diag2}
\text{diag}(\triangle_{R,r'_2, b'_2}L_1) = \begin{bmatrix}
\frac{2\lambda_3(b_2 + l_1)}{R^3} + \frac{2 \lambda_4 (l_1+l_2)}{R^3} \\
\\
-\frac{2 \lambda_2(R-r_1-r_2)(b_2 - b'_2)}{r_1(r'_2-r_2)^3} + \frac{2 \lambda_4(b_2 - b'_2)}{r^{'3}_2} - \frac{2\lambda_5(b_2 - b'_2)(R - r_2)}{r_1 (r'_2 - r_2)^3} \\
\\
0
\end{bmatrix}
\eeq

Similar as that for suboptimization~$1$, we have $\lambda_5^* = 0$. Substitute it into \Eqref{diag2}, we have
\beq\label{diag3}
\text{diag}(\triangle_{R,r'_2, b'_2}L_1) = \begin{bmatrix}
\frac{2\lambda_3(b_2 + l_1)}{R^3} + \frac{2 \lambda_4 (l_1+l_2)}{R^3} \\
\\
-\frac{2 \lambda_2(R-r_1-r_2)(b_2 - b'_2)}{r_1(r'_2-r_2)^3} + \frac{2 \lambda_4(b_2 - b'_2)}{r^{'3}_2} \\
\\
0
\end{bmatrix}
\eeq

Next we consider KKT's necessary conditions for the optimization under $\lambda_5^* = 0$. Remember that when solving sub-optimization~$2$, we only consider $R < r_1+ r'_2$. As before, we consider only $b'_2 < b_2$. Therefore, we have:

\beq\label{kkt2}
\resizebox{.95 \textwidth}{!} 
{
$
\begin{aligned}
& 1+\lambda_1 - \frac{\lambda_2(b_2- b'_2)}{r_1(r'_2-r_2)^2} - \frac{\lambda_3(b_2 + l_1)}{R^2} - \frac{\lambda_4 (l_1+l_2)}{R^2} - \lambda_6 = 0,
& & \frac{\lambda_2(R-r_1-r_2)}{r_1(r'_2-r_2)} - \frac{\lambda_4}{r'_2} - \lambda_8 + \lambda_9 = 0,\\
& \frac{\lambda_2(R-r_1-r_2)(b_2 - b'_2)}{r_1(r'_2-r_2)^2} - \frac{\lambda_4(b_2 - b'_2)}{r^{'2}_2} = \lambda_1 + \lambda_7, 
& & \lambda_i \geq 0, \text{ for }i = 1, ...9, \\
& R - r_1 - r'_2 < 0, 
& & \lambda_1 = 0, \\
& \frac{b_1 + b_2}{r_1} - \frac{(R- r_1 - r_2)(b_2 - b'_2)}{r_1(r'_2 - r_2)} -d_1 \leq 0, 
& & \lambda_2 \left(\frac{b_1 + b_2}{r_1} - \frac{(R- r_1 - r_2)(b_2 - b'_2)}{r_1(r'_2 - r_2)} -d_1 \right) = 0, \\
& \frac{b_2 + l_1}{R} - d_2 \leq 0, 
& &\lambda_3\left( \frac{b_2 + l_1}{R} - d_2\right) = 0, \\
& \frac{b_2 - b'_2}{r'_2} + \frac{l_1 + l_2}{R} - d_2 \leq 0,
& &\lambda_4 \left(\frac{b_2 - b'_2}{r'_2} + \frac{l_1 + l_2}{R} - d_2  \right) = 0, \\
& \frac{b_1 + b'_2}{r_1} - \frac{(R-r'_2)(b_2 - b'_2)}{r_1(r'_2 - r_2)} < 0, 
& &\lambda_5 =0, \\
& r_1 + r_2 -R \leq 0,
& &\lambda_6 (r_1 + r_2 -R) = 0,\\
& r_2 -r'_2 \leq 0, 
& &\lambda_7 (r_2 -r'_2) = 0, \\
& l_2 -b'_2 \leq 0,  
& &\lambda_8 (l_2-b'_2) = 0,\\
& b'_2 - b_2 < 0 
& &\lambda_9 = 0
\end{aligned}
$}
\eeq
\begin{flushleft}
From the conditions in \Eqref{kkt2}, we have:
\begin{itemize}
\item when $d_2 < \frac{b_2 + l_1}{r_1 + r_2}$, and $d_1 \in [\frac{d_2(b_2 - l_2)}{b_2 + l_1} + \frac{b_1 + l_2}{r_1}, \frac{d_2(b_1 + b_2)}{b_2 + l_1 - d_2 r_2})$, $R^*_2 = \frac{b_2 + l_1}{d_2} < R^*_0 $;
\item otherwise, $R^*_2 = R^*_0$.
\end{itemize}
\end{flushleft}

\begin{proof}
When $R = r_1 + r_2$, from $\frac{b_1 + b_2}{r_1} - \frac{(R-r_1-r_2)(b_2 - b'_2)}{r_1(r'_2 - r_2)} - d_1 \leq 0$ we have $\frac{b_1 + b_2}{d_1} \leq r_1$, which gives $\frac{b_1 + b_2}{d_1} + r_2 \leq r_1 + r_2$. From Corollary~\ref{improve}, we know that adding reprofilers cannot decrease the minimum required bandwidth. Therefore, we consider only $R > r_1 + r_2$ afterwards. Combine $R > r_1 + r_2$ with $R < r_1 + r'_2$, we have $r'_2 > r_2$. Hence, $\lambda_6 = \lambda_7 = 0$. 

As $\lambda_1 = 0$, we have $\frac{\lambda_2(R-r_1-r_2)(b_2 - b'-2)}{r_1(r'_2-r_2)^2} - \frac{\lambda_4(b_2 - b'_2)}{\lambda_2^{'2}} = 0$. Substitute it to $\frac{\lambda_2(R-r_1-r_2)}{r_1(r'_2-r_2)} - \frac{\lambda_4}{r'_2} - \lambda_8 + \lambda_9 = 0$, we have $-\frac{\lambda_4 r_2}{r^{'2}_2} - \lambda_8 + \lambda_9 = 0$. As $\lambda_9 = 0$, \ie $b'_2 \neq b_2$, we have $\lambda_2 = \lambda_4 = \lambda_8 = 0$. As $\lambda_1 = \lambda_6 = 0$, from $1+\lambda_1 - \frac{\lambda_2(b_2- b'_2)}{r_1(r'_2-r_2)^2} - \frac{\lambda_3(b_2 + l_1)}{R^2} - \frac{\lambda_4 (l_1+l_2)}{R^2} - \lambda_6 = 0$, we have $\lambda_3 > 0$, \ie $R = \frac{b_2 + l_1}{d_2}$. Note that when $\lambda_2 = \lambda_4 = 0$ and $\lambda_3 > 0$, from \Eqref{diag2}, we have $\text{diag}(\triangle_{R,r'_2, b'_2}L_1) \geq 0$. Hence, $R = \frac{b_2 + l_1}{d_2}$ is a local optimum. 

Hence, the constraints of the optimization reduce to 
\beq
\left\{
\begin{array}{ll}
r'_2 \in (R-r_1, R), \qquad b'_2 \in [l_2, b_2) \\
\\
\frac{b_1 + b_2}{r_1} - \frac{(R-r_1-r_2)(b_2-b'_2)}{r_1(r'_2-r_2)} \leq d_1 \implies \frac{b_2 - b'_2}{r'_2 - r_2} \geq \frac{b_1 + b_2 - d_1 r_1}{R-r_1-r_2}, \\
\\
\frac{b_2 - b'_2}{r'_2} + \frac{l_1 + l_2}{R} \leq d_2 = \frac{b_2 + l_1}{R}
\implies b_2 - b'_2 \leq \frac{r'_2(b_2 - l_2)}{R}, \\
\\
\frac{b_1 + b'_2}{r_1} - \frac{(R-r'_2)(b_2-b'_2)}{r_1(r'_2 - r_2)} < 0 \implies \frac{b_2 - b'_2}{r'_2 - r_2} > \frac{b_1 + b_2}{R-r_2},
\end{array}
\right.
\eeq

Basic algebraic manipulation gives $\frac{b_1 + b_2}{R-r_2} > \frac{b_1 + b_2 - d_1 r_1}{R-r_1-r_2}$ iff $R > \frac{b_1 + b_2}{d_1} + r_2 = R^*_0$. Hence, we only consider the case where  $\frac{b_1 + b_2}{R-r_2} < \frac{b_1 + b_2 - d_1 r_1}{R-r_1-r_2}$. Under such a condition, we have $b_2 - b'_2 \in [\frac{(b_1 + b_2 - d_1 r_1)(r'_2 - r_2)}{R-r_1-r_2}, \frac{r'_2(b_2 - l_2)}{R}]$.

Next we configure the conditions where: $$\exists \ r'_2 \in (R-r_1, R), \text{ s.t. } \mathalpha{S} := [\frac{(b_1 + b_2 - d_1 r_1)(r'_2 - r_2)}{R-r_1-r_2}, \frac{r'_2(b_2 - l_2)}{R}] \cap (0, b_2 - l_2] \neq \emptyset.$$
When $R^* < R^*_0$, it has $\frac{b_1 + b_2}{d_1} + r_2 > r_1 + r_2$, \ie $b_1 + b_2 - d_1r_1 > 0$. Hence we have $\frac{(b_1 + b_2 - d_1 r_1)(r'_2 - r_2)}{R-r_1-r_2} > 0$. From $r'_2 < R$, we have $\frac{r'_2(b_2 - l_2)}{R} < b_2 -l_2$. Therefore, $\mathalpha{S} \neq \emptyset$ is equivalent to 
$$\exists \ r'_2 \in (R-r_1, R) \text{ s.t. } \frac{(b_1 + b_2 - d_1 r_1)(r'_2 - r_2)}{R-r_1-r_2} \leq \frac{r'_2(b_2 - l_2)}{R}.$$
Define
$$g(x) = \frac{x(b_2 - l_2)}{R}- \frac{(b_1 + b_2 - d_1 r_1)(x - r_2)}{R-r_1-r_2}.$$
As $g(x)$ is linear w.r.t $x$, $\mathalpha{S} \neq \emptyset$ is equivalent to at least one of $g(R)$ and $g(R-r_1)$ is non-negative, which gives $d_1 > \frac{b_1 + l_2}{r_1}$ and $\frac{b_2 + l_1}{d_2} >  \frac{r_1 (b_2 - l_2)}{d_1 r_1 - b_1 - l_2}$ through basic algebraic manipulations. Combine it with $\frac{b_1 + b_2}{d_1} + r_2 > \frac{b_2 + l_1}{d_2} \geq r_1 + r_2$, we have:
$$\circ \ d_2 < \frac{b_2 + l_1}{r_1 + r_2}, \text{ and } d_1 \in [\frac{d_2(b_2 - l_2)}{b_2 + l_1} + \frac{b_1 + l_2}{r_1}, \frac{d_2(b_1 + b_2)}{b_2 + l_1 - d_2 r_2}).$$
\end{proof}

\twocolumn

\end{document}